\documentclass[prd,showpacs,preprintnumbers,amsmath,amssymb,nofootinbib]{revtex4}
\usepackage[T1]{fontenc} % if needed
\usepackage{hyperref}
\usepackage{latexsym}
\usepackage{graphicx}
\usepackage{color}

\newcommand{\be}{\begin{equation}}
\newcommand{\ee}{\end{equation}}
\newcommand{\bes}{\begin{equation*}}
\newcommand{\ees}{\end{equation*}}
\newcommand{\ba}{\begin{eqnarray}}
\newcommand{\ea}{\end{eqnarray}}
\newcommand{\ar}{\arrowvert} 
\newcommand{\ra}{\rangle}

\newcommand{\tr}[1]{\textrm{#1}}
\newcommand{\bw}{\begin{widetext}}
\newcommand{\ew}{\end{widetext}}

\newcommand{\Imag}{\mathop{\mathrm{Im}}}
\newcommand{\Real}{\mathop{\mathrm{Re}}}

\begin{document}
\title{Unitarity, analyticity, dispersion relations and resonances in strongly interacting \boldmath $W_L W_L$, $Z_L Z_L$ and $hh$ scattering}

\author{Rafael L. Delgado, Antonio Dobado and Felipe J. Llanes-Estrada}
\affiliation{Departamento de F\'isica Te\'orica I, Universidad Complutense de Madrid, 28040 Madrid, Spain}
\begin{abstract}
If the Electroweak Symmetry Breaking Sector turns out to be strongly interacting, the actively investigated effective theory for longitudinal gauge bosons plus Higgs can be efficiently extended to cover the regime of saturation of unitarity (where the perturbative expansion breaks down). This is achieved by dispersion relations, whose subtraction constants and left cut contribution can be approximately obtained in different ways giving rise to different unitarization procedures. We illustrate the ideas with the Inverse Amplitude Method, one version of the N/D method and another improved version of the K-matrix. In the three cases we get partial waves which are unitary, analytical with the proper left and right cuts and in some cases poles in the second Riemann sheet that can be understood as dynamically generated resonances. In addition they reproduce at Next to Leading Order (NLO) the perturbative expansion for the five partial waves not vanishing (up to J=2) and they are renormalization scale ($\mu$) independent. Also the unitarization formalisms are extended to the coupled channel case. Then we apply the results to the elastic scattering amplitude for the longitudinal components of the gauge bosons $V=W, Z$ at high energy. We also compute $h h \rightarrow h h$ and the inelastic process $VV\rightarrow h h$ which are coupled to the elastic $VV$ channel for custodial isospin $I=0$. We numerically compare the three methods for various values of the low-energy couplings and explain the reasons for the differences found in the $I=J=1$ partial wave. Then we study the resonances appearing in the different elastic and coupled channels in terms of the effective Lagrangian parameters.
\end{abstract}
\maketitle
% * <fllanes@fis.ucm.es> 2015-01-26T11:17:09.437Z:
%
% 
%
%%%%%%%%%%%%%%%%%%%%%%%%%%%%%%%%%%%%%%%%%%%%%%%%%%%%%%%%%%%%%%%%%%%%%%%%%%%%
\section{Introduction}
% * <fllanes@fis.ucm.es> 2015-01-26T11:17:28.174Z:
%
% 
%
% * <fllanes@fis.ucm.es> 2015-01-26T11:17:16.267Z:
%
% 
%
%%%%%%%%%%%%%%%%%%%%%%%%%%%%%%%%%%%%%%%%%%%%%%%%%%%%%%%%%%%%%%%%%%%%%%%%%%%%

The most outstanding discovery in particle physics during the last years is probably the finding that the LHC collaborations ATLAS~\cite{ATLAS} and CMS~\cite{CMS}  published in 2012 announcing  a new boson with scalar quantum numbers and couplings compatible with those of a Standard Model Higgs at about 125 GeV. However the first LHC run finished without any other finding~\cite{searches} up to an energy of 600-700 GeV (and higher yet for additional vector bosons). This lightness respect to any new physics could alternatively suggest that the Higgs is indeed an additional Goldstone boson (together with those giving rise to the $W$ and $Z$ masses) related with some global spontaneous symmetry breaking triggering the $SU(2)_L\times U(1)_Y\to U(1)_{em}$ gauge symmetry breaking \cite{GB}.
 
If that were the case, some effective description of the Symmetry Breaking Sector (SBS) of the Standard Model (SM) would be appropriate (see for example \cite{Englert:2015oga,TaoHan,Espriu:2013fia,Azatov:2012bz,Brivio:2013pma,Alonso:2012px,Pich:2013fba,Jenkins:2013zja,Degrande:2012wf,Buchalla:2013rka,Buchalla:2012qq}). The presence of that energy gap is also suggestive of a non-linear realization (the most general approach to the effective theory). The old Electroweak Chiral Lagrangian (ECL) \cite{Appelquist} technique, based on standard Chiral Perturbation Theory (ChPT) for QCD~\cite{ChPT}, can be extended to include the new found Higgs-like light particle $h$ as a scalar singlet.

In a recent work~\cite{Delgado:2013loa} we have shown that, for essentially any parameter choice except that of the Standard Model and perhaps other very carefully tuned sets, the interactions will generically become strong at sufficiently high energy, and have argued that a second, very broad scalar pole is expected. In  a more recent work \cite{Delgado:2013hxa} we performed the one-loop computation of the two-body scattering amplitudes among the $\omega$ Goldstone bosons and the $h$ scalar by using a generic effective Lagrangian, in the kinematic regime $M_h^2\ll s<4\pi v\simeq 3\,{\rm TeV}$.

In spite of the success of one-loop computations (for example in ordinary ChPT) it is clear that it can only be useful at very low energies. Moreover, the case in point of the ECL deals with the would-be Goldstone bosons $\omega$ that are eventually related to the longitudinal components of the gauge bosons only through the Equivalence Theorem (ET) \cite{ET}, which is valid only in the kinematic regime $M_W^2\sim M_Z^2\sim M_h^2\ll s$, that corresponds to a high energy limit. Thus even at low energies one could not expect the truncated series to apply in a context of strong interactions. This situation is not improved in any significant way by going further in the chiral expansion by computing two or more loops. Going to higher orders one has to deal with a very fast increasing number of chiral couplings and extremely complicated computations. 

On the other hand one can try a different strategy to extend the low energy regime by using dispersion relations (DR) compatible with analyticity and unitarity. This program has proved to be extremely useful in the original ChPT as applied to low energy hadron dynamics and it is quite possible that it could also be useful for the SBS of the strongly interacting sector of the SM if properly applied. This program is bringing much attention, because some predictions can be checked at the LHC Run-II~\cite{Englert:2015oga}, as we will show here for the one-loop computation. However, the use of unitarization methods for extending the applicability of ChPT has been criticized because they are considered arbitrary in some sense, and the results they provide depend on the particular method considered.

In this work we will try to show that the one-loop results, when properly complemented with DR, can provide an analytical and unitary description of higher energy dynamics which is essentially unique qualitatively; at least so up to the first resonances which can also be described as poles in second Riemann sheet due to the proper analytical behaviour of the amplitudes. The rationale for this is that any physically sensible amplitude must fulfill the appropriate DRs which are typically integral equations. In principle those equations have many solutions. However one can impose some particular dynamics by performing subtractions on the DR and fixing the values of the subtraction constants. In our case it is clear that these constants must be obtained from the effective Lagrangian and so we will introduce the different dynamics compatible with the low-energy expansion in the DR relation with an appropriate number of subtractions. In this way one expects to reduce enormously the space of possible solutions of the integral equations, at least up to the first poles. If this is really the case, different unitarization methods will provide qualitatively similar results and the differences could be attributed to the different approximations used for solving the DR equations. 
%In order to have some hint about this possibility we will consider in the following different unitarisation methods based on DR. In particular 

We study in detail the Inverse Amplitude Method (IAM, formerly called the Pad\'e method) \cite{Truong:1988zp,Dobado:1992ha}, the N/D \cite{N/D} method as applied to the SBS of the SM and also the so called improved K-matrix method (see \cite{Kmatrix},\cite{Dicus} and \cite{Kilian} for exposition and some uses of the non-improved method). The main novelty in the case of the electroweak chiral descriptions compared with standard ChPT applied to hadrons is that now the GB are really massless. We then pay attention, not only to the usual ultraviolet (UV) divergences appearing in the DR integrals but also to the infrared (IR) ones. 

Thus we rederive the IAM method for massless particles using a twice-subtracted DR instead of the original derivation that used three subtractions \cite{Dobado:1992ha}. For our construction of the N/D method we introduce a renormalization-scale invariant splitting of the NLO amplitude into a left and a right parts each bearing the corresponding left and right cuts. Then we write a thrice-subtracted DR for the denominator function and solve the corresponding integral equation by iteration (in fact one is sufficient to get a sensible result). Finally we consider also an improved version of the K-matrix method \cite{Dobado:2001rv,Delgado:2013loa} that produces partial waves having a proper RC that allows for analytic continuation to the second Riemann sheet in the search for poles (resonances), thus fixing the typical absence of these poles of this unitarization method.

We follow the natural order of presentation, with the Effective Lagrangian briefly recounted in subsec.~\ref{sec:eff_Lag}, followed by a short discussion on the elastic and inelastic scattering amplitudes in subsec.~\ref{subsec:amplitudes}. A part of this work can be found in our earlier article\cite{Delgado:2013hxa}. We have however calculated the fifth, non-vanishing NLO amplitude with angular momentum and custodial isospin 2, a new result not commonly quoted in analogous hadron systems. Thus, we have now exhausted the massless low-energy Lagrangian to NLO, by computing all nonvanishing channels and interchannel couplings. The calculated amplitude coefficients, their behavior under scale changes, and our conventions for the partial wave amplitudes, are all given in the appendix to make this subsection more readable.

We dedicate section~\ref{sec:IAM} to the Inverse Amplitude Method, both for single and coupled channels, and provide a new derivation based on twice-subtracted dispersion relations especially useful for massless particles.
Section~\ref{sec:NoverD} in turn is dedicated to generically describing the N/D method, but also to construct a new approximate solution thereof that has the same desirable physical properties of the IAM (and unsurprisingly, both coincide where both are applicable).
More so, we explore the Improved-K matrix method in subsection~\ref{subsec:K} and compare all three methods extensively. 

The computer evaluations of all three methods are exposed in sections~\ref{sec:numericcomp} and~\ref{sec:sysIAM} and we conclude that the unitarization methods are qualitatively robust and a reliable guide in the search for strongly-interacting new physics, with little model dependence. 

After a terse summary in section~\ref{sec:summary}, we dedicate four appendices to technical details of the perturbative amplitude calculations, their partial waves, the coupled-channel IAM, the (one-iteration) solution of the N/D method, and the numerical extraction of poles in the complex plane (resonances if in the second Riemann sheet or tachyons if in the first Riemann sheet).

%%%%%%%%%%%%%%%%%%%%%%%%%%%%%%%%%%%%%%%%%%%%%%%%%%%%%%%%%%%%%%%%%%%%%%%%%%%%
\section{Electroweak Chiral Lagrangian and scattering amplitudes with a Higgs} 
%%%%%%%%%%%%%%%%%%%%%%%%%%%%%%%%%%%%%%%%%%%%%%%%%%%%%%%%%%%%%%%%%%%%%%%%%%%%
We have already presented the Lagrangian density and perturbative LO amplitudes in~\cite{Delgado:2013loa} for $VV$ and $hh$ scattering. Here we quickly remind of the basic equations with reduced discussion and settle to a more standard notation than we previously used. Also in \cite{Delgado:2013hxa} we obtained the one-loop scattering amplitudes between the (massless) $\omega$ and $h$. This section is divided in two subsections, one~\ref{sec:eff_Lag} dedicated to exposing the effective Lagrangian and another subsection~\ref{subsec:amplitudes} dealing with the scattering amplitudes. Some further detail is relegated to appendix~\ref{app:L}.

%%%%%%%%%%%%%%%%%%%%%%%%%%%%%%%%%%%%%%%%%%%%%%%%%%%%%%%%%%%%%%%%%%%%%%%%%%%%
\subsection{Effective Lagrangian\label{sec:eff_Lag}}
%%%%%%%%%%%%%%%%%%%%%%%%%%%%%%%%%%%%%%%%%%%%%%%%%%%%%%%%%%%%%%%%%%%%%%%%%%%%

There are several equivalent forms of the universal Electroweak Chiral Lagrangian employing only the experimentally known particles. At leading order we adopt the gauged $SU(2)_L \times SU(2)_R/SU(2)_C = SU(2) \simeq S^3$ Non-linear Sigma Model (NLSM) coupled to a scalar field $h$ as
\be 
\label{genericLagrangian}
{\cal L}_0=\frac{v^2}{4}{\cal F}(h)\tr(D_\mu U)^\dag
D^\mu U+\frac{1}{2}\partial_\mu h \partial^\mu
h-V(h) 
\ee 
with $U=\sqrt{1-\tilde \omega^2/v^2}+i\tilde\omega/v$; $\tilde\omega = \omega_a\tau^a$ parameterizing the would-be Goldstone boson (WBGB) field. 
Since we will neglect the coupling to transverse gauge bosons, $D_\mu\simeq \partial_\mu$ in this article's computations.

The constant $v$ is well known from Fermi's weak constant, $v^2:=1/(\sqrt{2}G_F)=(246\,{\rm GeV})^2$.  
The scalar field interacts through ${\cal F}$, an arbitrary analytical function; in the effective-theory approach we need only the first terms of its Taylor expansion
\be \label{gexpansion}
{\cal F}(h)= 1 + 2a\frac{h}{v} + b\left(\frac{h}{v}\right)^2 + \dots
\ee 
widely used in the literature%
\footnote{%
One can alternatively employ
$
{\cal F}(h)= 1 + 2\alpha\frac{h}{f} +\beta\left(\frac{h}{f}\right)^2+\dots
$,
where $f$ is an arbitrary, new-physics energy scale, as we have done in recent work. This is perhaps more natural if the Higgs happens to be the Goldstone boson of a higher symmetry broken at the scale $f$, but in this article we adopt the more widely used convention of employing $v$, the SM symmetry-breaking scale. Obviously $a=\alpha v/f$ and $b=\beta v^2/v^2$.}.

Reference~\cite{bounds} provides some recent experimental bounds on $a$ and $b$ that we employed in~\cite{Delgado:2013loa}. Finally $V$ is an arbitrary analytical potential for the scalar field, that is of no further reference in this work,
\ba \label{potential}
V(h) =  \sum _{n=0}^{\infty}V_n h^n
\equiv  V_0 + 
\frac{1}{2}M_h^2 h^2 + d_3\frac{M_h^2}{2v} h^3 + d_4 \frac{M_h^2}{8 v^2}h^4 + \dots 
\ea

At NLO in the chiral expansion we need to add the four-derivative terms
\begin{eqnarray} \label{higherorderL}
{\cal L}_4 & = &  a_4(tr V_\mu V_\nu)^2 +  a_5(tr V_\mu V^\mu)^2 \\ \nonumber
 & + &\frac{g}{v^4} (\partial_\mu h \partial^\mu h)^2
 +  \frac{d}{v^2} (\partial_\mu h \partial^\mu h)tr(D_\nu U)^\dag D^\nu U
+\frac{e}{v^2} (\partial_\mu h \partial^\nu h)tr(D^\mu U)^\dag D_\nu U+...
\end{eqnarray}
where $V_\mu= D_\mu U U^\dagger$. We have explicitly written only the five terms strictly needed to renormalize the one-loop elastic WBGB scattering amplitudes (for $s\gg M_W^2$) and the coupled-channel processes $\omega\omega \rightarrow hh$ and $hh\rightarrow hh$. These terms produce additional contributions to the amplitudes which are of order $s^2$. 

The $a_4$ and $a_5$ chiral parameters multiply the operators $O_{D1}$ and $O_{D2}$ in the classification of~\cite{Buchalla:2013rka}. Those, as well as the additional ones $g$, $d$ and $e$, encode the dependence on the possible underlying dynamics triggering the spontaneous symmetry breaking of electroweak interactions. They all vanish in the MSM. The operators with coefficient $d$ and $e$ correspond to $O_1$ and $O_2$ as classified by Azatov {\it et al.}~\cite{Azatov:2012bz}, or to $P_{19}$ and $P_{20}$ in~\cite{Brivio:2013pma}, and are NLO equivalent to $O_{D7}$, $O_{D8}$ in~\cite{Buchalla:2013rka}
\footnote{The two operators multiplying $d$ and $e$ are of dimension 6 in what would concern transverse gauge-boson inelastic scattering $W_TW_T\to hh$, but are of dimension 8 for the longitudinal ones, as seen upon expanding $U$ as we will show shortly in Eq.~(\ref{bosonLagrangian}).
}.%end footnote 

The operator associated with $g$ is denoted as $P_H$ in~\cite{Brivio:2013pma} and $O_{D11}$ in~\cite{Buchalla:2013rka}. 
In line with recent literature~\cite{Brivio:2013pma,Contino:2011np}, we are not much interested in the dynamics of the pure Higgs sector, since the process $hh\to hh$ will hardly be measured in the foreseeable future. But since the NLO renormalization of our effective Lagrangian requires this one operator, we will assess its numeric effect on the $\omega\omega$ channel in figure~\ref{fig:ga} below, where we see that for it to be sizeable the values of $g$ have to be quite unnaturally large.

We have given the renormalization of these operators in~\cite{Delgado:2013hxa}, and we rewrite it in the new notation in appendix~\ref{sec:renorm}. An off-shell analysis that covers a larger number of operators has also recently appeared~\cite{Gavela:2014uta}.

The Lagrangian in Eq.~(\ref{genericLagrangian}) and Eq.~(\ref{higherorderL}) contains the more general low-energy physics of the ESBS for any conceivable dynamics having at least an approximate $SU(2)$ custodial isospin symmetry in the limit $g=g'=0$. 

The easiest example is the Minimal SM (MSM) \cite{GWS}, which corresponds to the parameter selection $a=b=1$ and $a_4=a_5=g=d=e=0$. 
The Higgs field $H$ is just the scalar field $h$, so that $M_H^2=M_h^2= 2\lambda v^2$, and the scalar self-couplings are $d_3= \lambda v$, $d_4 = \lambda/4$ (both proportional to $M_h^2$) and $d_i=0$ for $i \ge 4$. 

Another interesting class of models are the dilaton models~\cite{Grinstein} where $h$ would represent the dilaton field and we have $a^2=b=v^2/f^2$ with $f$ being the scale of the scale symmetry breaking. The potential and NLO parameters depend on the particular dilaton model but in any case the $d_i$ are also of order $M^2_h$ for any $i$. 

Third, the popular $SO(5)/SO(4)$ Minimally Composite Higgs Model~\cite{SO(5)} (MCHM) also provides an example where $a^2= 1-v^2/f^2$ and $b=1-2v^2/f^2$ (where $f$ is in this case the scale of the $SO(5)/SO(4)$ symmetry breaking) 
while the scalar-boson self-couplings $d_i$, contingent on model details, are of order $M_h^2$ too.

Finally it is also possible to reproduce the old Higgsless Electroweak Chiral Lagrangian (EWChL) in \cite{Appelquist} by the simple parameter choice $a=b=g=d=e=0$.  

To address the high-energy (i.~e., for $\sqrt{s}\gg 100\,{\rm GeV}$) elastic scattering of the longitudinal components of the electroweak bosons, we can apply the ET~\cite{ET}, which relates the WBGB amplitudes with the corresponding amplitudes involving longitudinal components on the electroweak bosons at high energies. For example one has:
\be
T(W_L^aW_L^b \rightarrow W_L^cW_L^d ) =
T(\omega^a\omega^b \rightarrow \omega^c\omega^d )  
 + O\left(\frac{M_W}{\sqrt{s}}\right)\ , 
\ee 
Thus the ET allows to carry the computations out with the simpler WBGB dynamics. This theorem applies for any renormalizable gauge, but for the Landau gauge (where the WBGB are formally massless) it is especially useful and transparent.  Since the transverse degrees of freedom are weakly coupled to the longitudinal sector, to explore just the latter we will set $g=g'=0$.

The remaining active degrees of freedom are the massless (Landau-gauge) WBGB, and the Higgs-like scalar $h$ that will be considered massless in the following as we are interested in the high energy region. According to ATLAS and CMS, $M_h \simeq 125$ GeV. Then $M_h \sim M_W \sim M_Z \sim 100\,{\rm GeV}$ and consistency requires to consider the massless $h$ limit, i.e., $M_h \simeq 0$ if one is only interested in the energy region where the ET can be applied. Consequently we concentrate on WBGB scattering for $M^2_h, M^2_W, M^2_Z \simeq 0\ll s < \Lambda^2$ where $\Lambda$ is some ultraviolet (UV) cutoff of about $3\,{\rm TeV}$, setting the limits of applicability of the effective theory. 

According to the results of LHC Run-I~\cite{searches,Datos}, no new physics has been discovered up to an energy of about 600-700~GeV. However, the center-of-mass energy of the LHC is going to be increased from 7-8~TeV to 13-14~TeV (and the luminosity will be much higher) at Run-II. Thus, the applicability limit $M^2_h\ll s < (3\,{\rm TeV})^2$ of the theory is within the new range of energy. Actually, LHC Run-II is going to be a great opportunity to check strongly interacting EWSBS theories controlled by unitarity~\cite{Englert:2015oga}.

We will also assume that the $d_i$ self-potential parameters of the $h$-scalar are of order $M_h^2$ so that we can neglect them altogether, as is natural in the three particular models just mentioned, MSM, Dilaton, and CHM. 

Under these kinematics, Eqs.~(\ref{genericLagrangian}) through~(\ref{higherorderL}) yield a Lagrangian density 
\ba \label{bosonLagrangian}
{\cal L} & = & \frac{1}{2}\left(1 +2 a \frac{h}{v} +b\left(\frac{h}{v}\right)^2\right)
\partial_\mu \omega^a
\partial^\mu \omega^b\left(\delta_{ab}+\frac{\omega^a\omega^b}{v^2}\right)   
\nonumber +\frac{1}{2}\partial_\mu h \partial^\mu h \nonumber  \\
 & + & \frac{4 a_4}{v^4}\partial_\mu \omega^a\partial_\nu \omega^a\partial^\mu \omega^b\partial^\nu \omega^b +
\frac{4 a_5}{v^4}\partial_\mu \omega^a\partial^\mu \omega^a\partial_\nu \omega^b\partial^\nu \omega^b  +\frac{g}{v^4} (\partial_\mu h \partial^\mu h )^2  \nonumber   \\
 & + & \frac{2 d}{v^4} \partial_\mu h\partial^\mu h\partial_\nu \omega^a  \partial^\nu\omega^a
+\frac{2 e}{v^4} \partial_\mu h\partial^\nu h\partial^\mu \omega^a \partial_\nu\omega^a.
\ea

%%%%%%%%%%%%%%%%%%%%%%%%%%%%%%%%%%%%%%%%%%%%%%%%%%%%%%%%%%%%%%%%%%%%%%%%%%%
\subsection{The WBGB scattering amplitude in EWChPT at the one-loop level}
\label{subsec:amplitudes}
%%%%%%%%%%%%%%%%%%%%%%%%%%%%%%%%%%%%%%%%%%%%%%%%%%%%%%%%%%%%%%%%%%%%%%%%%%%

Concentrating first on elastic scattering, the custodial symmetry of the SBS of the SM in the limit $g=g'=0$ allows to write the WBGB amplitude $\omega_a\omega_b \rightarrow \omega_c\omega_d$ as
\be
{\mathcal A}_{abcd}= A(s,t,u)\delta_{ab}\delta_{cd}+A(t,s,u)\delta_{ac}\delta_{bd}+A(u,t,s)\delta_{ad}\delta_{bc}\ .
\ee 
Because of crossing symmetry for four identical particles, only one function of the Mandelstam variables $A$ is needed.
In terms of the charge states $\omega^{\pm}=(\omega^1\mp i \omega^2)/\sqrt{2}$ and $z=\omega^0$ the amplitudes can be written as:
\begin{eqnarray}
A(\omega^+\omega^- \rightarrow zz) & = & A(s,t,u)   \\ \nonumber
A(\omega^+\omega^- \rightarrow \omega^+\omega^-) & = & A(s,t,u)+A(t,s,u)   \\ \nonumber
A(zz \rightarrow zz) & = & A(s,t,u)+A(t,s,u)+A(u,t,s)   
\end{eqnarray}
(the remaining charge combinations can be obtained from these by crossing symmetry).
The $A(s,t,u)$ amplitude can be expanded in a similar way to ordinary ChPT. Quoting the NLO tree-level and one-loop subamplitudes yields
\be 
\label{loopexpansion}
A = A^{(0)} + A^{(1)} \dots =  A^{(0)} + A^{(1)}_{\rm tree} + A^{(1)}_{\rm loop} \dots
\ee

The next two-body processes to consider are the channel coupling  $\omega_a\omega_b \rightarrow h h$ between two $\omega$ WBGB and a scalar boson pair and $hh\rightarrow \omega_a\omega_b $, that are needed to obtain one-loop unitarity in $\omega\omega$ scattering. Obviously both processes have the same amplitude because of time reversal invariance. With $h$ being an isospin singlet, the amplitude takes the form
\begin{equation}
{\mathcal M}_{ab}(s,t,u) = M(s,t,u) \delta_{ab}.
\end{equation}

We also consider the amplitude for elastic scattering $hh \rightarrow hh$,
\be
{\mathcal T}(s,t,u) =
 T^{(0)} + T^{(1)}_{\rm tree} + T^{(1)}_{\rm loop} \dots
\ee
All these amplitudes are explicitly given in appendix~\ref{app:amplitude}.

The unitarity of these three scattering amplitudes is best exposed in terms of the isospin- and spin-projected partial waves; this requires projecting over custodial-isospin and angular momentum.
For elastic WBGB scattering there are three custodial-isospin $A_I$ amplitudes ($I=0,1,2$), analogous to those in pion-pion scattering in hadron physics, 
\begin{eqnarray}
A_0(s, t, u)  & = &  3 A(s, t, u) + A(t, s, u) + A(u, t, s)  \\ \nonumber
A_1(s, t, u)  & = & A(t, s, u) - A(u, t, s)     \\ \nonumber
A_2(s, t, u)  & = & A(t, s, u) + A(u, t, s)\ .
\end{eqnarray}

The projection over definite orbital angular momentum (the WBGBs carry zero spin) is then
\begin{equation} \label{Jprojection}
A_{IJ}(s)=\frac{1}{64\,\pi}\int_{-1}^1\,d(\cos\theta)\,P_J(\cos\theta)\,
A_I(s,t,u)\ .
\end{equation}

These partial waves also accept a chiral expansion
\be
A_{IJ}(s)=A^{(0)}_{IJ}(s)+A^{(1)}_{IJ}(s)+\dots ,
\ee
that take the general form
\begin{eqnarray}\label{expandpartialwave}
\nonumber   A^{(0)}_{IJ}(s)     & = & K s   \\
    A^{(1)}_{IJ}(s) & = & \left( B(\mu)+D\log\frac{s}{\mu^2}+E\log\frac{-s}{\mu^2}\right) s^2\ .
\end{eqnarray}
The constants $K$, $D$ and $E$ and the function $B(\mu)$ depend on the different channels $IJ=00,11,20,02,22$, as shown below in appendix~\ref{app:amplitude}, and we will use the same notation for the inelastic and pure-$h$ scattering reactions. 

As $A_{IJ}(s)$ must be scale independent we have
\begin{equation} \label{Bruns}
B(\mu)=
 B(\mu_0)+(D+E)\log\frac{\mu^2}{\mu_0^2}\ ;
 \end{equation}
This $B(\mu)$ function depends on the NLO chiral constants (with certain proportionality coefficients $p_4$ and $p_5$ that can be read off Eq.~(\ref{partial00}) and following)
\be \label{Bfunction}
B(\mu)=B_0 +p_4 a_4(\mu)+p_5 a_5(\mu)\ ,
\ee
where $B_0$ also depends on $a$ and $b$ and from now on we omit the superindices $r$ on the renormalised coupling constants for simplicity.

Since the ``Higgs'' boson is assigned zero custodial isospin, $\omega\omega \rightarrow hh$ and $hh \rightarrow hh$ occur only in the isospin zero channel $I=0$. 

The normalisation of the $\ar \omega\omega\ra_{I=0}$ state introduces a factor $1/\sqrt{3}$ and the sum over the three contributing charge combinations $(+-,-+,00)$ a factor 3, so that
for the inelastic amplitude we have $M_0(\omega\omega  \rightarrow h h)=\sqrt{3} M(s,t,u)$.
For the scalar-scalar interaction there is no such factor and $T_0(h h \rightarrow h h)= T(s,t,u)$. Omitting the isospin subindices (which take only the value 0) and proceeding to the angular momentum projections, we find the chiral expansions equivalent to the $\omega\omega$ elastic one in Eq.~(\ref{expandpartialwave}).
They read
\begin{eqnarray}
\nonumber   M_{J}(s)     & = & K' s+  \left( B'(\mu)+D'\log\frac{s}{\mu^2}+E'\log\frac{-s}{\mu^2}\right)s^2+ \dots \\
    T_{J}(s)     & = & K'' s  +  \left( B''(\mu)+D''\log\frac{s}{\mu^2}+E''\log\frac{-s}{\mu^2}\right) s^2+\dots
\end{eqnarray}
(with $J$ subindex omitted in the constants). The functions $B'(\mu)$ and $B''(\mu)$ are in all analogous to
$B(\mu)$ in Eq.~(\ref{Bfunction}), renormalization is carried out by $d$, $e$ (for $B'$) and $g$ (for $B''$) involving the $h$ boson.

The partial-wave amplitudes $A_{IJ}(s)$, $M_{J}(s)$ and $T_{J}(s)$ are all analytical functions of complex Mandelstam-$s$, having the proper left and right (or unitarity) cuts, shortened to LC and RC respectively.
The physical values of their argument are $s=\it{E}_{\rm CM}^2+i\epsilon$ (i.e. on the upper lip of the RC), where $\it{E}_{\rm CM}$ is the total energy in the center of mass frame.
For these physical $s$ values, exact unitarity requires a set of non-trivial relations between the different partial waves that we now spell out.
For the problem of $\omega \omega $ scattering considered here the reaction matrix is block-diagonal:
\be
F(s)=\begin{pmatrix}
F_{00} & 0&0&0&0 \\
0& F_{02} &0&0&0 \\
0&0&F_{11} & 0&0 \\
0&0& 0&F_{20} & 0 \\
0 & 0&0&0&... \\
\end{pmatrix},
\ee
where $F_{IJ}(s)$ are the partial-waves matrices. For example for $I=0$ we have:
\be
F_{00}(s)=\begin{pmatrix}
A_{00}(s) & M_0(s) \\
M_0(s)& T_0(s) \\
\end{pmatrix}
\ee
and
\be
F_{02}(s)=\begin{pmatrix}
A_{02}(s) & M_2(s) \\
M_2(s)& T_2(s) \\
\end{pmatrix}.
\ee
For $I \ne 0$ there is no mixing with the $hh$ channel and the $F_{IJ}(s)$ matrices have just one single element:
\be
 F_{IJ}(s)=A_{IJ}(s)
\ee
Now unitarity requires that on the right cut:
\be
\Imag F(s) = F(s)F^\dagger(s).
\ee
This equation produces a set of  relations concerning the different partial waves. For $I=0$ and either $J=0$ or $J=2$ we have:
\begin{eqnarray}{}
 \Imag A_{0J} &  = & \lvert A_{0J} \rvert ^2 +  \lvert M_J \rvert ^2 \\
 \nonumber    \Imag M_J &  = & A_{0J} M_{J}^*+  M_J T_J^* \\
 \nonumber    \Imag T_J &  = & \lvert M_J \rvert ^2+  \lvert T_J \rvert ^2\   .
\end{eqnarray}         
These relations are not exactly respected by perturbation theory, but are instead satisfied only to one less order in the expansion than kept in constructing the amplitude. At the one-loop level one has
\begin{eqnarray}{}
\nonumber \Imag A^{(1)}_{0J} &  = &\lvert A^{(0)}_{0J}\rvert^2+  \lvert M^{(0)}_{J}\rvert ^2 \\
 \nonumber   \Imag M^{(1)}_{J} &  = & A^{(0)}_{0J} M^{(0)}_{J}+ 
 M^{(0)}_{J}T^{(0)}_{J} \\
 \nonumber    \Imag T^{(1)}_{J} &  = & \lvert M^{(0)}_J\rvert ^2 +  \lvert T^{(0)}_{J}\rvert^2   .
\end{eqnarray}    
     
For the remaining channels with $I=J=1$ and $I=2$, $J=0$ the $\omega\omega \rightarrow \omega\omega$ reaction is elastic and the unitarity condition is just
\be \label{unitarity1channel}
 \Imag A_{I J}   =  \lvert A_{IJ} \rvert ^2  \ \ \ I\neq 0
\ee
and at the NLO perturbative level,
\be
 \Imag A^{(1)}_{I J}   =  \lvert A^{(0)}_{IJ} \rvert ^2  \ \ \ I\neq 0\ .
\ee

There are in all nine independent one-loop perturbative relations, that can also be obtained by applying the Landau-Cutkosky cutting rules and directly checked in each of the partial waves for the three reactions, providing a very good, non-trivial check of our amplitudes.

Therefore the perturbative reaction matrix
\be
  F_{IJ}=F_{IJ}^{(0)}+F_{IJ}^{(1)}+...
\ee
fulfils
\be
\Imag  F_{IJ}^{(1)} = F_{IJ}^{(0)}F_{IJ}^{(0)}
\ee
since the $F_{IJ}^{(0)}$ elements are real.

%%%%%%%%%%%%%%%%%%%%%%%%%%%%%%%%%%%%%%%%%%%%%%%%%%%%%%%%%%%%%%%%%%%%%
\section{The Inverse Amplitude Method for massless particles}\label{sec:IAM}
%%%%%%%%%%%%%%%%%%%%%%%%%%%%%%%%%%%%%%%%%%%%%%%%%%%%%%%%%%%%%%%%%%%%%
\subsection{Derivation for one channel}
%%%%%%%%%%%%%%%%%%%%%%%%%%%%%%%%%%%%%%%%%%%%%%%%%%%%%%%%%%%%%%%%%%%%%
The Inverse Amplitude Method (IAM)~\cite{Truong:1988zp} was developed for ordinary ChPT for mesons~\cite{Dobado:1992ha,JRA} and it was also applied to the unitarization of the one-loop WBGB scattering amplitudes, at the time without a light Higgs resonance (see \cite{DHD} and third reference in~\cite{Truong:1988zp}). Its standard derivation is valid for one or several channels of particle pairs \emph{all of which have equal mass}. For different masses there are technical complications (such as overlapping left and right cuts) that have hindered a rigorous derivation.

In the context where we wish to apply it, for energies $E\gg M_W,M_h$, both masses can be taken as equal and negligible. Yet for massless particles, the standard derivation is also problematic, since the dispersion relation is thrice subtracted and the factors $1/s^3$ cause infrared divergences.

Since it would be nice to have a derivation valid for massless particles, we now address a twice-subtracted dispersion relation that avoids infrared problems. The price to pay is that, with chiral amplitudes, the large circle at infinity to close the contour in the complex plane will give a contribution that needs to be calculated. As we will see in this section, this is feasible for elastic scattering of massless particles. 

\begin{figure}
\includegraphics[width=7cm]{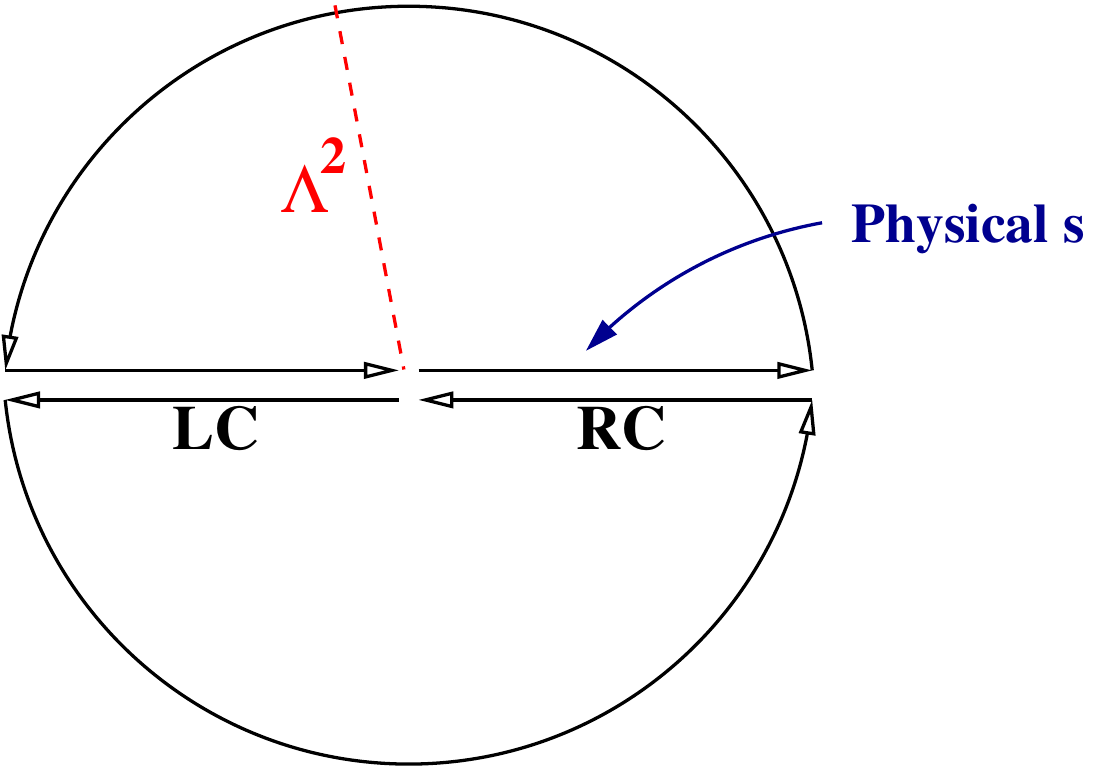}
\caption{\label{fig:disprel}%
Contour to apply Cauchy's theorem in the presence of a right-hand cut (RC) due to elastic intermediate states in the $s$-channel, and a left-hand cut (LC) due to angular integration over $t$, $u$-channel exchanges. In the massless limit $M\to 0$, the contour encloses only the upper half plane. The radius of the large circle is $\Lambda^2$.}
\end{figure}

We start by writing a twice-subtracted dispersion relation (DR) for a generic elastic partial wave amplitude $A(s)$ (we suppress the $I$ and $J$ indices), that has both left- and right-hand cuts as shown in figure~\ref{fig:disprel},
\be \label{disprelamp}
A(s)=K s+\frac{s^2}{\pi}\int_0^\infty  \frac{ds' \Imag A(s')}{s'^2(s'-s-i\epsilon)} +\frac{s^2}{\pi}\int_{-\infty}^{0}  \frac{ds' \Imag A(s')}{s'^2(s'-s-i\epsilon)}
\ee
An introduction to dispersion relations can be found on refs.~\cite{Dispers1,Dispers2}. To sum up, the derivation of Eq.~\ref{disprelamp} is based on the Cauchy theorem and on the analyticity of $A(s)$ for $\Imag s > 0$ (first Riemann sheet), as well as on the analytic properties of $A(s)$. Note that, according to Eq.~\ref{expandpartialwave}, our computations have a left-cut (i.e., they are not analytic on the real axis for $s<0$). So, \emph{forward} dispersion relations, like those commonly introduced on text books, cannot be used.

Because $A(s)$ describes the scattering of Goldstone bosons, there are two simplifying properties. The first is that there is an Adler zero. In the massless limit this is located at $s=0$ and guarantees $A(0)=0$. 
Accordingly, we set the first subtraction constant to zero and the first term is linear in $s$.
The second is that there are no (subthreshold, bound-state) poles of $A(s)$ in the first or physical Riemann sheet (which does not make sense for Goldstone bosons that interact with weak strength at low energies). So only the two cuts contribute as written since $A(s)$ is analytic in the rest of the upper half plane.

We will obtain a second dispersion relation for the partial-wave amplitude expanded to NLO in the EChL, that is, truncated up to order $s^2$,
$A^{\rm NLO}(s)=A^{(0)}(s) +A^{(1)}(s)$, that has the generic form:
\begin{eqnarray}
\nonumber  
 A^{(0)}(s)     & = & K s   \\ \label{amppertth}
    A^{(1)}(s) & = & \left( B(\mu)+D\log\frac{s}{\mu^2}+E\log\frac{-s}{\mu^2}\right)s^2\ .
\end{eqnarray}   

To derive the dispersion relation, we will first introduce the auxiliary function
 \ba \label{faux}
 f(s) &\equiv& \frac{A^{\rm NLO}(s)-A^{(0)}(s) }{s^2}  = \frac{A^{(1)}(s)}{s^2}   \\\nonumber 
&=& B(\mu)+D\log\frac{s}{\mu^2}+E\log\frac{-s}{\mu^2}
 \ea       
  
Therefore $f(s)$ is analytic in the whole complex plane except for the left (LC) and right cuts (RC) along the negative and positive real axis respectively. Cauchy's theorem provides an unsubtracted dispersion relation for $f(s)$:  
 \be
f(s)=\frac{1}{\pi}\int_0^{\Lambda^2}  \frac{ds' \Imag f(s')}{s'-s-i\epsilon} +\frac{1}{\pi}\int_{-\Lambda^2}^0  \frac{ds' \Imag f(s')}{s'-s-i\epsilon}+
\frac{1}{2 \pi i}\int_{C_\Lambda}  \frac{ds'  f(s')}{s'-s}
\ee
where $C_{\Lambda}$ is a circumference of radius $\Lambda^2$ oriented anticlockwise and $\Lambda$ is an UV regulator which will be sent to infinity 
at the end (see figure~\ref{fig:disprel}).  

Returning to Eq.~(\ref{faux}), we see that this dispersion relation can easily be turned into one for $A^{\rm NLO}(s)$,
\be \label{disprelNLO}
A^{\rm NLO}(s)= Ks + \frac{s^2}{\pi}\int_0^{\Lambda^2}  \frac{ds' \Imag A^{(1)}(s')}{s'^2(s'-s-i\epsilon)} +\frac{s^2}{\pi}\int_{-\Lambda^2}^0  \frac{ds' \Imag A^{(1)}(s')}{s'^2(s'-s-i\epsilon)}+
\frac{s^2}{2 \pi i}\int_{C_\Lambda}  \frac{ds'  A^{(1)}(s')}{s'^2(s'-s)}.
\ee
Comparing this dispersion relation for the NLO amplitude with that
for the exact amplitude $A(s)$ in Eq.~(\ref{disprelamp}), we notice that the difference is the contribution of the circle at infinity, a term due to the 
divergent UV behavior of $A^{\rm NLO}(s)\propto s^2$. 
Taking now $\Lambda^2\gg s$ beyond the region where the amplitude is considered,
the three integrals may easily be computed,
\begin{eqnarray}
 \frac{s^2}{\pi}\int_0^{\Lambda^2}  \frac{ds' \Imag A^{(1)}(s')}{s'^2(s'-s-i\epsilon)}  & = & s^2 E \log\frac{-s}{\Lambda^2} \nonumber \\
 \frac{s^2}{\pi}\int_{-\Lambda^2}^0  \frac{ds' \Imag A^{(1)}(s')}{s'^2(s'-s-i\epsilon)}  & = & s^2 D \log\frac{s}{\Lambda^2} \nonumber \\
\frac{s^2}{2 \pi i}\int_{C_\infty}  \frac{ds'  A^{(1)}(s')}{s'^2(s'-s)} & = &  s^2\left( B(\mu)+D\log\frac{\Lambda^2}{\mu^2}+E\log\frac{\Lambda^2}{\mu^2}\right)
\end{eqnarray}     
so that the dispersion relation for $A^{\rm NLO}(s)$ in Eq.~(\ref{disprelNLO})
reproduces Eq.~(\ref{amppertth})
\be
A^{\rm NLO}(s)=K s+ \left(B(\mu)+D \log\frac{s}{\mu^2}+E \log\frac{-s}{\mu^2}\right) s^2\ .
\ee
This is a consistency check of the dispersion relation and also shows its nice interplay with renormalized chiral couplings; the integral over the large circle trades the UV-cutoff scale $\Lambda$ for the arbitrary renormalization scale $\mu$.

So far we have an elastic, exact, but not too useful, dispersion relation for $A(s)$ in Eq.~(\ref{disprelamp}) and another in Eq.~(\ref{disprelNLO}) for $A^{(1)}(s)$ which is known anyway from chiral perturbation theory. The practical use of the technique comes from its application to the following auxiliary function,
\be \label{gdef}
  w(s)\equiv \frac{[A^{(0)}(s)]^2}{A(s)}\ .
\ee 
This construction has the same analytic structure than $A(s)$ up to possible poles coming from zeroes of $A(s)$, excluding the Adler zero (cancelled by the numerator). In addition $w(0)=0$, $w(s)= K s + O(s^2)$ and on the RC one has $\Imag w(s)=-[A^{(0)}(s)]^2$. Therefore, neglecting the possible pole contribution%
\footnote{A more careful treatment in the massive case that includes subthreshold poles found that their effect is very small, at the permille level or less in the physical zone~\cite{GomezNicola:2007qj}.}, the twice-subtracted dispersion relation for this function reads
\be
w(s)=K s+\frac{s^2}{\pi}\int_0^{\Lambda^2}  \frac{ds' \Imag w(s')}{s'^2(s'-s-i\epsilon)} +\frac{s^2}{\pi}\int_{-\Lambda^2}^{0}  \frac{ds' \Imag w(s')}{s'^2(s'-s-i\epsilon)}  + \frac{s^2}{2 \pi i}\int_{C_\infty}  \frac{ds' w (s')}{s'^2(s'-s)}                                             \ .
\ee
The careful choice of definition for $w(s)$ in Eq.~(\ref{gdef}) makes possible to compute the elastic-RC integral \emph{exactly} since 
$\Imag w(s)=-K^2s^2= E \pi s^2$ there. This is important because it is the nearest complex-plane feature to the physical zone (which is the upper lip of this cut, in the first Riemann sheet).

The LC integral cannot be obtained exactly, so we choose to compute it in perturbation theory: its contribution on the physical zone is down by 
$\lvert s'-s\rvert$ with $s'\sim u$ respect to the RC, so it is small when perturbation theory deteriorates at $u \ll 0$. Then, it is a fair approximation to take
\be
 \Imag w(s) \simeq - \Imag A^{(1)}(s) .
\ee
Then one finds:
\be
 w(s) \simeq K s-D s^2\log\frac{s}{\Lambda^2}-E s^2\log\frac{-s}{\Lambda^2} + \frac{s^2}{2 \pi i}\int_{C_\Lambda}  \frac{ds'  w(s')}{s'^2(s'-s)}  .      
\ee

It is easy to check that this approximate integral equation is solved by 
$w(s)=A^{(0)}(s)-A^{(1)}(s)$. This is quite remarkable since $w(s)$ in Eq.~(\ref{gdef}) is defined from the exact amplitude. 
Again, the only used approximations are the absence of poles in the inverse amplitude and the perturbative treatment of the LC integral. 
It stands out that, from the very definition of $w(s)$, 
we can write down the IAM amplitude as:
\be \label{IAM1channel}
 A(s)\simeq A^{\rm IAM}(s) = \frac{[A^{(0)}(s)]^2}{A^{(0)}(s)-A^{(1)}(s)} \ .
\ee
This IAM amplitude obtained from the ChPT expansion has many interesting properties. First it has the proper analytic structure which, in particular, makes poles on the second Riemann sheet possible (that can be understood as dynamically generated resonances).
Second, it is $\mu$-invariant, depending only on the renormalized chiral constants $a_4$, $a_5$, $e$, $d$ or $g$ that encode higher energy dynamics. 
It satisfies elastic unitarity, so that on the RC,
\be
 \Imag  A^{\rm IAM} = A^{\rm IAM}(A^{\rm IAM})^*\ .
\ee
Finally, if expanded at low energy, it coincides with the NLO-ChPT amplitude,
\be
A^{\rm IAM}(s) =A^{\rm NLO}(s) + O(s^3) 
\ee
It is important to stress once more that the IAM amplitude has been obtained here by using a twice-subtracted dispersion relation, whereas previous derivations used a thrice-subtracted DR. 
Therefore we needed to carefully take into account the contribution of the circumference at infinity $C_\infty$, which is not present with three subtractions. 

This was necessary to avoid the infrared problems that would otherwise  appear in the derivation of the IAM amplitude for massless particles, having all the LC and RC thresholds located at $s=0$.
We restate that the only approximations used  was  taking $\Imag w(s) \simeq -\Imag A^{(1)}(s)$ on the LC integral and assuming that $w(s)$ has no poles, whereas  the numerically more important RC integral is computed exactly. 
A posteriori these assumptions have been validated in low energy meson-meson scattering where the IAM method has proven to be extremely successful, as with a very small set of parameters it describes many different channels including their first resonances.

%%%%%%%%%%%%%%%%%%%%%%%%%%%%%%%%%%%%%%%%%%%%%%%%%%%%%%%%%%%%%%%%%%%%%
\subsection{Coupled-channel Inverse Amplitude Method}\label{subsec:coupledIAM}
%%%%%%%%%%%%%%%%%%%%%%%%%%%%%%%%%%%%%%%%%%%%%%%%%%%%%%%%%%%%%%%%%%%%%

The IAM method can also be extended to the coupled-channel case provided the masses of the particle appearing in the different channels are all the same, to avoid overlapping left and right cuts. This is the case here since we are considering the WBGB and the $h$ particle massless. 
The demonstration is an immediate extension of the single-channel case and we relegate it to appendix~\ref{app:IAM}.

The multichannel matrix with adequate properties can be constructed from the perturbative expansion
\ba
F_{IJ}&=&F_{IJ}^{(0)}+F_{IJ}^{(1)}+\dots \\ 
F^{(0)}(s)&=& K s \nonumber\\
F^{(1)}(s)&=& \left(B(\mu) + D \log\frac{s}{\mu^2}+ E \log \frac{-s}{\mu^2} \right)s^2
\label{pertmultichannel}
\ea 
where now $K, B(\mu), D$ and $E$ have to be considered as (two by two) matrices. For example
$K_{11}=K, K_{12}=K_{21}=K'$ and $K_{22}=K''$ (Notice that $K$ refers in different formulae to $K_{11}$ or to the matrix $K$). Finally, $F_{IJ}$ is found to be
\be \label{IAMforF}
 F_{IJ}^{\rm IAM}=F_{IJ}^{(0)}(F_{IJ}^{(0)}-F_{IJ}^{(1)})^{-1}F_{IJ}^{(0)}
\ee
that satisfies exact elastic unitarity on the RC
\be \label{multichannelUnit}
\Imag  F_{IJ}^{\rm IAM} = F_{IJ}^{\rm IAM}(F_{IJ}^{\rm IAM})^\dagger  .
\ee
The various amplitudes (matrix elements of $F_{IJ}^{\rm IAM}$) enjoy all the already mentioned desirable properties of the elastic IAM method. 
The coupled-channel IAM method is particular useful in the isoscalar channels ($I=0$ and $J=0, 2,\dots$) where the $\omega\omega$ and $hh$ channels can be strongly coupled.
We dedicate section~\ref{sec:sysIAM} to a detailed numerical analysis of the method based on Eq.~(\ref{IAM1channel}) and~(\ref{IAMforF}).

%%%%%%%%%%%%%%%%%%%%%%%%%%%%%%%%%%%%%%%%%%%%%%%%%%%%%%%%%%%%%%%%%%%%%%%%%%%%%%
\section{The N/D method}\label{sec:NoverD}
%%%%%%%%%%%%%%%%%%%%%%%%%%%%%%%%%%%%%%%%%%%%%%%%%%%%%%%%%%%%%%%%%%%%%%%%%%%%%%
\subsection{Elastic \boldmath $\omega\omega$ scattering} \label{subsec:NDelastic}
%%%%%%%%%%%%%%%%%%%%%%%%%%%%%%%%%%%%%%%%%%%%%%%%%%%%%%%%%%%%%%%%%%%%%%%%%%%%%%
The IAM is a reliable unitarization method, but to assess the systematic error introduced by approximating the left cut in perturbation theory, it is recommendable to compare with a different unitarization method applicable to the one-loop results for the $\omega\omega$ and $hh$ scattering amplitudes. 
A well-known alternative that we consider here is the N/D method. This can be applied in many different ways depending on the problem at hand. When the $\omega\omega$ is purely elastic $(J \ne 0)$ the starting point is
an ansatz for the scattering partial waves, from which the method is named,
\be
A(s)=\frac{N(s)}{D(s)}\
\ee
where the numerator function $N(s)$ has only a LC and the denominator function $D(s)$ only a RC, so that $A(s)$ has the expected analytical structure. Therefore $\Imag N(s)=0$ on the RC and $\Imag D(s)=0$ on the LC. In addition, elastic unitarity, $\Imag A(s)=\lvert A(s)\rvert^2$ requires $\Imag D(s)=-N(s)$ on the RC and we also have $\Imag N(s)=D(s)\Imag A(s)$ on the LC. It is then possible in principle to write two coupled dispersion relations for $N(s)$ and $D(s)$. The normalization $D(0)=1$ may be chosen by making $N(0)=A(0)$, so
\begin{eqnarray}\label{NsobreD}
D(s) & = & 1- \frac{s}{\pi}\int_0^\infty  \frac{ds' N(s')}{s'(s'-s-i\epsilon)}\\ \label{NsobreD2}
N(s) & = & \frac{s}{\pi}\int_{-\infty}^{0}  \frac{ds' D(s') \Imag A(s')}{s'(s'-s-i\epsilon)}   \ .
\end{eqnarray}
More generally, one needs an $n$-times subtracted DR,
which is useful to input the particular low-energy dynamics to be described:
\be\label{N/D_D_n_subtr}
D(s) = 1+ h_1s+h_2s^2+...+h_{n-1}s^{n-1}- \frac{s^n}{\pi}\int_0^\infty  \frac{ds' N(s')}{s'^n(s'-s-i\epsilon)}
\ee
The coupled equations for $N(s)$ and $D(s)$ can be solved in principle by using some recursive method. For example, starting from some approximate $N_0(s)$ function featuring a LC (typically a tree level result) we can obtain $D_0(s)$ by integration on the RC. 
Then a first approximation for the partial wave would be $A_0(s)=N_0(s)/D_0(s)$. To continue the procedure one can now insert $D_0(s)$ in the second coupled equation to get the new a $N_1(s)$ yielding $A_1(s)=N_1(s)/D_1(s)$  and so on. Presumably in this way it should be possible to approach as much as needed the real solution for some given subtraction constants, provided the original guess for $N_0(s)$ is appropriate enough. Even more, in many cases the simplest and crude approximation $A(s) \simeq N_0(s)/D_0(s)$ could be considered a sensible estimate of the exact solution. For example, taken $N_0(s)=A^{(0)}(s)$ and regularizing the integrals with IR and UV cutoffs $m^2$ and $\Lambda ^2$, one gets
\be
D_0(s)=1+\frac{A^{(0)}(s)}{\pi}\log\frac{-s}{\Lambda^2}
\ee
so that
\be
A(s) \simeq \frac{A^{(0)}(s)}{1+\frac{  A^{(0)}(s)}{\pi}\log   \frac{-s}{\Lambda^2}   } \ .
\ee
We do not find this approximation satisfactory though, at least when compared with the Inverse Amplitude Method in section~\ref{sec:IAM}. In particular, because the equation for $N$ has not been iterated yet, the amplitude only has a RC but not a LC. It is unitary and depending on the UV scale $\Lambda$ and also not compatible with the NLO result to order $s^2$.
The reason for this is that we are not yet taking into account the information coming from the NLO term $A^{(1)}$ containing the one-loop effects and the chiral couplings.
However, introducing these NLO effects in the N/D method is far from trivial for various reasons.

For one, it is not obvious how to choose the starting function $N_0(s)$: remember that the NLO partial-waves have the general form $ A(s)=A^{(0)}(s)+A^{(1)}(s)+\dots$,
with the general form given in Eq.~(\ref{expandpartialwave}). Thus $A^{(1)}(s)$ contains a logarithm with a LC and another one with a RC that, taken independently, are scale-dependent:
the scale-independence of $A(s)$ is achieved with the compensating dependence of $B(\mu)$ in Eq.~(\ref{Bruns}).
%\begin{equation}
%B(\mu)=
% B(\mu_0)+(D+E)\log\frac{\mu^2}{\mu_0^2}.
% \end{equation}
Thus a naive choice for $N_0(s)$ featuring a LC will not be in general $\mu$-invariant and that makes the N/D method less attractive.
% thus spoiling the good properties of the N/D method when applied to the case considered here.

To solve this problem we split $A^{(1)}(s)$ in two pieces, one having only a RC and the other only a LC and both $\mu$-independent, by adequately splitting the function $B(\mu)$,
\begin{eqnarray}\label{splitLR}
\nonumber             A_L(s)      & \equiv & \left(\frac{B(\mu)}{D+E}+\log\frac{s}{\mu^2}\right) D s^2 \\
                       A_R(s)      & \equiv & \left(\frac{B(\mu)}{D+E}+\log\frac{-s}{\mu^2}\right) E s^2   .
\end{eqnarray}

The cut structure is obviously as advertised, $A^{(1)}(s)= A_L(s)+A_R(s)$ is also trivially verified, and the scale-independence follows from Eq.~(\ref{Bruns}). 
In addition, on the RC (the physical region), perturbative unitarity reads $\Imag  A^{(1)} =\Imag A_R= (A^{(0)})^2$.
The split in Eq.~(\ref{splitLR}) is not usable in the $IJ=11$ channel in the particular parameter case $a^2=b$ because of a coincidence\footnote{It is known that in this elastic vector-isovector amplitude the NLO amplitude on the physical cut is a polynomial due to canceling logarithms, so the combination of chiral constants $(a_4-2 a_5)$ in Eq.~(\ref{partial11}) is $\mu$-invariant by itself.} in Eq.~(\ref{partial11}), that yields $E=-D$. In all other circumstances the denominator is finite and does not give any problem.

It is illustrative to express $A_L$ and $A_R$ in terms of an auxiliary ``loop'' function
\be \label{defgwithB}
g(s)=\frac{1}{\pi}\left(\frac{B(\mu)}{D+E}+\log\frac{-s}{\mu^2}\right)    \ .
\ee
This function, as the notation suggests, is $\mu$-independent (as is easily checked). Furthermore, it is analytical on the whole complex plane but for a RC. On this RC (i.e. for $s=E^2+i\epsilon$) we have $\Imag g(s)= -1$. Then,
%This function is analytical on the whole complex plane but for a RC. 
%As the notation suggests, it is $\mu$-independent (as is easily checked) and finally, on the RC (i.e. for $s=E^2+i\epsilon$) we have $\Imag g(s)= -1$. Then,
\ba
A_L(s) & = & \pi g(-s) D s^2 \nonumber\\
A_R(s) & = & \pi g(s) E s^2
\ea
so that perturbatively,
\be
A(s)=A^{(0)}(s)+A_L(s)-[A^{(0)}(s)]^2g(s)+O(s^3) \ .
\ee
We have now the ingredients to apply the N/D method to the NLO computation: the useful starting point is the function 
\be \label{numeratordef}
N_0(s)\equiv A^{(0)}(s)+A_L(s) \ . 
\ee
Notice that this function contains the LC, information about the chiral parameters and additionally it is $\mu$ independent. 

The inconvenience now is that the UV behavior of the integral for $D_0(s)$ in Eq.~(\ref{NsobreD}) is even worse than with the tree-level ansatz, since a term $s^2$ is included in $N_0$.
To obtain a UV-finite integral three subtractions are required, 
at the prize of a chiral coupling of order $s^3$ (see appendix~\ref{app:NoverD}). Thus we can write:
\be
D_0(s) = 1+h_1s+h_2s^2- \frac{s^3}{\pi}\int_0^\infty \frac{ds' [A^{(0)}(s)+A_L(s)]}{s'^3(s'-s-i\epsilon)}.
\ee
As further shown in appendix~\ref{app:NoverD}, the N/D partial wave in this approximation can be written as
\be\label{N/D_A}
A(s) \simeq A^{\rm N/D}(s)  = \frac{N_0(s)}{D_0(s)}. 
\ee \be\label{N/D__D0_g}
D_0(s)=1-\frac{A_R(s)}{A^{(0)}(s)}+\frac{\pi}{2} [(g(s)]^2 D s^2.
\ee
By using the $A_L(s)$ and $A_R(s)$ definitions in Eq.~(\ref{splitLR}) this denominator can also be written as:
\be\label{N/D_D0}
D_0(s)=1-\frac{A_R(s)}{A^{(0)}(s)}+\frac{1}{2}g(s)A_L(-s)=1-\frac{A_R(s)}{A^{(0)}(s)}-\frac{A_L(-s)A_R(s)}{2(A^{(0)})^2}.
\ee

This amplitude in Eqs.~ (\ref{numeratordef}), (\ref{N/D_A}) and~(\ref{N/D_D0}) has many interesting properties.  First it is UV finite, the IR divergences have been removed and it is $\mu$ independent. Second, it has the right analytical structure and it satisfies elastic unitarity exactly:
\be
\Imag A^{\rm N/D}(s) = \lvert A^{\rm N/D}(s) \rvert^2
\ee
on the RC. Finally it is compatible with the NLO computation up to order $s^2$ since:
\be
A^{\rm N/D}(s) =A^{(0)}(s)+A^{(1)}(s)+O(s^3).
\ee
All these properties are shared with the Inverse Amplitude Method. In Eq.~(\ref{Atogether}) below we show that this amplitude converges to the IAM amplitude whenever $A_L \ll A^{(0)}$.

%%%%%%%%%%%%%%%%%%%%%%%%%%%%%%%%%%%%%%%%%%%%%%%%%%%%%%%%%%%%%%%%%%%%%%%%%%%
\subsection{Coupled \boldmath $\omega\omega-hh$ channels}
%%%%%%%%%%%%%%%%%%%%%%%%%%%%%%%%%%%%%%%%%%%%%%%%%%%%%%%%%%%%%%%%%%%%%%%%%%%

Just as for the IAM, it is possible to generalize the N/D method to the multichannel case needed for the $I=0$ ($J=0,\ 2$) cases where the $\omega\omega $ state couples to the $hh$ channel. Following~\cite{Bjorken} we introduce two matrices, a numerator one $N$ and a denominator $D$ so that
\be
F(s)= [D(s)]^{-1}N(s)\ .
\ee 
To generalize our previous result for the single channel case, we start again from the perturbative expansion at NLO, Eq.~(\ref{pertmultichannel}),
Again the $\mu$ evolution of $B(\mu)$ is  given by Eq.~(\ref{Bruns}), now a matrix equation .

Thus we can introduce the $\mu$-independent matrix
\be             \label{Gmatrix}
G(s)= \frac{1}{\pi}\left(B(\mu)(D+E)^{-1}+\log\frac{-s}{\mu^2}\right)
\ee
and the (also $\mu$-invariant) left and right matrices
\ba
F_L(s)&=&\left(B(\mu)(D+E)^{-1}+\log\frac{s}{\mu^2}\right)  Ds^2 \\ \nonumber
      &=& \pi G(-s) D s^2 \\
F_R(s)&=&\left(B(\mu)(D+E)^{-1}+\log\frac{-s}{\mu^2}\right) Es^2 \\ \nonumber
      &=& \pi G(s) E s^2\ .
\ea
On the RC cut perturbative unitarity reads:
\be
\Imag F^{(1)}(s) = \Imag F_R(s) = F^{(0)}(s)^2 = K^2 s^2
\ee
which implies:
\be
E =-\frac{1}{\pi}  K^2
\ee
and therefore
\be
F_R(s)= -G(s)[F^{(0)}(s)]^2 .
\ee
Now we can follow essentially the same steps that we took in the single-channel case in subsection~\ref{subsec:NDelastic}, taking into account the matrix character of the different amplitudes and of $K, B(\mu), D$ and $E$. Like in the case of the IAM, this produces a sensible result because all particles involved, the WBGB and the Higgs-like particle, are massless and therefore we are not overlapping the LC and the RC in any unitarized partial wave. Then we get
\be \label{NDcoupled}
F^{\rm N/D}(s)=[D_0(s)]^{-1}N_0(s)
\ee
where
\be
N_0(s)=F^{(0)}(s)+F_L(s)
\ee
and
\be
D_0(s)=1- F_R(s) [F^{(0)}(s)]^{-1}+ \frac{\pi}{2}[G(s)]^2 Ds^2   
\ee
that can also be written as:
\be
D_0(s)=1- F_R(s)[F^{(0)}(s)]^{-1}+ \frac{1}{2}G(s)F_L(-s) =
1- F_R(s)[F^{(0)}(s)]^{-1}- \frac{1}{2}F_R(s) [F^{(0)}(s)]^{-2}F_L(-s) 
\ee
It is not difficult to check that these partial waves in Eq.~(\ref{NDcoupled}) fulfill exact elastic unitarity on the RC,
\be
\Imag F^{\rm N/D}=F^{\rm N/D}\left(F^{\rm N/D}\right)^\dagger 
\ee
and also reproduce the low-energy expansion to NLO:
\be
F^{\rm N/D}(s)=F^{(0)}(s)+F^{(1)}(s)+...
\ee

Thus the $F^{\rm N/D}(s)$ partial-wave amplitudes have all the required properties including unitarity and analyticity. They have a LC and RC and they can be extended to the second Riemann sheet, and in some cases have poles there that could be understood as resonances. 

Interesting cases where the N/D method has the advantage are those in which $K=E=0$ such as the $IJ=02, 22$ waves. The vanishing of the leading-order term proportional to $K$ makes the IAM yield zero at this order, and one needs the NNLO IAM or an approximation thereof, which we have not developed here but see~\cite{Dobado:2001rv}. However the N/D method can be safely applied to these situations too, as it is very easy to check since $g(s)$ or $G(s)$ are well defined even for $K=E=0$. 
%%%%%%%%%%%%%%%%%%%%%%%%%%%%%%%%%%%%%%%%%%%%%%%%%%%%%
\section{Other Unitarization Methods, a comparison among them, and their Resonances}
%%%%%%%%%%%%%%%%%%%%%%%%%%%%%%%%%%%%%%%%%%%%%%%%%%%%%
\subsection{The K-matrix and the Improved K-matrix} \label{subsec:K}
%%%%%%%%%%%%%%%%%%%%%%%%%%%%%%%%%%%%%%%%%%%%%%%%%%%%%

Finally we will briefly comment on some other unitarization methods which have also been considered for the scattering of the would-be GB in the context of the SBS of the SM. One of the most popular unitarization procedures is the so called K-matrix method \cite{Kmatrix} (see also %\cite{Gupta} and
\cite{Kilian} for a recent review in the  context of this work). The K-matrix is defined in terms of the $S$ matrix as:
\be\label{SofK}
S=\frac{1- i K/2}{1+i K/2}.
\ee
With this parametrization $S$ is unitary {\it if and only if} $K$ is Hermitian. Eq.~\ref{SofK} can be inverted to give $K$ in terms of $S$:
\be
K= \frac{i(S-1)}{1+(S-1)/2}.
\ee
In practice the $S$ matrix is obtained in the form of some expansion:
\be
S= 1 + S^{(1)} +S^{(2)}+\dots
\ee
However the truncation of this series usually produces an approximate $S$ matrix which is not unitary. However, if we truncate instead an expansion of $K$,
\be
K=K^{(1)}+K^{(2)}+\dots\ ,
\ee
and introduce this (truncated) series into Eq.~(\ref{SofK}) to find a new series for $S$,
\be
S=1+ \tilde S^{(1)}+\tilde S^{(2)}+\dots\ ,
\ee
this is exactly unitary at any order.

In terms of a partial-wave amplitude for some unspecified elastic process $A(s)$, this amounts to the following.
One starts from some approximate estimation $A_0(s)$  real in the physical region and therefore not unitary. Then one defines the K-matrix unitarized partial wave:
\be \label{oldKamplitude}
A_0^K(s)=\frac{A_0(s)}{1-i A_0(s)}
\ee
Clearly, unitarity is satisfied again  in the physical region,
\be
\Imag A_0^K= \left\lvert A_0^{K} \right\rvert^2 = \frac{A_0^2}{1 + A_0^2}\ .
\ee
 However it is very important to stress that this K-matrix partial wave {\emph{is not analytical}} (in the first Riemann sheet) and consequently 
%it is lacking the proper analytical structure required to the 
it is not a proper partial wave  $A(s)$ 
%by unitarity and analyticity 
compatible with microcausality. For example, even if $A_0(s)$ has a LC, the corresponding $A^K(s)$ does not show any RC and then it cannot define a second Riemann sheet. Most of the criticisms on the unitarization methods appearing in the literature are based in the fact that some of them (for example the IAM or the N/D) can produce resonances (poles in the second Riemann sheet) but others (typically the K-matrix) cannot. However it is obvious, from the discussion above, that discrepancy is coming from the limitations (lack of proper analytic structure) of the K-matrix. The $A_0^K(s)$ partial-wave is defined only in the physical region and cannot be extended to the whole complex plane. 

So we insist here that this naive K-matrix has no RC, cannot be extended to the second Riemann sheet, and therefore it cannot produce poles that could be understood as resonances. However, from our experience in unitarization methods in hadron physics, we know that such poles frequently appear and describe well known hadronic resonances \cite{Dobado:1992ha,JRA}. The original K-matrix method cannot reproduce these hadronic resonances and should be considered as less appropriate than other methods that are, not only unitary,  but also analytical, as it is the case of the IAM or N/D methods. 

Nevertheless the K-matrix method can be improved as follows: we can introduce the analytical function
\be
g(s)=\frac{1}{\pi}\left(C+\log\frac{-s}{\mu^2}\right) 
\ee
where $C$ is an arbitrary constant and $\mu$ is also an arbitrary scale. One interesting possibility is to define $C$ 
as in Eq.~(\ref{defgwithB}) so that $g(s)$ becomes $\mu$ independent (which is the one we will be using in the rest of the paper). In any case this function is analytical in the whole complex plane but for a RC. In the physical region on this RC we have:
\be
g(s)=\frac{1}{\pi}\left(C+\log\frac{s}{\mu^2}-i\pi\right) 
\ee
and thus its imaginary part is simply
\be
\Imag g(s)= -1.
\ee
Therefore it is tempting to perform the formal substitution: $-i \rightarrow g(s)$ in the K-matrix method to get what we will call ``improved K-matrix'' (IK) amplitude:
\be
A^{\rm IK}(s)  = \frac{A_0(s)}{1+g(s)A_0(s)}.
\ee
This new amplitude is, not only unitary, but also analytical on the whole complex plane but for a RC that allows 
for analytical continuation to the second Riemann sheet, making possible the existence of poles 
as in the IAM or N/D methods. To apply this improved K-matrix method to our $\omega\omega$ amplitudes, we can start by taking $A_0(s)=A^{(0)}(s)$ to get
\be
A^{\rm IK}(s)=\frac{A^{(0)}(s)}{1+g(s)A^{(0)}(s)}\ .
\ee
Interestingly enough this amplitude may also be obtained from the twice-subtracted N/D method by setting in Eq.~(\ref{N/D_D_n_subtr})
\be
 h_1=h_1(\mu)=\frac{B(\mu)K}{\pi(D+E)}\ .
\ee
A more accurate result can be obtained by defining $A_0(s)=A^{(0)}(s)+A_L(s)$ which leads to 
\be
A^{\rm IK}(s)=\frac{A^{(0)}(s)+A_L(s)}{1+g(s)[A^{(0)}(s)+A_L(s)]} \ .
\ee
or:
\be
A^{\rm IK}(s)=\frac{A^{(0)}(s)+A_L(s)}
{1-\frac{A_R(s)}{A^{(0)}} -\frac{A_L(s)A_R(s)}{(A^{(0)})^2}   }
\ee
This amplitude has the proper analytical behavior, is unitary and  reproduces the NLO result up to order  $s^2$ since $A_R(s)=- g(s) (A^{(0)})^2$. 

In addition, this improved K-matrix method can also be extended to the coupled-channel case, simply taking
\be \label{Kcoupled}
 F^{\rm IK}(s) = \left(1 + G N_0\right)^{-1} N_0
\ee
where again:
\be
N_0(s)=F^{(0)}(s)+F_L(s).
\ee
and $G$ is defined in Eq.~(\ref{Gmatrix}).

%%%%%%%%%%%%%%%%%%%%%%%%%%%%%%%%%%%%%%%%%%%%%%%%%%%%%%%%%%%%%%%%%%%%%%%%%%
\subsection{The large N method}
%%%%%%%%%%%%%%%%%%%%%%%%%%%%%%%%%%%%%%%%%%%%%%%%%%%%%%%%%%%%%%%%%%%%%%%%%%%

Finally another interesting way to improve  the unitarity behavior of the amplitudes is the so called large-$N$ limit. It is based on the observation that our coset space for the electroweak SBS is 
$SU(2)_L\times SU(2)_R/SU(2)_{L+R}=SO(4)/SO(3)=S^3$.  
This suggests considering a generalization to $SO(N+1)/SO(N)=S^N$ and computing the WBGB scattering amplitudes in the non-perturbative large-$N$ limit. These amplitudes were studied in detail in~\cite{Dobado:1996pp} for the case of the minimal SM and one of their main properties is their unitarity up to NLO corrections in the $1/N$ expansion.

However there is a limitation to the $1/N$ expansion as applied as an unitarization method: 
all channels happen to be $1/N$-suppressed with respect  to the $IJ=00$. Therefore this approximation is not appropriate to describe models in which other channels could be relevant, for example, those showing vector-meson dominance (such as Composite Higgs Bosons with low-energy $W'$ and $Z'$ resonances). Thus we will not consider this approach here, but we have shown an example of its use in~\cite{Delgado:2013loa}. 

%%%%%%%%%%%%%%%%%%%%%%%%%%%%%%%%%%%%%%%%%%%%%%%%%%%%%%%%%%%%%%%%%%%%%%%%
\subsection{Summary of the various unitarization methods}
%%%%%%%%%%%%%%%%%%%%%%%%%%%%%%%%%%%%%%%%%%%%%%%%%%%%%%%%%%%%%%%%%%%%%%%%

It has now become clear that of the several unitarization methods considered above, three stand out as acceptable, the IAM in sec.~\ref{sec:IAM}, the version of the N/D method obtained here in sec.~\ref{sec:NoverD} and the IK method from subsec.~\ref{subsec:K}. 
Let us gather their expressions for the elastic channels, writing them all in terms of $A^{(0)}$, $A_L$, $A_R$ from Eq.~(\ref{splitLR}) and $g(s)$ from Eq.~(\ref{defgwithB}), for easy comparison;
\begin{eqnarray} \label{Atogether}
A^{\rm IAM}(s) &=&  \frac{[A^{(0)}(s)]^2}{A^{(0)}(s)-A^{(1)}(s)} \\
 &=& \frac{A^{(0)}(s)+A_L(s)}{1-\frac{A_R(s)}{A^{(0)}(s)} -\left(\frac{A_L(s)}{A^{(0)}(s)}\right)^2 +g(s)A_L(s)}\nonumber\\
A^{\rm N/D}(s)& = &   \frac{A^{(0)}(s)+A_L(s)}{1-\frac{A_R(s)}{A^{(0)}(s)} +\frac{1}{2}g(s)A_L(-s)}\nonumber\\
A^{\rm IK} (s)& = &   \frac{A^{(0)}(s)+A_L(s)}{1-\frac{A_R(s)}{A^{(0)}(s)} +g(s)A_L(s)}. \nonumber
\end{eqnarray}
All three amplitudes are IR and UV finite, $\mu$ independent, unitary, have the proper analytical structure, can be generalized to the coupled-channel case [see the corresponding formulae in Eqs.~(\ref{IAMforF}), (\ref{NDcoupled}) and (\ref{Kcoupled})] and they reproduce the NLO predictions of  EWChPT. This attribute means that they differ from each other only at $O(s^3)$,
\be
A^{NLO}(s)=A^{0}(s)+A^{(1)}(s)=A^{\rm IAM}(s)+ O(s^3) =A^{\rm N/D}(s)+ O(s^3)=A^{\rm IK}(s)+ O(s^3)\ .
\ee
Thus these three unitarization methods each provide a consistent UV completion of the low energy chiral amplitudes. Unfortunately, as energy grows their predictions will start differing.
Then, which of them is a better description of reality?
% In section~\ref{sec:numericcomp} we will study numerically these three unitarization procedures and we will compare their results to try to answer this question. 
In principle all of them are consistent but their domain of applicability will be different.

First notice that the IAM method is the only one that does not really require the splitting of $A^{(1)}$ into $A_L$ and $A_R$ (or the use of the $g(s)$ or $G(s)$ function). This splitting is in fact in some way arbitrary, since we can always add and subtract a quadratic term $C s^2$ to $A_L$ and $A_R$ respectively without changing their fundamental properties.  Notice also that the splitting is not possible at all whenever $D+E=0$ (as in the $I=J=1$ channel for the particular parameter choice $a=b$) and the N/D and IK methods cannot be constructed for that case. Hence, for the vector-isovector channel, the IAM is most appropriate. Since for $D+E$ small, $A_L\sim A_R$, the three methods are not expected to be equivalent, and we see that there are sound theoretical reasons to choose the IAM over the other two.

Conversely the IAM method cannot be applied in the cases where $K=E=0$ which  happens in the $J=2$ channels (because they start at NLO in the effective theory, so $K=0$, and then perturbative unitarity forces $E=K^2$ that also vanishes). In that case the IAM is not usable and the N/D method comes to the fore.

In section~\ref{sec:numericcomp} we will provide numerical results for the various situations to illustrate how the three unitarization methods work in the different channels and to try a comparison when all are applicable. For a brief summary, see table~\ref{tab_val_methods}.
\begin{table}[ht]
{\renewcommand{\arraystretch}{1.4}%
\newcommand{\bb}[1]{%
\begin{minipage}[c]{.1\textwidth}%
\vspace{.5\baselineskip}#1\vspace{.5\baselineskip}%
\end{minipage}}%
\begin{tabular}{|c|*{5}{c}|}
\hline
$IJ$    &   00  & 02    & 11    & 20    & 22    \\
\hline
  \bb{Method\\ of\\ choice}
& \bb{Any}
& \bb{N/D\\ IK}
& \bb{IAM}
& \bb{Any}
& \bb{N/D\\ IK} \\
\hline
\end{tabular}}
\caption{\label{tab_val_methods}%
Unitarization methods usable in each $IJ$ channel. See section~\ref{sec:numericcomp}.}
\end{table}

%%%%%%%%%%%%%%%%%%%%%%%%%%%%%%%%%%%%%%%%%%%%%%%%%%%%%%%%%%%%%%%%%%%%%%%%
\subsection{Resonances}
%%%%%%%%%%%%%%%%%%%%%%%%%%%%%%%%%%%%%%%%%%%%%%%%%%%%%%%%%%%%%%%%%%%%%%%%

As already mentioned, one of the more interesting properties of the IAM, N/D and IK partial waves is the possibility of finding poles in the second Riemann sheet under the real axis. This interest arises because these poles have the we can use the simple of dynamical resonances, at least when they lie close enough to the real axis in the complex $s$ plane.

For the amplitudes considered here the non trivial analytical behavior is coming exclusively from the logs which are defined in the first Riemann sheet as usual ($log(z)=\log (\lvert z \rvert) + i\arg(z)$ with the $\arg(z)$ cut lying along the negative real axis).
To find a pole in the second Riemann sheet, an option is to extend all the logarithms to it, through the simple equation
\be
\log^{II}(-z)=\log (\lvert z \rvert) + i[\arg(z)-\pi] 
\ee
 and then find zeroes of the denominators of the amplitudes $A^{II}$ or $F^{II}$ for coupled channels. This is the strategy that we followed in~\cite{Delgado:2013loa}.

An alternative is to observe that given some analytical elastic amplitude $A(s)=A^{I}(s)$ representing the physical (first) Riemann sheet, the second Riemann sheet in the quadrant under the physical region can be obtained as (see for example~\cite{Gribov}):
\be
\label{anapro}
A^{II}(s)=\frac{A(s)}{1- 2 i A(s)}.
\ee 
Therefore resonances under the real, physical $s$ axis (the right cut) are located at points $s_R$ solving the resonance equation
\be
\label{resoneq}
A(s_R)+\frac{i}{2}=0
\ee
so that the extension of the logarithms is unnecessary.

The mass $M$ and width $\Gamma>0$ of the resonance can be extracted from its position, $s_R=M^2- i \Gamma M$. 
Equivalently we have $s_R= \lvert s_R \rvert e^{-i \theta}$ with $\theta > 0$ and $\tan\theta = \gamma = \Gamma/M$. The resonance equation~(\ref{resoneq}) obviously takes a different form for each of the unitarization methods, which we now show in turn. 
For the IAM method,
\be \label{IAMRES}
A^{(0)}(s_R)-A^{(1)}(s_R)- 2 i [A^{(0)}(s_R)]^2=0
\ee
whilst for the N/D method we find
\be
A^{(0)}(s_R)-A_R(s_R)+ \frac{1}{2}g(s_R)A^{(0)}(s_R)A_L(-s_R)- 2 iA^{(0)}(s_R) [A^{(0)}(s_R)+A_L(s_R)]  =0
\ee
and for the IK method,
\be \label{IKresonanceeq}
A^{(0)}(s_R)-A_R(s_R)+ g(s_R)A^{(0)}(s_R)A_L(s_R)- 2 iA^{(0)}(s_R) [A^{(0)}(s_R)+A_L(s_R)] =0.
\ee
These resonance equations are all $\mu$  independent through cancellation of their  explicit  and implicit (through the renormalized chiral parameters) dependence on $\mu$. As expected they are different, but decreasingly so in the limit $A_L(s_R) \ll 1$, since $A^{(1)}(s_R)=A_R(s_R)+A_L(s_R)$. 

If we find a solution $s_R$ for some given channel $IJ$ and some given unitarization method $X=$ IAM, N/D, IK in the appropriate region $M, \Gamma > 0$ this solution will be a $\mu$ invariant function of the $a, b$ and the renormalized chiral parameters, i.e.
\ba
M &=&M(a,b, a_4(\mu),  a_4(\mu), d(\mu), e(\mu), g(\mu) ; \mu) \\ \nonumber
\Gamma&=&\Gamma (a,b, a_4(\mu),  a_5(\mu), d(\mu), e(\mu), g(\mu) ; \mu).
\ea 
These functions trivially fulfill the observable renormalization group equations
\ba
\frac{d M}{d  \mu}&=&\frac{\partial  M}{\partial  \mu}+\frac{\partial  M}{\partial  a_4}\frac{d a_4}{d  \mu}+\frac{\partial  M}{\partial  a_5}\frac{d a_5}{d  \mu}+... =0  \\ \nonumber
\frac{d \Gamma}{d  \mu}&=&\frac{\partial  \Gamma}{\partial  \mu}+\frac{\partial  \Gamma}{\partial  a_4}\frac{d a_4}{d  \mu}+\frac{\partial  \Gamma}{\partial  a_5}\frac{d a_5}{d  \mu}+...=0.
\ea
If we set a scale and fix the chiral couplings at that scale $\mu_0$, so that $a_4 = a_4(\mu_0),\, a_5 = a_5(\mu_0),...$,  the resonance position becomes a function of the chiral couplings evaluated at this scale only,
\ba
M=M(a,b, a_4,  a_4, d, e, g ) \\ \nonumber
\Gamma=\Gamma (a,b, a_4,  a_5, d, e, g)\ .
\ea

When there is channel coupling, the amplitude matrix elements $F_{ij}(s)$ correspond to different reactions having the same quantum numbers $IJ$. Obviously if there is a resonance at some point $s_R$ in any of them, it should appear also at the same point in the rest of the matrix elements. In other words, the $F_{ij}(s)$ are all different as analytical functions but all of them have the same resonances at the same points since physically these resonances can be produced in any of the $j \rightarrow i$ reactions. 

This property is guaranteed for the three unitarization methods now at hand.
This is because in all of them we need to invert some matrix. Thus the unitarized amplitudes $F_{ij}(s)$ for some given $I$ and $J$  contain always a common denominator which is a determinant depending on the unitarization method. The roots of this determinant in the second Riemann sheet will define the pole position for all the different processes simultaneously.

Once we have obtained the unitarized amplitude $F_{ij}(s)$ by using some coupled unitarization method, and extended it to the corresponding second Riemann sheet $F_{ij}^{II}(s)$, we can find the position of any pole (resonance) in the quadrant below the physical region. In the next sections we will study numerically the different channels as a function of the $a$ and $b$ parameters and the renormalized chiral couplings for the three unitarization methods considered here and we will compare the results obtained.

%%%%%%%%%%%%%%%%%%%%%%%%%%%%%%%%%%%%%%%%%%%%%%%%%%%%%
\subsection{Spurious resonances} \label{subsec:spurious}
%%%%%%%%%%%%%%%%%%%%%%%%%%%%%%%%%%%%%%%%%%%%%%%%%%%%%

In addition to the \emph{bona-fide} resonances in the second Riemann sheet, for certain sets of parameters a given unitarization method can yield a pole in the complex $s$ plane that lies on the \emph{first} Riemann sheet. As recalled below in appendix~\ref{app:numeric}, because of Schwarz's reflection principle, these poles always come in pairs one above and one below the real $s$-axis (see fig.~\ref{fig:poles_b_g} that represents the situation very graphically).

But causality demands that the scattering amplitude be analytic in the upper half-plane, whence the pole in the first Riemann sheet is tachyonic. So, poles appearing in the first or physical Riemann sheet are not acceptable and therefore they set limits on the applicability of the method or even on the validity of the given parameter set (it might be that no underlying theory is compatible with the effective theory with certain parameter values). To search for these poles on the first sheet we just need to find zeroes of the denominators in the representations of $A^{\rm IAM}$, $A^{\rm N/D}$ and $A^{\rm IK}$ in Eq.~(\ref{Atogether}). 

When we find such situation we conclude that the unitarized amplitude with the given parameters is in violation of causality; either there is no underlying theory that can provide such set of parameters~\footnote{Other authors speak of ``negative width resonances'', presumably because of Eq.~(\ref{NLOparamsfromres}), but we do not favor this concept, as it seems a linguistic contradiction in terms.} or the unitarization method is at its limit of validity for such set, and one can take the real part of the corresponding $s$ where such pole appears as a point beyond which the theory is not applicable at all.

Sometimes one can detect this breakup of causality in repulsive phase shifts (such as the isotensor channel) that vary quickly and break Wigner's bound.

In practice, we will consider the regions of parameter space where this phenomenon occurs as excluded. Some examples can be found in figures~\ref{fig:poles_a4_a5}, \ref{fig:poles_ab}, \ref{fig:poles_a2b_a4}, \ref{fig:poles_de_et} and  \ref{fig:poles_b_g} below.  The regions where we find a pole in the first Riemann sheet, are automatically excluded from our parameter space.

%%%%%%%%%%%%%%%%%%%%%%%%%%%%%%%%%%%%%%%%%%%%%%%%%%%%%
\section{Numeric comparison of the three methods} \label{sec:numericcomp}
%%%%%%%%%%%%%%%%%%%%%%%%%%%%%%%%%%%%%%%%%%%%%%%%%%%%%

%%%%%%%%%%%%%%%%%%%%%%%%%%%%%%%%%%%%%
\subsection{The \boldmath $I=J=0$ channel}\label{sec_IJ_00}
%%%%%%%%%%%%%%%%%%%%%%%%%%%%%%%%%%%%%

\begin{figure}
\includegraphics[width=0.32\textwidth]{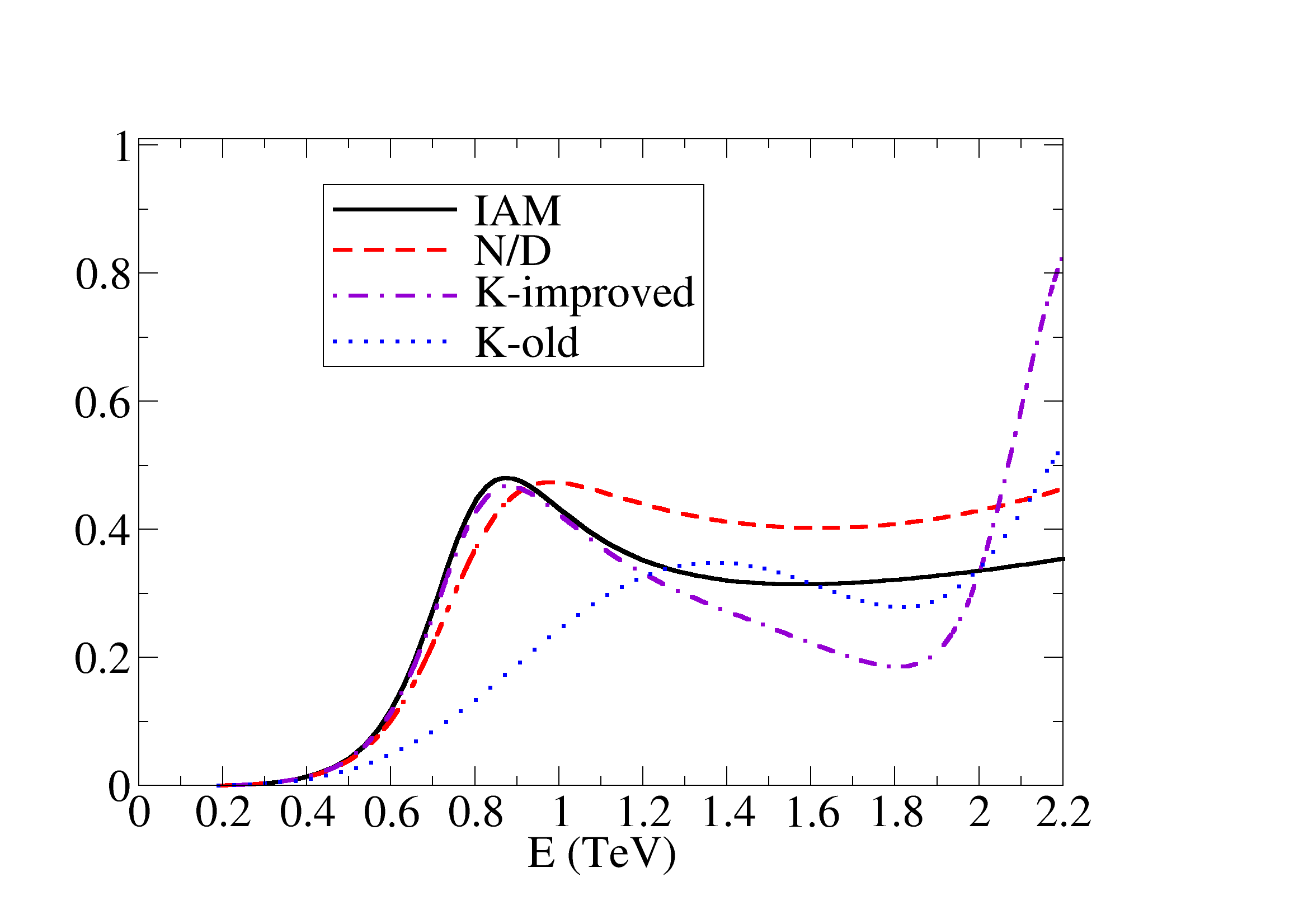}
\includegraphics[width=0.32\textwidth]{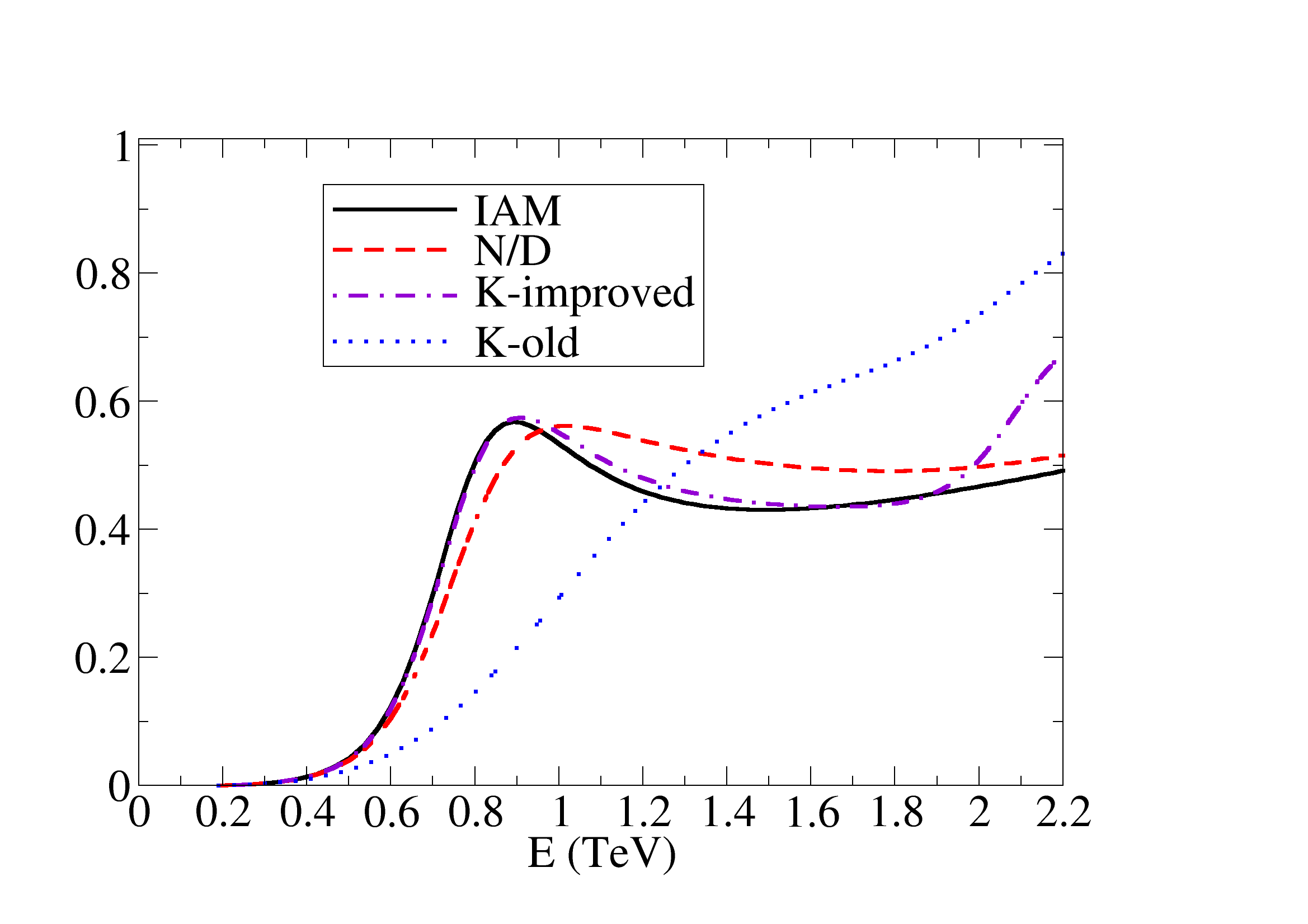}
\includegraphics[width=0.32\textwidth]{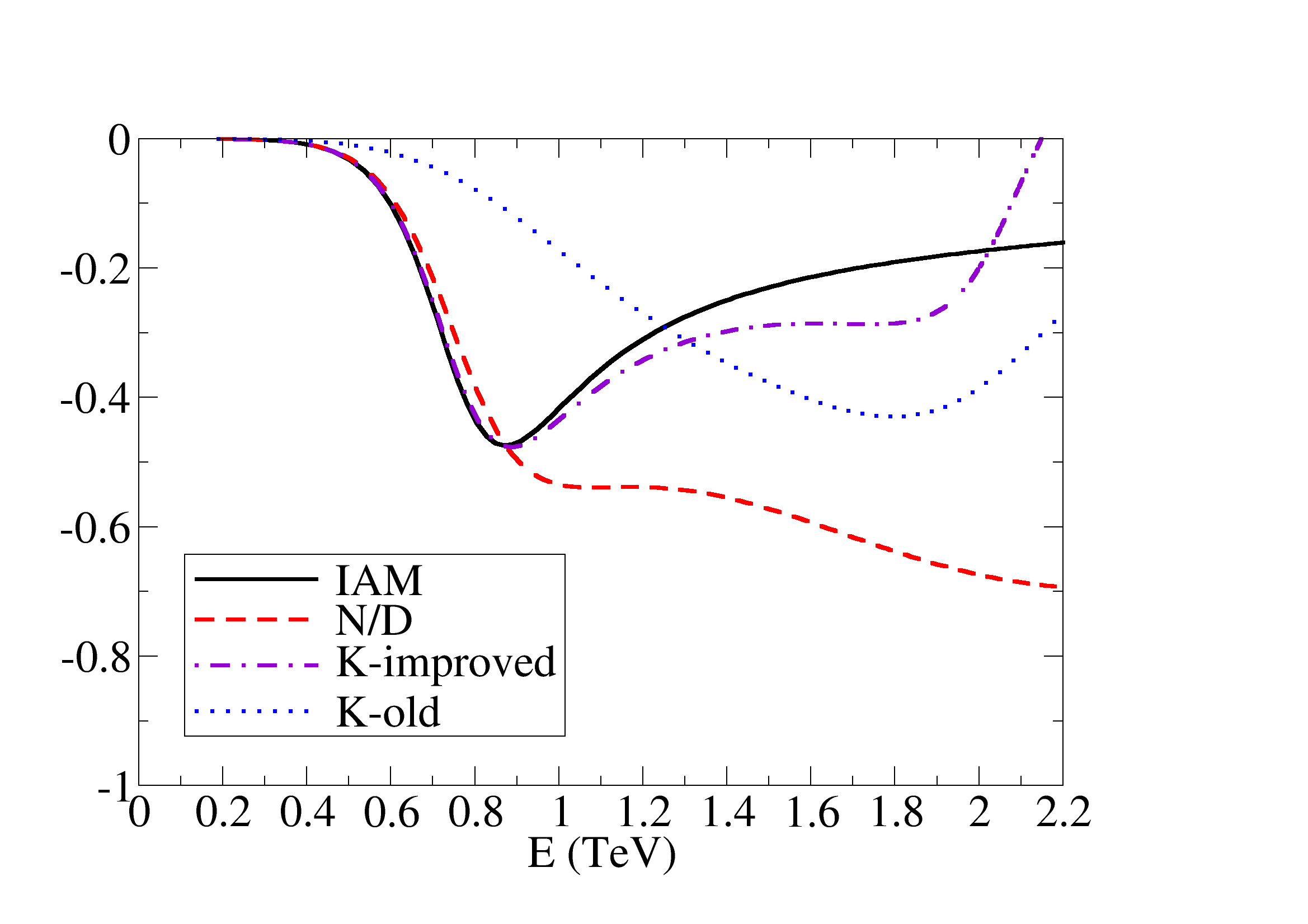}
\caption{\label{fig:compJ0}%
Scalar-isoscalar amplitudes (from left to right, elastic $\omega\omega$, elastic $hh$, and cross-channel $\omega\omega\to hh$), for $a=0.88$, $b=3$, and all NLO parameters set to 0 at a scale $\mu=3$ TeV. Note that, as explained on sec.~\ref{sec_IJ_00}, the old K-matrix method gives different results because its complex-s plane analytic structure is not the correct one. It will be discarded from now on.}
\end{figure}

The scalar-isoscalar channel is a coupled-channel problem with the $\omega\omega$ and $hh$ elastic and crossed reaction forming a symmetric two by two matrix. We represent the two diagonal and the off-diagonal matrix elements as functions of $s$ in figure~\ref{fig:compJ0} for four different  methods, all of which satisfy exact unitarity.
 
The three methods with the correct analyticity properties (IAM, N/D and IK) agree in predicting a scalar resonance that is visible in all three amplitudes between 0.8 and 0.9 TeV. The old K-matrix method gives somewhat different results, as known from the literature, but its complex-s plane analytic structure is not the correct one, as visible in Eq.~(\ref{oldKamplitude}). We therefore discard the old K-matrix method from now on. 

The other three methods are practically in perfect agreement up to the first elastic resonance and they start deviating quantitatively only for higher energies. The reason that there is good agreement between the various methods was discussed under Eq.~(\ref{IKresonanceeq}): since we have set the NLO terms to 0, $A_L$ is small, and the three resonance equations become dominated by the tree-level and right-cut parts of the amplitude, which suggests similar masses for all the methods. 

Note that in~\cite{Delgado:2014dxa} we have shown that the resonance found in figure~\ref{fig:compJ0} appears even if we set $a=1$ (its SM value with one Higgs): it is sufficient that the coupled-channel dynamics is strong through $a^2-b\neq 0$ for it to appear.  Moreover, with the values chosen to prepare the figure this $a^2-b$ is negative, so the cross-channel amplitude $M_J$ shown in the right-most plot is also negative as dictated by Eq.~(\ref{Mscalar}).
At last, observe that the resonance appears in all three elastic or inelastic amplitudes in the same position (though of course, with different shapes due to different backgrounds).

%%%%%%%%%%%%%%%%%%%%%%%%%%%%%%%%%%%%%
\subsection{The \boldmath $I=J=1$ channel}\label{sec_IJ_11}
%%%%%%%%%%%%%%%%%%%%%%%%%%%%%%%%%%%%%
\begin{figure}
\includegraphics[width=0.4\textwidth]{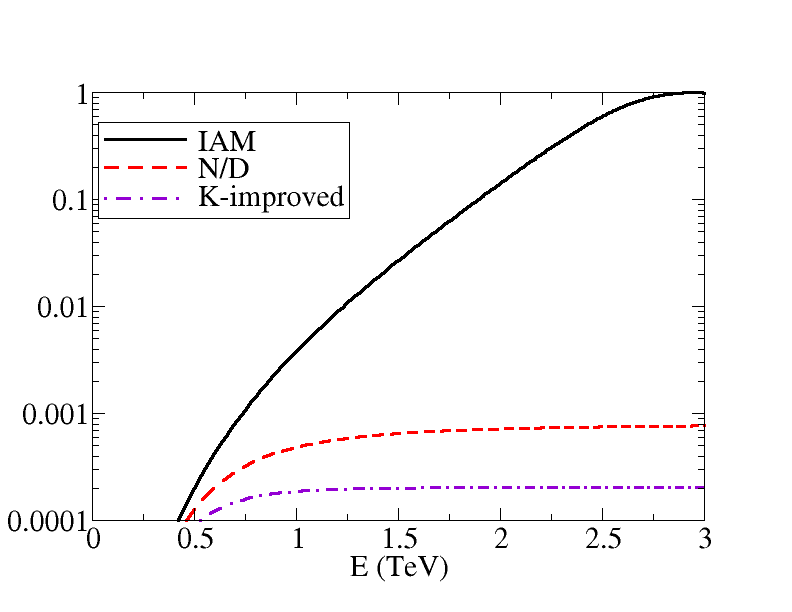}
\includegraphics[width=0.4\textwidth]{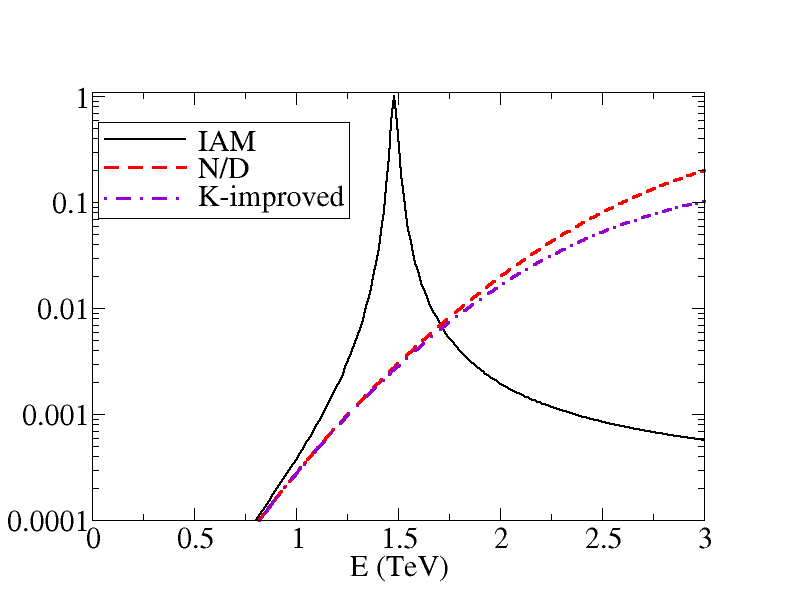}
\caption{\label{fig:compJ1}%
Vector-isovector partial wave. We have taken $a=0.88$ and $b=1.5$, but while for the left plot all the NLO parameters vanish, for the right plot we have taken $a_4=0.003$, known to yield an IAM resonance from the work of the Barcelona group~\cite{Espriu:2013fia}. Note that the N/D and K-improved methods are not reliable in this channel, as explained below on section~\ref{sec_IJ_11}. They are included to show the lack of agreement with the IAM.}
\end{figure}
The comparison between the three methods IAM, N/D and IK for the vector-isovector channel is shown in figure~\ref{fig:compJ1}. First we set all the NLO parameters to 0 (left plot). Clearly, there is no good agreement between the three IAM, N/D and IK methods. Moreover, if we introduce one NLO counterterm with an appropriate value to generate a resonance in the IAM, here  $a_4=0.003$ as an example (right plot), the N/D and IK methods do not react in the same way as the former, and fail to yield a vector resonance.

In order to understand the discrepancy found in this elastic channel we notice that the possibility of defining the N/D and the IK methods depends dramatically on having $D+E \ne 0$ since otherwise we cannot define the $g(s)$ function in Eq.~(\ref{defgwithB}) nor the $A_L$ and $A_R$ splitting in Eq.~(\ref{splitLR}). 

But here comes the coincidence, in the $I=J=1$ channel we have
\be \label{DnearE}
D_{11}+E_{11}=\frac{3}{(96)^2\pi^3v^4}(a^2-b)^2
\ee
which vanishes for $a^2=b$. This is in particular the case of the SM where $a=b=1$, which is not very important for our discussion because there are no strong interactions to start with. 
More importantly, $a^2=b$ is also satisfied by the Higgsless electroweak chiral perturbation theory, characterized by $a=b=0$.
This situation is already ruled out by the discovery of the light Higgs-like particle, 
but it is still interesting because it is equivalent to two-flavor low-energy QCD in the chiral limit with $v$ playing the role of $f_\pi$ and the WBGB being the pions.

\begin{figure}
\includegraphics[width=0.45\textwidth]{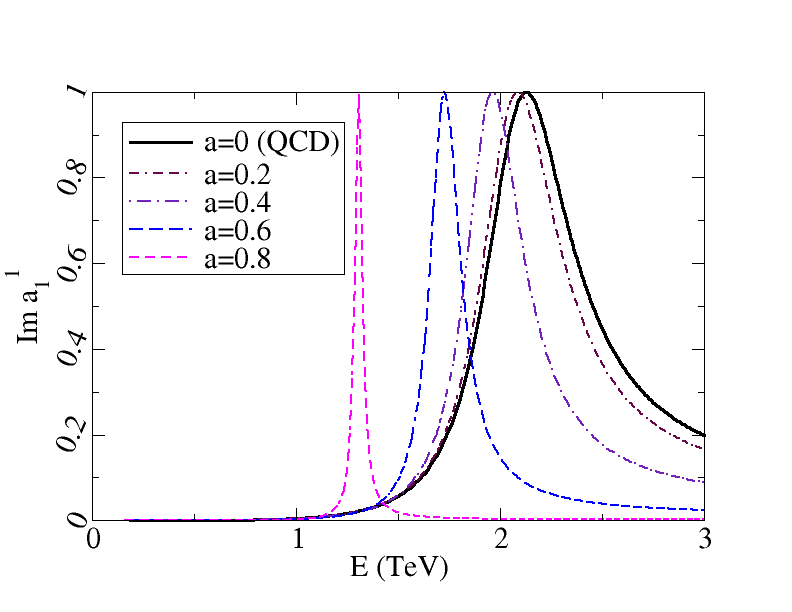}
\caption{\label{QCDcomparison}%
We show the vector-isovector resonance with NLO $a_4$, $a_5$ parameters taken from large-$N_c$ QCD, $b=a^2$ and $a$ as shown in the legend. The right-most solid line is the rescaled QCD case, towards the left we approach the EWSBS with a Higgs, where the resonance is narrow and relatively light for these $a_4$, $a_5$.}
\end{figure}

Within $a=0=b$, we know that a vector resonance (the $\rho$) appears in the spectrum (because we can look up the answer in QCD), and know what the low-energy parameters are, with good approximation. Figure~\ref{QCDcomparison} shows the result of the calculation with the IAM (solid line). We have there taken $a=b=0$ and $a_4=-2a_5=\frac{3}{192\pi^2}$, the large-$N_c$ prediction for these NLO constants (others taken to 0). The $\rho$ vector-isovector resonance then comes with reasonable parameters (to see it, substitute
$v=246\,{\rm GeV}$ by $f=92\,{\rm MeV}$ in the scale; this amounts to $m_\rho\simeq 2.1\,{\rm TeV}\to 0.79\,{\rm GeV}$, just slightly above the actual $0.775\,{\rm GeV}$ in the hadron spectrum).

The other lines in figure~\ref{QCDcomparison} have been computed by increasing $a$ towards 0.88, the value taken for figure~\ref{fig:compJ1}. One sees without doubt how the QCD-like resonance becomes narrower and lighter (this depends on the interplay of $a$ with the NLO parameters $a_4$, $a_5$), matching the calculation of figure~\ref{fig:compJ1}.
We find that the IK and N/D methods fail to provide a resonance.
Therefore, the IAM is the method of choice for the vector-isovector channel, given that the other two fail at least over the $a^2=b$ parameter election, while the IAM yields a resonance that can be continuously matched to the one we know is there for that parameter set.

The resonance may be exactly fit to data with an adequate choice of the $a_4$ and $a_5$ chiral parameters to adjust its mass and width. Beyond trial and error, an elegant method is to couple the resonance to the Chiral Lagrangian in a chiral invariant way and then integrate the resonance at tree level as done for example in~\cite{Donoghue:1988ed} (see also the early treatment by~\cite{Pham:1985cr} and the more formal one in~\cite{Ecker:1988te}, as well as that in the context of Composite Higgs Models in~\cite{Contino:2011np}). The tree-level chiral couplings obtained take the general form  
\be \label{couplingsfromresonances}
a_i^{\rm tree}=\eta_i \gamma^{\rm tree} \left(\frac{v}{M^{\rm tree}}\right)^4
\ee
where $i=4,5$, $\eta_4= -\eta_5= 12 \pi$ and $\gamma^{\rm tree}=\Gamma^{\rm tree}/M^{\rm tree}$ with $M^{\rm tree}$, $\Gamma^{\rm tree}$ being the tree-level vector-resonance parameters. Thus the tree-level $s^2$ term induced by the resonance is
\be
A_{11}^{\rm tree}(s)=s^2 \left( p_4 a_4^{\rm tree}+p_5 a_5^{\rm tree}  \right)
\ee
where the $p_4$ and $p_5$ constants are obtained from Eq.~(\ref{Bfunction}), 
$B_{11}(\mu)=B_0+p_4 a_4(\mu)+p_5 a_5(\mu) $,
and are given by $p_4=1/(24 \pi v^4)$ and $p_5 = -2 p_4$. Following~\cite{Donoghue:1988ed}
we can now obtain the contribution to the renormalized chiral couplings induced by the resonance by matching the $O(s^2)$ tree level amplitude with the NLO result at the point $s=M^{\rm tree\ 2}$, i. e.
\be
A_{11}^{\rm tree}(M^{\rm tree\ 2})=\Real A_{11}^{(1)}(M^{\rm tree\ 2}).
\ee
This identification leads us to
\be \label{NLOparamsfromres}
a_i(M^{\rm tree})=\eta_i \gamma^{\rm tree} \left(\frac{v}{M^{\rm tree}}\right)^4-\frac{B_0}{p_4+p_5}
\ee
for $i=4,5$. Therefore we get
\be
A^{(1)}_{11}(s) =s^2 \left( \frac{3 \gamma^{\rm tree}}{2M^{\rm tree\ 4}}+ D_{11}\log \frac{s}{M^{\rm tree\ 2}}  +
E_{11}\log \frac{-s}{M^{\rm tree\ 2}} \right).
\ee
Then the IAM resonance equation~(\ref{IAMRES}) leads us to the second Riemann-sheet resonance parameters in the narrow-resonance limit $\gamma =\Gamma/M\ll 1$:
\ba
\left(M^{\rm IAM}\right)^2&=&\frac{K_{11}}{B_{11}(M^{\rm tree})} \\ \nonumber
\Gamma^{\rm IAM} &=&\frac{K_{11}^2 M^{\rm IAM}}{B_{11}(M^{\rm tree})}
\ea
which implies the $M^{\rm tree}$-independent result  
$\gamma^{\rm IAM} = K_{11} \left(M^{\rm IAM}\right)^2$ or:
\be
\Gamma^{\rm IAM} =\frac{\left(M^{\rm IAM}\right)^3}{96 \pi v^2}(1-a^2)
\ee 
which is recognizable as a version of the KSFR relation~\cite{Kawarabayashi:1966kd,Riazuddin:1966sw} (slightly generalized to $a\neq 0$). This is here a restriction arising from the constraint of exact unitarity, that has been discussed in~\cite{Truong:1988zp} and references therein and is a non-trivial relation between three observable quantities.

Also we have the equation:
\be
M^{\rm IAM} = M^{\rm tree}\left(\frac{2 \Gamma^{\rm IAM}}{3\Gamma^{\rm tree}} \right)^{1/4}
\ee
which relates the resonance parameters with the tree level ones. This is a very consistent result showing that the IAM method properly predicts a vector resonance whenever $M^{\rm tree}, \Gamma^{\rm tree} >0$, 
in which case the chiral parameters receive a contribution and may be dominated by a vector resonance.
For example $M^{\rm IAM}=M^{\rm tree}$ implies $\Gamma^{\rm IAM} = (3/2)\Gamma^{\rm tree}$ which is a quite reasonable result taken into account the tree-level nature of the vector field integration performed to estimate the chiral parameters.

However the N/D and IK unitarization methods fail to predict this resonance for the appropriate values of the chiral parameters. First they are not even defined for $a=b$. For $a \ne b $ but still  in the parameter region close to the SM where $a \sim b \sim 1$ we have $D_{11}+E_{11} \sim 0$. In this case the methods are well defined but then $A_L \sim A_R$ which means that the IAM method is very different from the N/D and IK methods. Thus, as the IAM method works pretty well in this channel according to the previous discussion, we have to conclude that the other two methods are not appropriate to describe the vector  channel.

%%%%%%%%%%%%%%%%%%%%%%%%%%%%%%%%%%%%%%%%%%%%%%%%%%%%%
\subsection{Scalar-Isotensor \boldmath $J=0$, $I=2$ channel}
%%%%%%%%%%%%%%%%%%%%%%%%%%%%%%%%%%%%%%%%%%%%%%%%%%%%%
\begin{figure}
\includegraphics[width=0.4\textwidth]{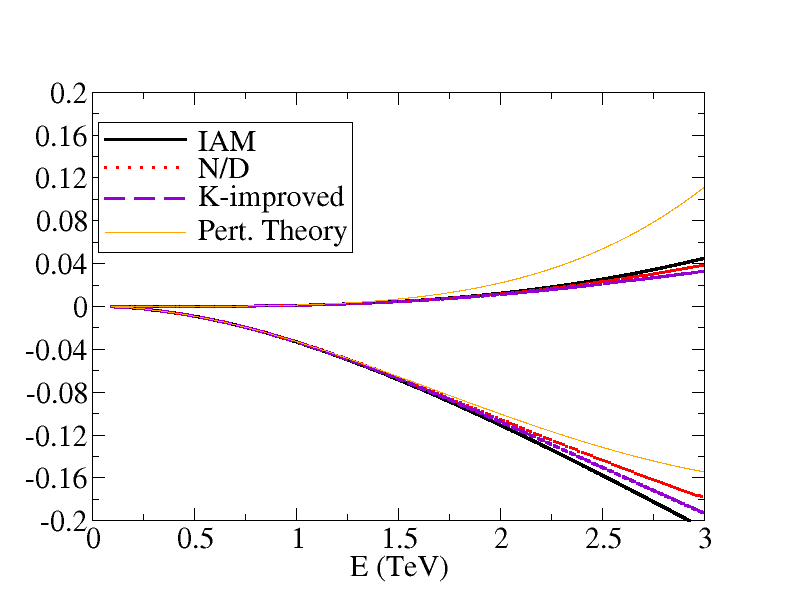}
\caption{\label{fig:isotensorcomp}%
Scalar-isotensor amplitudes for $a=0.88$, $b=a^2$, and the NLO parameters set to 0. All three unitarization methods agree qualitatively and with the perturbative amplitude too, as loop corrections are small. Here we plot both the imaginary part (top set of lines) and the real part (bottom set). That the real part is negative reflects the repulsive interaction in this channel given by $-(1-a^2)<0$ in Eq.~(\ref{partial20}).}
\end{figure}

We now consider the isotensor channel (where a resonance, if there ever was one, would distinctly appear for example in equal-charge $w^+w^+$ spectra). Figure~\ref{fig:isotensorcomp} shows the resulting amplitude for $a=0.88$, $b=a^2$ and all NLO parameters set to 0.

We plot both the real and the imaginary parts of the three unitarized amplitudes and obtain a moderately weak, repulsive partial wave that does not bind a resonance (as seen from the negative real part). All three unitarization methods give a consistent picture: the unitarized interaction has a slightly larger real part and slightly smaller imaginary part than the (unitarity-violating) perturbative one.

\begin{figure}
\includegraphics[width=0.4\textwidth]{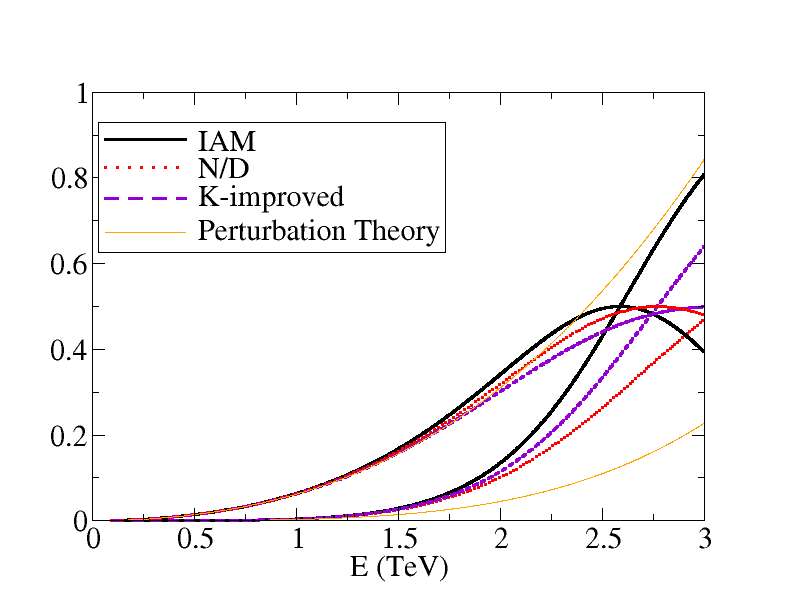}
\caption{\label{fig:isotensorcomp2}%
Scalar-isotensor amplitudes for $a=1.15$, $b=a^2$, and the NLO parameters set to 0. All three unitarization methods agree qualitatively once again, even though now the amplitudes are strong. The real part (corresponding to the set of lines larger at low-$E$, since it receives a tree-level contribution unlike the imaginary part) is now positive because of the sign reversal of $(1-a^2)$ respect to figure~\ref{fig:isotensorcomp}.}
\end{figure}

In figure~\ref{fig:isotensorcomp2} in turn we plot the same isotensor amplitude for 
$a=1.15$. Now the real part has opposite sign (attractive interaction) and grows more rapidly, with all the unitarization methods agreeing and once more tracking perturbation theory until about the end of our energy interval at $3\,{\rm TeV}$.

%%%%%%%%%%%%%%%%%%%%%%%%%%%%%%%%%%%%%%%%%%%%%%%%%%%%%
\subsection{Tensor isoscalar channel with \boldmath $J=2$, $I=0$}
%%%%%%%%%%%%%%%%%%%%%%%%%%%%%%%%%%%%%%%%%%%%%%%%%%%%%
In hadron physics there is a well known $f_2(1270)$ resonance that is broad and visible in $\pi^+\pi^-$ spectra. Its mass is well above the $775\,{\rm MeV}$ of the vector $\rho$, which is natural because the $d$-wave is smaller than the $p$-wave due to the $p^{l}$ suppression factor near threshold.

\begin{figure}
\includegraphics[width=0.4\textwidth]{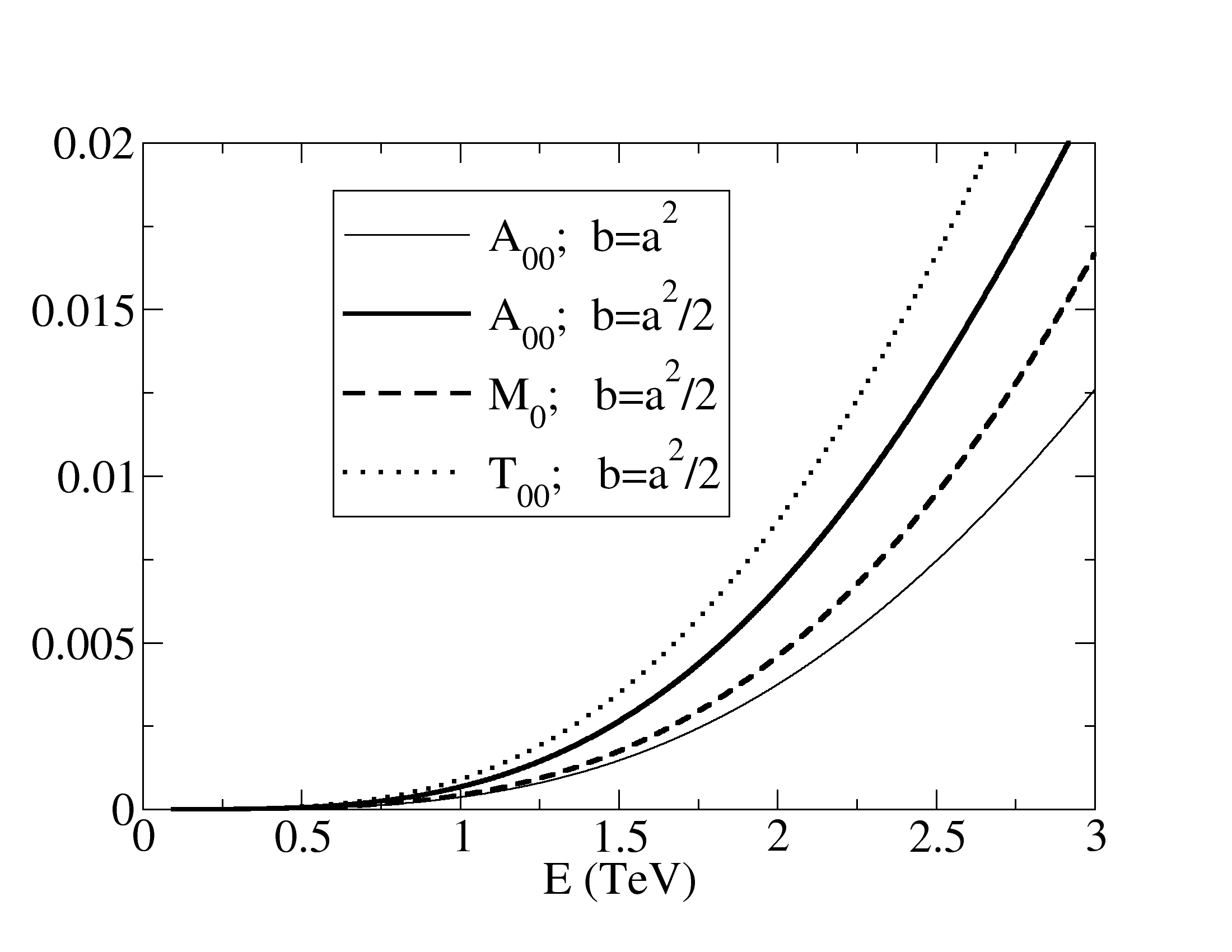}
\caption{\label{fig:tensorpert}%
Tensor-isoscalar amplitude for $a=0.88$, $b$ as shown, and the NLO parameters set to $0$. The amplitude is real.}
\end{figure}

In figure~\ref{fig:tensorpert} we show the tensor-isoscalar channel in perturbation theory, which is indeed small, with all the NLO parameters set to 0, and $a=0.88$, $b$ as shown in the figure legend. This is once equal to $a^2$ to show the elastic amplitude, and once equal to $a^2/2$ to see the other, inelastic and $hh$ amplitudes. All are of course real and quadratic in $s$ (because $K_{02}=0$, the LO $O(s)$ vanishes).

\begin{figure}
\includegraphics[width=0.4\textwidth]{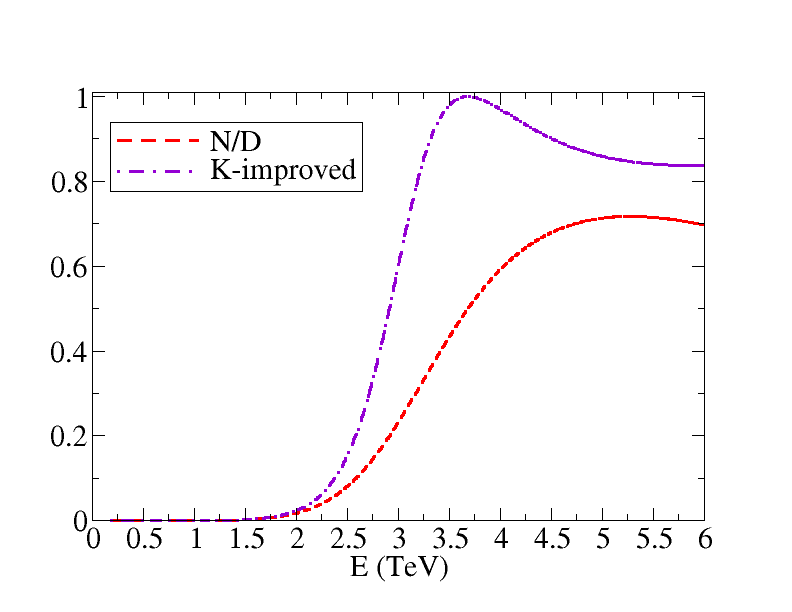}
\includegraphics[width=0.4\textwidth]{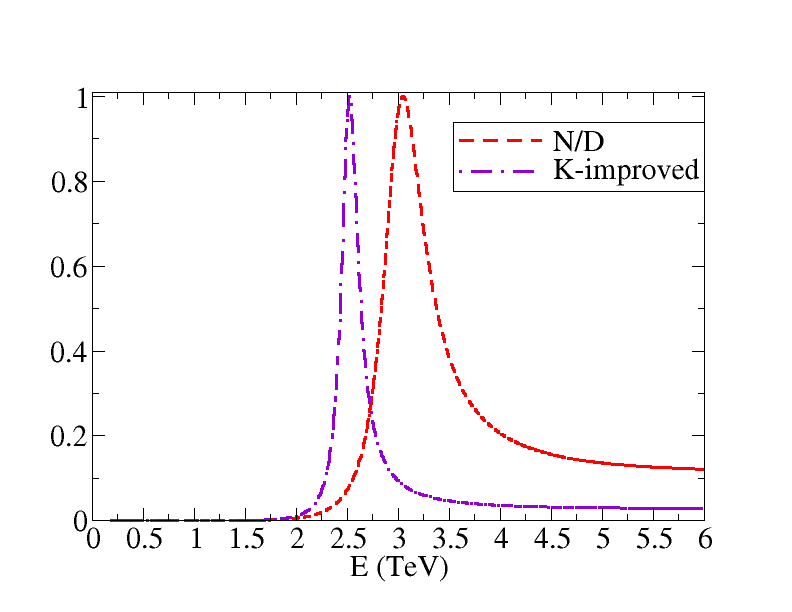}
\caption{\label{fig:tensorcompare_1ch}%
Comparison of the two available methods of unitarization for the isoscalar-tensor channel $I=0$, $J=2$ with $b=a^2$ (only one channel). The $a_4$, $a_5$ constants have been fixed to their values in large-$N_c$ gauge theory, so the left plot with $a$=0 reproduces the QCD situation with a broad, heavy $f_2$-like resonance. The right plot for $a=0.88$ shows how this becomes narrow. Both methods agree well. 
}
\end{figure}

Next we show, in figure~\ref{fig:tensorcompare_1ch}, the comparison between the N/D and IK method in unitarizing the partial wave with $I=0$, $J=2$. The IAM method vanishes and cannot be used without information from NNLO, because here the LO in perturbation theory is zero ($K_{02}=0$).

In the left plot we have set $a=b=0$ and $a_4=-2a_5=\frac{3}{192\pi^2}$
as in figure~\ref{QCDcomparison}.
The IK method clearly shows, and the N/D method is suggestive of, a QCD-like $f_2$ resonance (rescaling again $v=246\,{\rm GeV}$ to $f_\pi=92\,{\rm MeV}$, the $3.5\,{\rm TeV}$ resonance mass becomes $1.3\,{\rm GeV}$, in very good agreement with the experimental $1.27\,{\rm GeV}$ $f_2$ resonance in the hadron spectrum; and this with no free NLO parameters, since they are taken from large-$N_c$). 

In the right plot we have now increased $a=0.88$, with $b=a^2$ still fixed to avoid the coupled-channel situation. The resonance is seen to become lighter and narrower, and both unitarization methods qualitatively agree in predicting the resonance though the mass is slightly different.

\begin{figure}
\includegraphics[width=0.32\textwidth]{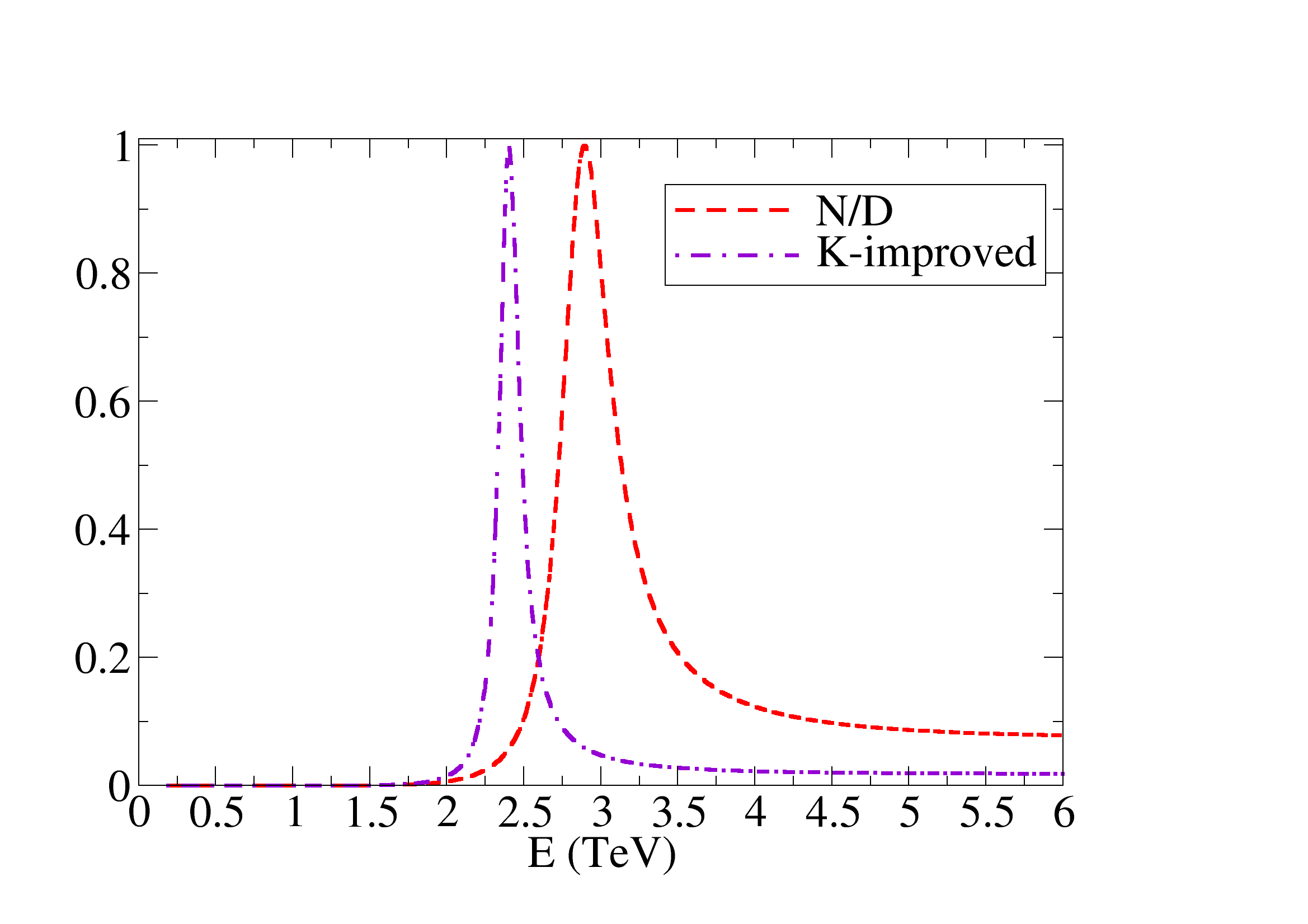}
\includegraphics[width=0.32\textwidth]{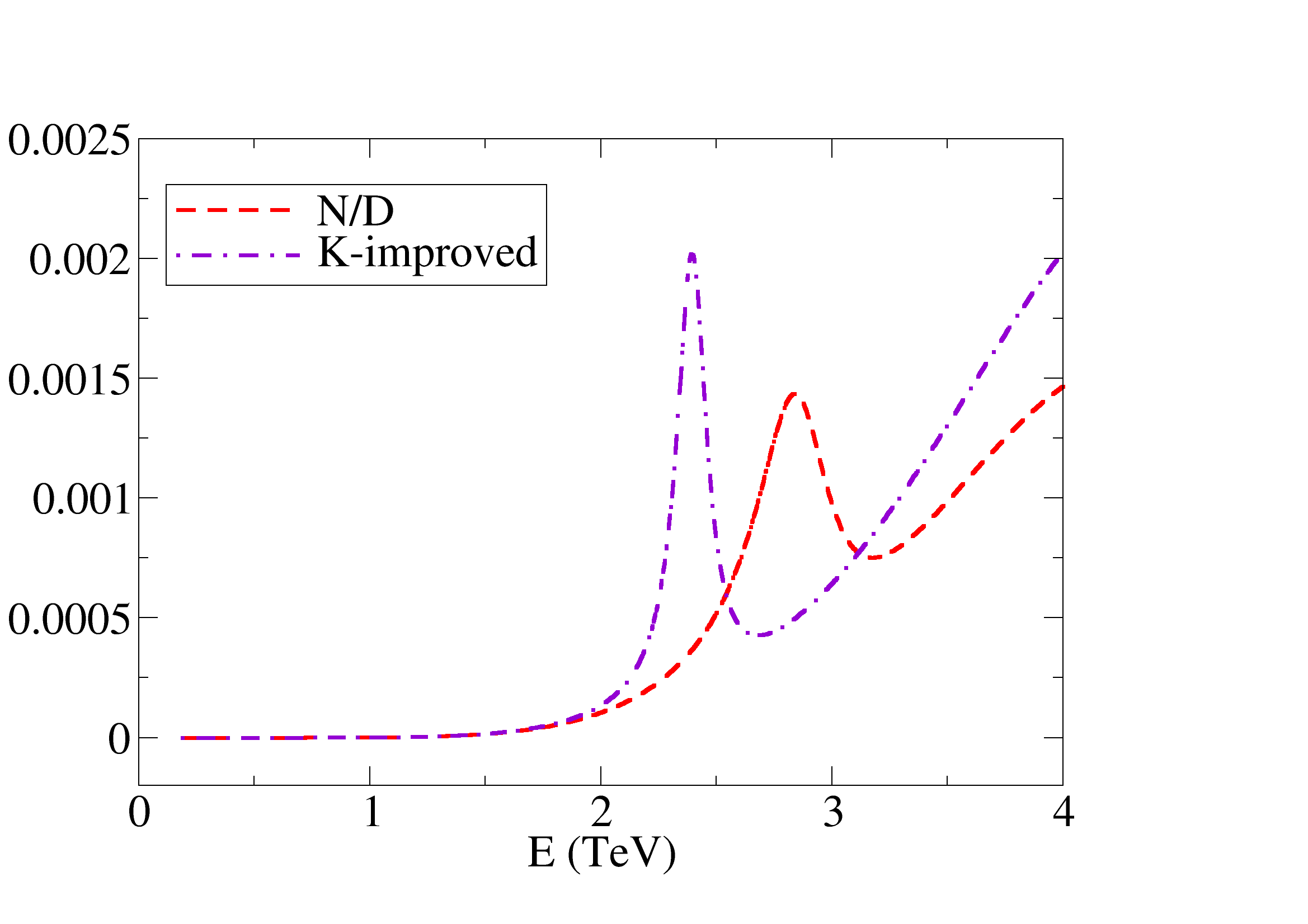}
\includegraphics[width=0.32\textwidth]{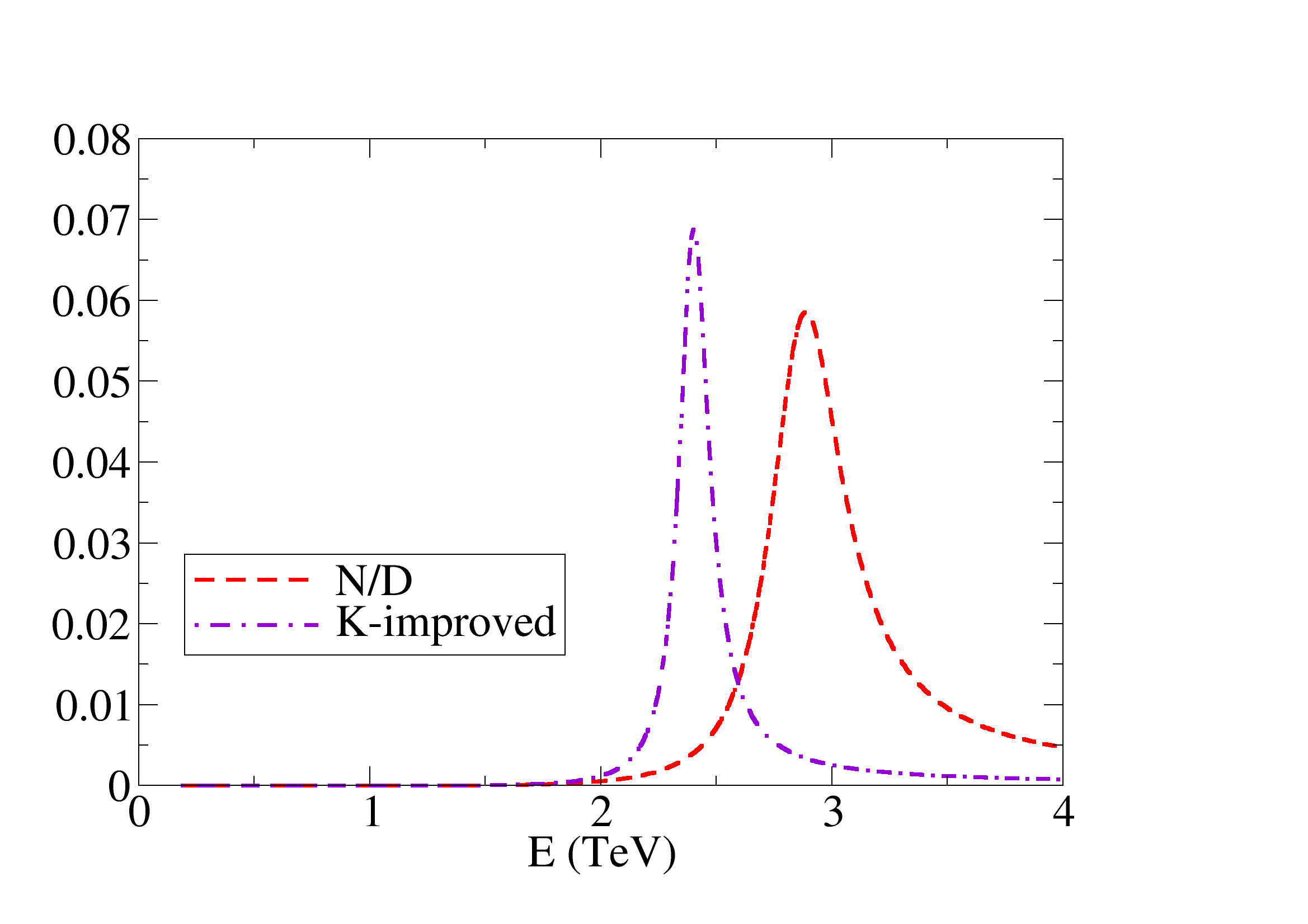}
\caption{\label{fig:2chanJ2I0}%
Isoscalar-tensor amplitudes (imaginary parts) for $a=0.88$, $b=a^2/2$, and the NLO parameters set to 0. From left to right: elastic $\omega \omega$, elastic $hh$ and cross-channel amplitudes.}
\end{figure}

If we now lift the $b=a^2$ requirement, because this is an isoscalar channel the $hh$ system becomes coupled to $\omega\omega$. Then the resonance should be visible in both particle spectra, and also in the channel-coupling amplitude;  all three are shown in figure~\ref{fig:2chanJ2I0} where the now inelastic resonance is clearly visible. 

Its mass is very similar to the purely elastic case, and both unitarization methods continue being in qualitative agreement.

\begin{figure}
\centering
\includegraphics[width=0.4\textwidth]{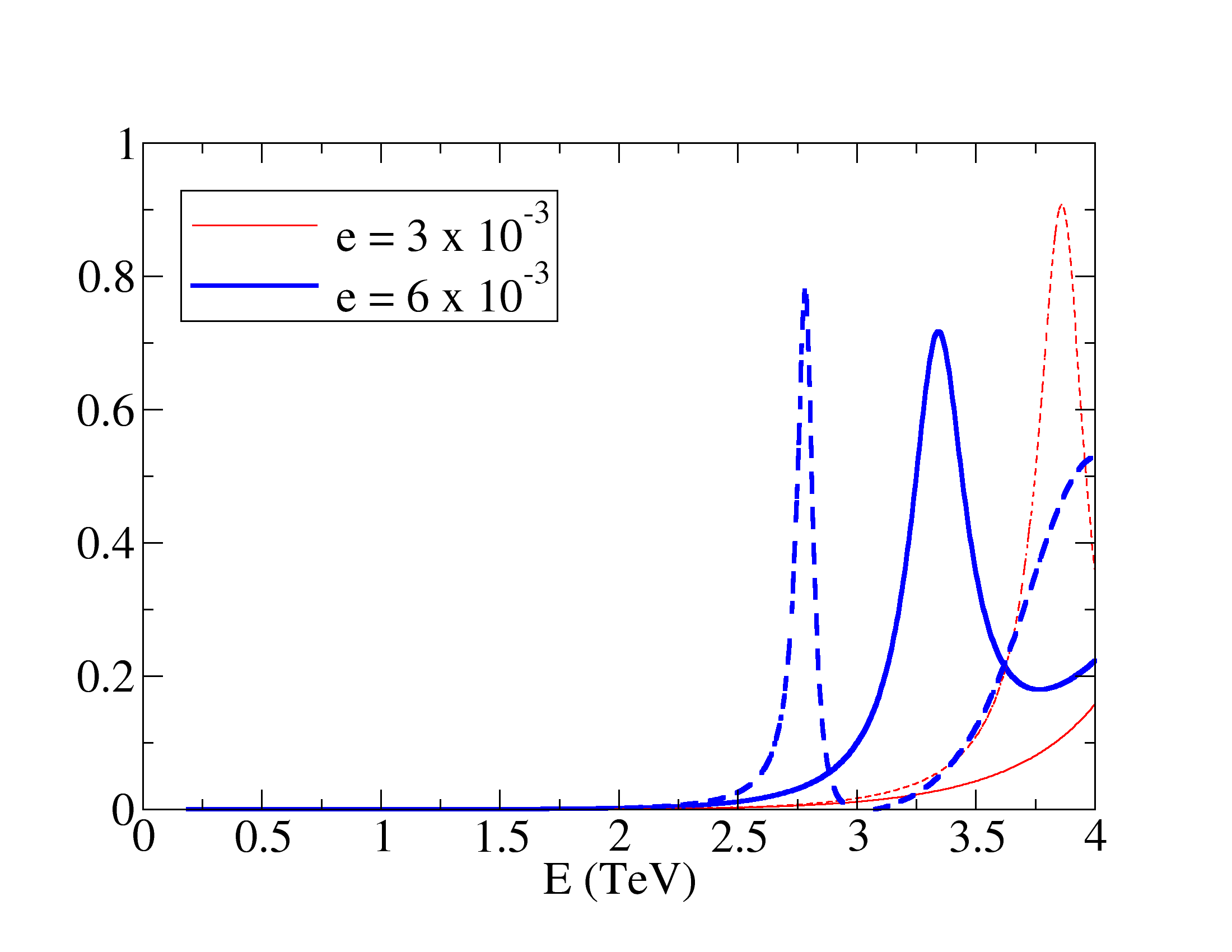}
\caption{\label{fig:e_dep}%
The tensor-isoscalar $J=2$, $I=0$ coupled channels analyzed with both IK (dashed lines) and N/D (solid line) methods can show a resonance induced by the parameter $e$.}
\end{figure}

We use the opportunity to show the appearance of this resonance also as a consequence of the channel coupling induced by the parameter $e$ of the effective Lagrangian. The IAM below does not capture the tensor channel, and the scalar one that the IAM does capture is only sensitive to the combination $d+e/3$ which does not allow to disentangle the two parameters. To see the separate effect of $e$ we need to examine the tensor channel\footnote{This arises naturally because the $\partial_\mu h\partial^\mu h$ contraction that multiplies $d$ in Eq.~(\ref{bosonLagrangian}) is a scalar, while the $\partial_\mu h \partial^\nu h$ one that accompanies $e$ has both scalar and tensor components.} as seen in Eq.~(\ref{Mtensor}), and this can be carried out with the N/D or IK methods.
We show the result of the analysis in figure~\ref{fig:e_dep}.
To prepare the figure we have taken $a=0.95$ and $b=a^2/2$. If all the NLO parameters vanish, there is no low-energy resonance in this tensor-isoscalar channel.
Adding $e$ at the level of $3-4\times 10^{-3}$ or more causes a resonance to enter the low-energy region.

%%%%%%%%%%%%%%%%%%%%%%%%%%%%%%%%%%%%%%%%%%%%%%%%%%%%%
\subsection{Tensor-isotensor channel with \boldmath $J=2$, $I=2$}
\begin{figure}
\includegraphics[width=0.4\textwidth]{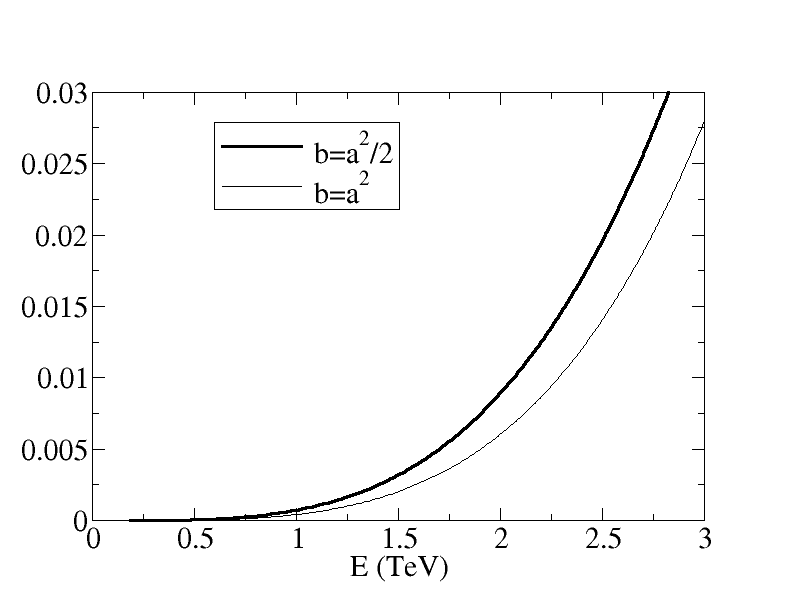}
\caption{\label{fig:J2I2perttheory}%
The real tensor, isotensor $I=J=2$ amplitude in NLO perturbation theory for $a=0.88$ and two values of $b$.}
\end{figure}

%%%%%%%%%%%%%%%%%%%%%%%%%%%%%%%%%%%%%%%%%%%%%%%%%%%%%
The last partial wave that does not vanish at one-loop order in perturbation theory, and that to our knowledge has not been considered in the literature, is the tensor-isotensor channel. 
Here again $K_{22}=E_{22}=0$ so that the amplitude in perturbation theory is real for physical energy. The non-vanishing constants, $B_{22}$ and $D_{22}$ are given in Eq.~(\ref{partial22}) below and the amplitude is drawn in figure~\ref{fig:J2I2perttheory} in perturbation theory.

\begin{figure}
\includegraphics[width=0.4\textwidth]{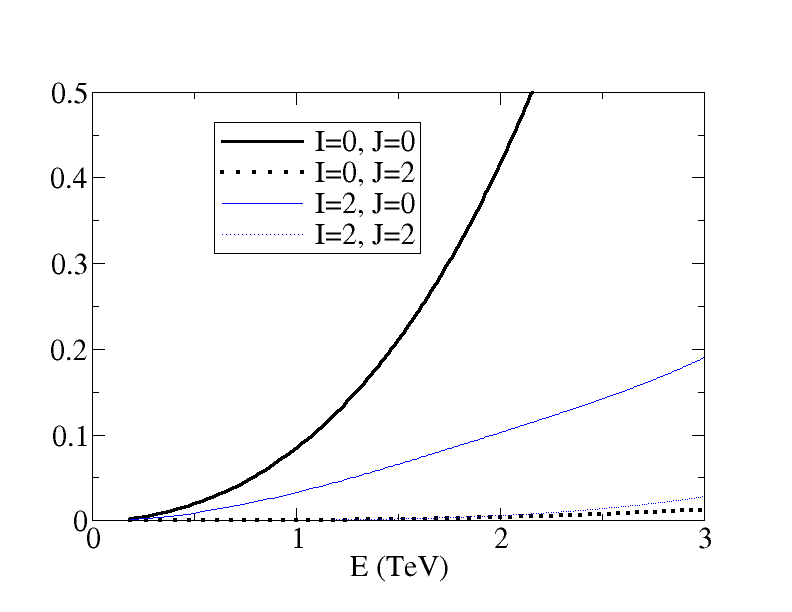}
\caption{\label{fig:convperttheory}%
Moduli of the isoscalar and isotensor NLO perturbative amplitudes theory for $a=0.88$ and $b=a^2$, showing good convergence of the partial wave expansion in the low energy region (the $J=2$ waves are much smaller than the two $J=0$ waves).}
\end{figure}

Moreover, figure~\ref{fig:convperttheory} shows this computation in perturbation theory for the case $b=a^2$ together with the isotensor-scalar one and also the two isoscalar amplitudes. Comparing those of equal $I$ we see that larger $J$ is suppressed below $4\pi v\sim 3\,{\rm TeV}$ (more so for the scalar channel, since the scalar-isoscalar amplitude is strongly interacting). 
Curiously, for $J=2$ the isotensor wave is stronger than the isoscalar one.

\begin{figure}
\includegraphics[width=0.45\textwidth]{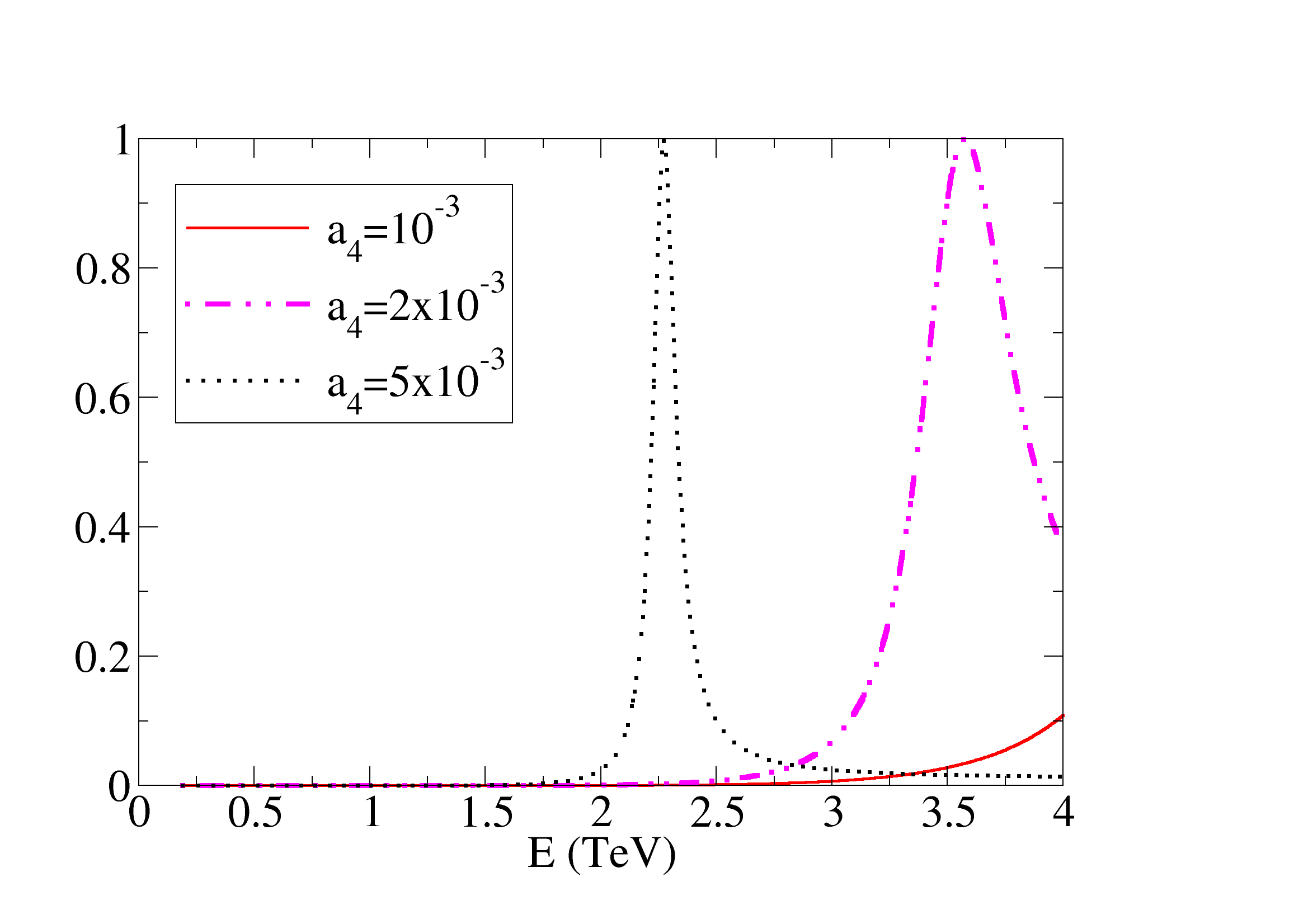}
\includegraphics[width=0.45\textwidth]{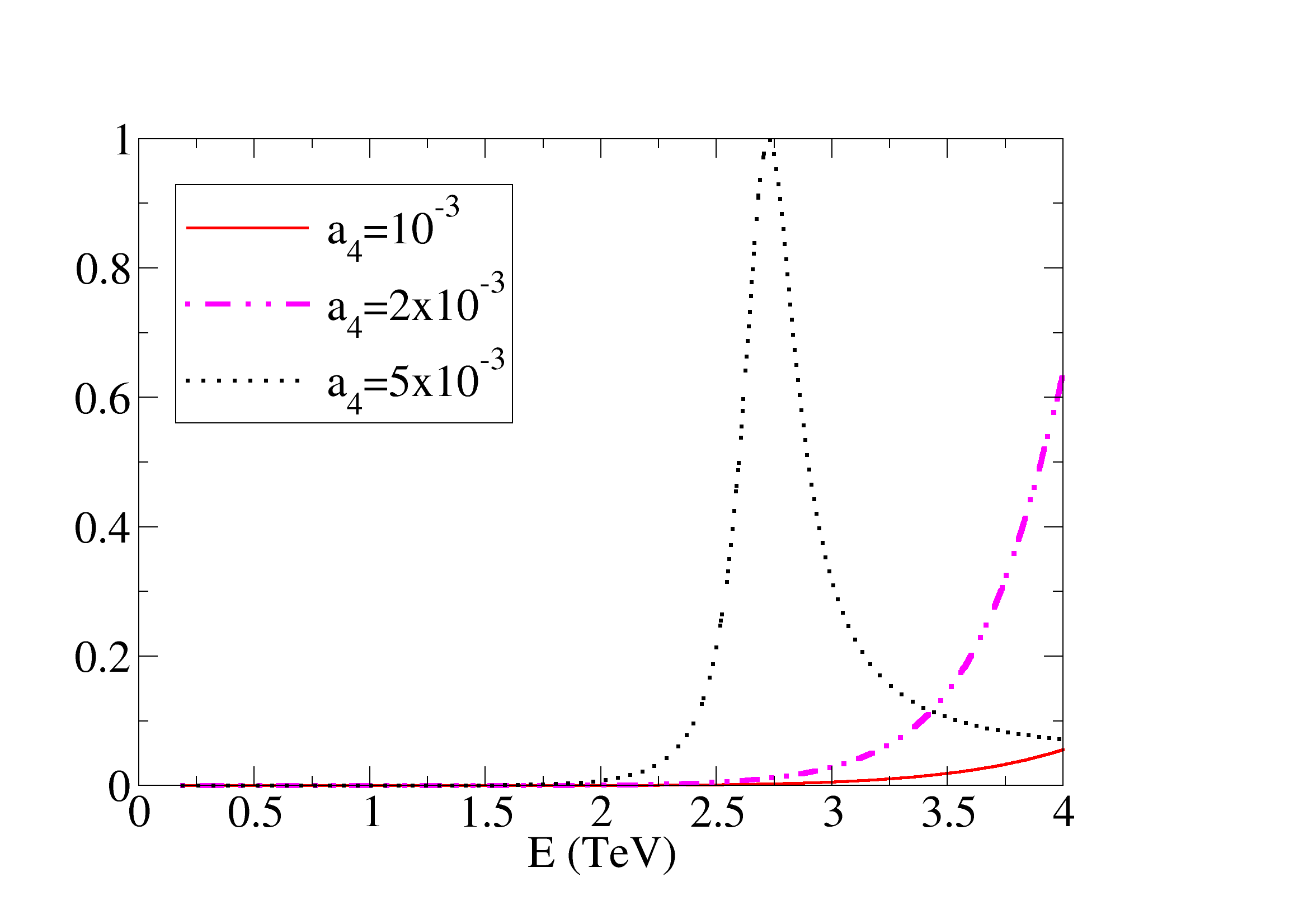}
\caption{\label{fig:resonantJ2I2}%
Tensor-isotensor resonance as function of the NLO $a_4$ parameter for the IK-matrix (left plot) and the N/D method (right plot). }
\end{figure}
The unitarization of the $J=I=2$ channel is not possible in the IAM method because $K_{22}=0$, but 
both IK and N/D methods concur in the presence of a resonance, as seen in figure~\ref{fig:resonantJ2I2}, when the $a_4$ NLO parameter is large enough. It is worth remarking that, for a given $a_4$, $m_{11}<m_{22}$ so that having this resonance in the 2-3~TeV region entails the presence of the vector-isovector ($\rho$-like one) in the 1-2~TeV energy interval.

As we have established that the convergence of the partial wave expansion is very good by comparing the $J=2$ and $J=0$ amplitudes, and that the order of the spectrum of resonances is the natural one, with those of lower angular momentum appearing at lower energy, we concentrate in the following on the three cases that are accessible to the NLO-IAM, the $00$, $11$ and $20$ channels; only the first one requires the coupled-channel treatment.

%%%%%%%%%%%%%%%%%%%%%%%%%%%%%%%%%%%%%%%%%%%%%%%%%%%%%
\section{Systematic numerical study of the IAM}
\label{sec:sysIAM}
%%%%%%%%%%%%%%%%%%%%%%%%%%%%%%%%%%%%%%%%%%%%%%%%%%%%%
In this section we undertake the systematic study of the IAM with the help of a computer. The calculations are very straightforward and involve simple algebraic formula (no integrations, as the dispersion relation has been analytically solved) and the inversion, at most, of dimension-two matrices.  The IAM cannot handle, without NNLO information, the higher partial waves with $J=2$ or beyond, but we have seen in figure~\ref{fig:convperttheory} that, under natural conditions, these are quite smaller in the low-energy region. For the three dominant low-energy amplitudes, the IAM based on NLO perturbation theory is reliable and powerful, so we proceed with it alone. 

First, in subsection~\ref{subsec:numIAMelas} we address the one-channel IAM in Eq.~(\ref{IAM1channel}) for the $W_LW_L$ elastic scattering, with the help of the equivalence theorem, of course. This involves setting $b=0$ and studying the behavior of the amplitudes upon varying each of the active parameters $a$, $a_4$ and $a_5$. These results are just reassuring as they are known to a large extent. Then  subsection~\ref{subsec:numIAMinelas} addresses the coupled channels, by means of Eq.~(\ref{IAMforF}) and it is here that we make a totally new contribution. 

One of our findings is a coupled-channel resonance akin to the low-energy $\sigma$ meson but that can be generated by purely $ww-hh$ interactions independently of the elastic potential strength between two $w$s  or two $h$s. We have chosen to highlight this curious object in a companion letter~\cite{Delgado:2014dxa} so we do not focuse on it so much here.

%%%%%%%%%%%%%%%%%%%%%%%%%%%%%%%%%%%%%%%%%%%%%%%%%%%%%
\subsection{Purely elastic scattering with \boldmath $b=a^2$}\label{subsec:numIAMelas}
%%%%%%%%%%%%%%%%%%%%%%%%%%%%%%%%%%%%%%%%%%%%%%%%%%%%%
\begin{figure}
\includegraphics[width=0.32\textwidth]{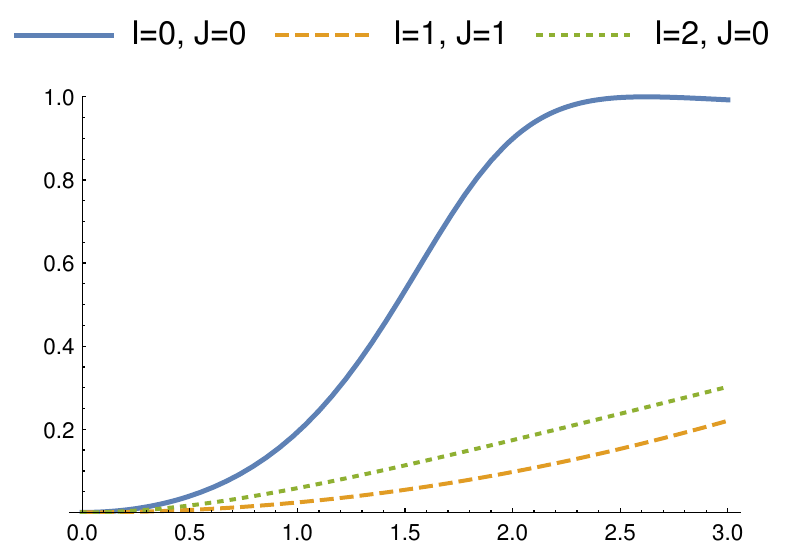}
\includegraphics[width=0.32\textwidth]{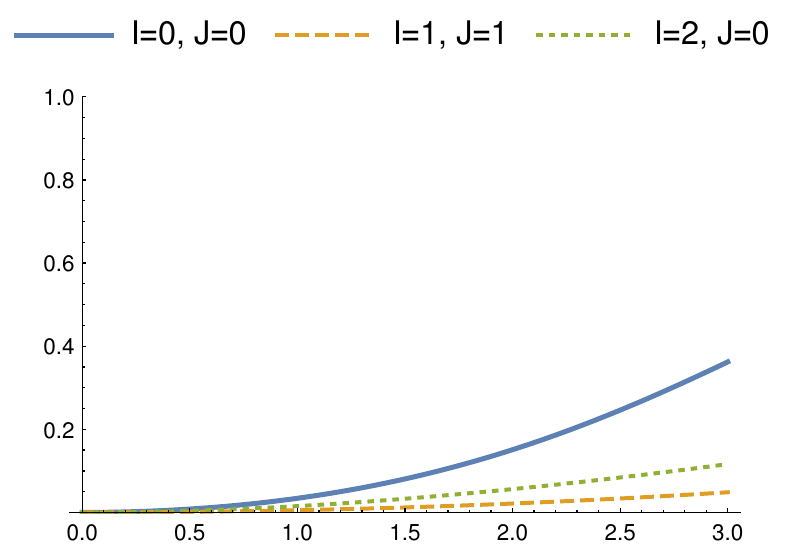}
\includegraphics[width=0.32\textwidth]{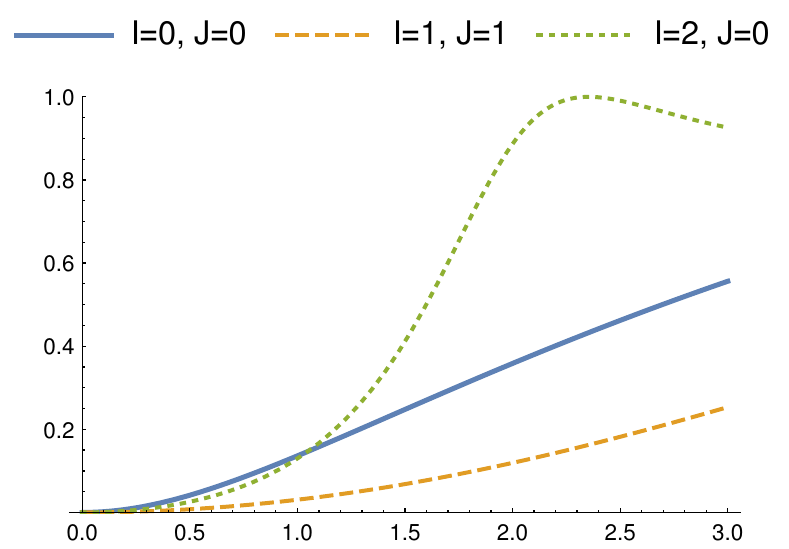}
\caption{\label{fig:senstoa}%
Moduli of the lowest elastic $\omega\omega\to \omega\omega$ partial waves in the IAM for $b=a^2$ (no coupled channels) as function of $a$. We will take the middle plot as reference for the parameter exploration in the next graphs. From left to right, $a=0.75$, $0.95$, $1.25$.}
\end{figure}
The current $2\sigma$ bounds on the $a$ parameter are, from CMS, $a\in(0.88-1.15)$, and $a\in (0.96-1.34)$ from the ATLAS collaboration~\cite{Datos}. We will take as reference a fixed value of $a=0.95$ with NLO parameters set to 0, and later exemplify the sensitivity to each parameter ($a$ is better chosen different from 1 because of the factor $(1-a^2)$ that enters the leading order amplitudes). In any case, the sensitivity to $a$ is displayed in figure~\ref{fig:senstoa}. Generally speaking, for $a<1$ (left plot) there is a broad scalar resonance akin to the $\sigma$ in hadron physics, and the other channels are nonresonant. For $a>1$ we can see a different situation in which the scalar strength significantly diminishes, but instead the isotensor wave becomes strong and possibly resonant (because the factor $1-a^2$ changes sign, so its normally repulsive amplitude becomes attractive). 

\begin{figure}
\null\hfill\includegraphics[width=0.4\textwidth]{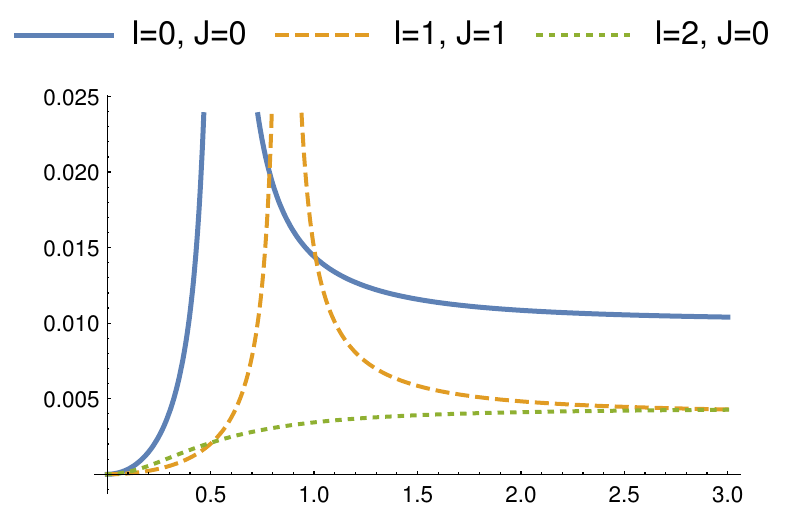}\hfill\hfill %
\includegraphics[width=0.4\textwidth]{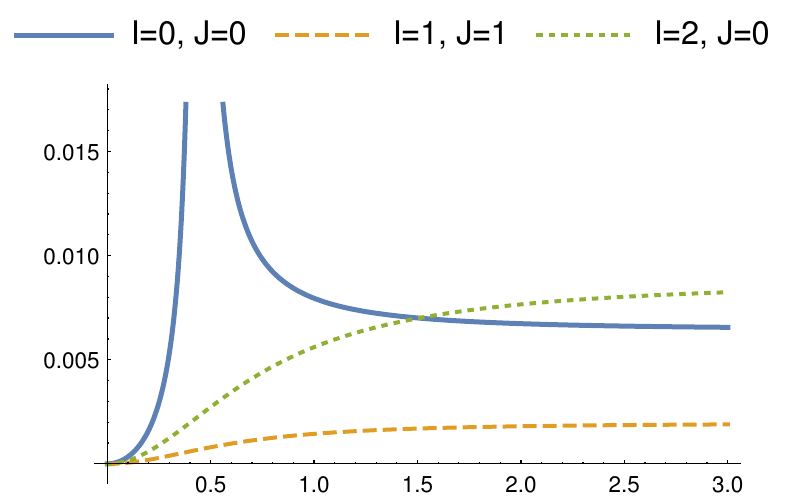}\hfill\null
\caption{\label{fig:dep_a4_a5}%
Moduli of the lowest elastic $\omega\omega\to \omega\omega$ partial waves in the IAM for $b=a^2$ (no coupled channels) showing the effect of $a_4$ (left) and $a_5$ (right) both positive and alternatively equal to 0.002. Here $a=0.95$. We see a light scalar-isoscalar resonance, a vector-isovector resonance around a TeV in the left plot (that moves to higher masses for smaller values of the positive $a_4$ that induces it), and an inconspicuous isotensor amplitude.}
\end{figure}

We now take the middle plot in figure~\ref{fig:senstoa} and add an NLO term proportional to either $a_4$ or $a_5$, with the outcome plotted in figure~\ref{fig:dep_a4_a5}.
The effect of $a_4$ of order $10^{-3}$ (left plot) is to produce a very narrow vector-isovector resonance, and narrowing plus making lighter the scalar-isoscalar one.
The effect of $a_5$ (right plot) at this same level of intensity is only dramatic in the scalar-isoscalar channel, while the vector one remains of moderate intensity and hardly resonant at all. This is in agreement with the observation in~\cite{Espriu:2013fia}.
\begin{figure}
\null\hfill%
\includegraphics[width=0.4\textwidth]{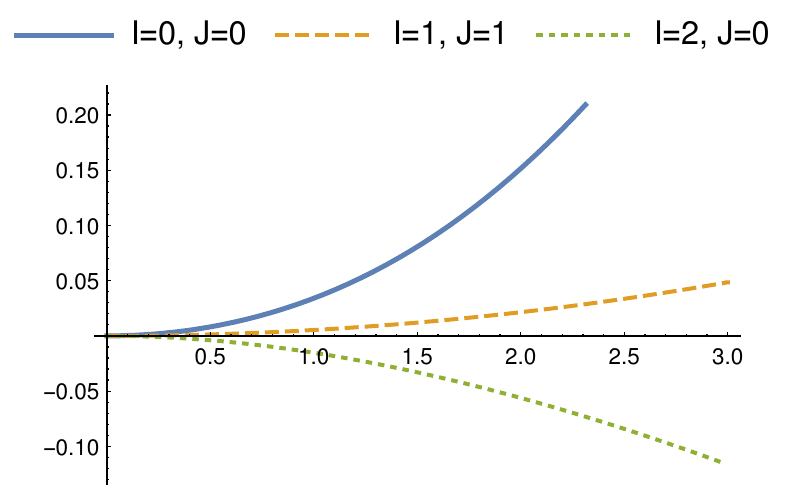}\hfill%
\includegraphics[width=0.4\textwidth]{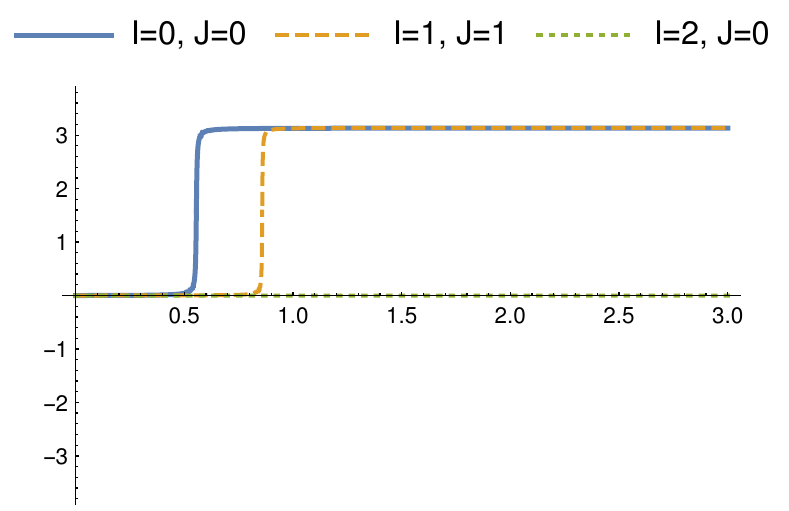}%
\hfill\null
\caption{\label{fig:phase1}%
Scattering phase shift of the lowest elastic $\omega\omega\to \omega\omega$ partial waves in the IAM for $b=a^2=0.95^2$ (no coupled channels), $a_4=0$ (left plot) and $a_4=0.002$ (right plot).  We can see how indeed the addition of an $a_4$ at the level of $10^{-3}$ generates phase motion crossing $\pi/2$ in the right plot corresponding to a resonance in both the scalar and vector channels.}
\end{figure}
\begin{figure}
\includegraphics[width=0.32\textwidth]{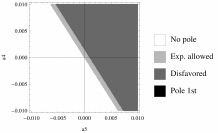}
\includegraphics[width=0.32\textwidth]{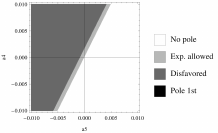}
\includegraphics[width=0.32\textwidth]{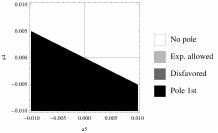}
\caption{\label{fig:poles_a4_a5}%
From left to right, isoscalar ($IJ=00$), isovector ($IJ=11$) and isotensor ($IJ=20$) channels in elastic $\omega\omega\rightarrow\omega\omega$ scattering. For $a=0.90$ (different from our base scenario so we may compare with other groups), $b=a^2$, we show the $a_4$-$a_5$ parameter map, setting the other NLO parameters to zero. Note the appearance of a pole on the first Riemann sheet for $IJ=20$ and negative enough values of both $a_4$ and $a_5$. The comparison with ref.~\cite{Espriu:2013fia} is very satisfactory.}
\end{figure}
The vector resonance induced by positive $a_4$ can also be seen in the scattering phase shift in figure~\ref{fig:phase1}. The left plot shows the phase motion in the three lowest-$E$ channels with all NLO parameters set to 0. No resonance is seen, in agreement with the middle plot of figure~\ref{fig:senstoa}. The right plot shows clear resonant phase motion corresponding to the resonances in figure~\ref{fig:dep_a4_a5}, where we study the effect of both $a_4$ and $a_5$.  The good agreement with~\cite{Espriu:2013fia} is remarkable, both works agreeing on the appearance of a pole on the first Riemann sheet in the isotensor channel for negative enough values of either $a_4$ or $a_5$.

This feature is shown in figure~\ref{fig:poles_a4_a5}, is in full agreement with the results of~\cite{Espriu:2013fia} and, as discussed in subsection~\ref{subsec:spurious}, excludes this parameter space within the IAM. 
The computational method to produce this and the following maps in parameter space is described in appendix~\ref{app:numeric}.

In figure~\ref{fig:poles_a4_a5} we call the \emph{experimentally disfavored} regions so because poles appear with $\lvert s \rvert\leq (700\,{\rm GeV})^2$ (scalar-isoscalar and isotensor channels) and $(1.5\,{\rm TeV})^2$ (vector-isovector channel).

The vector-isovector channel is here exceptional in that the two variables enter with opposite signs, in the combination $a_4-2a_5$, see Eq.~(\ref{partial11}), whereas in all other four NLO amplitudes they come with equal sign. Thus, the slant in the middle plot is opposite to the other two.

\begin{figure}
\includegraphics[width=0.45\textwidth]{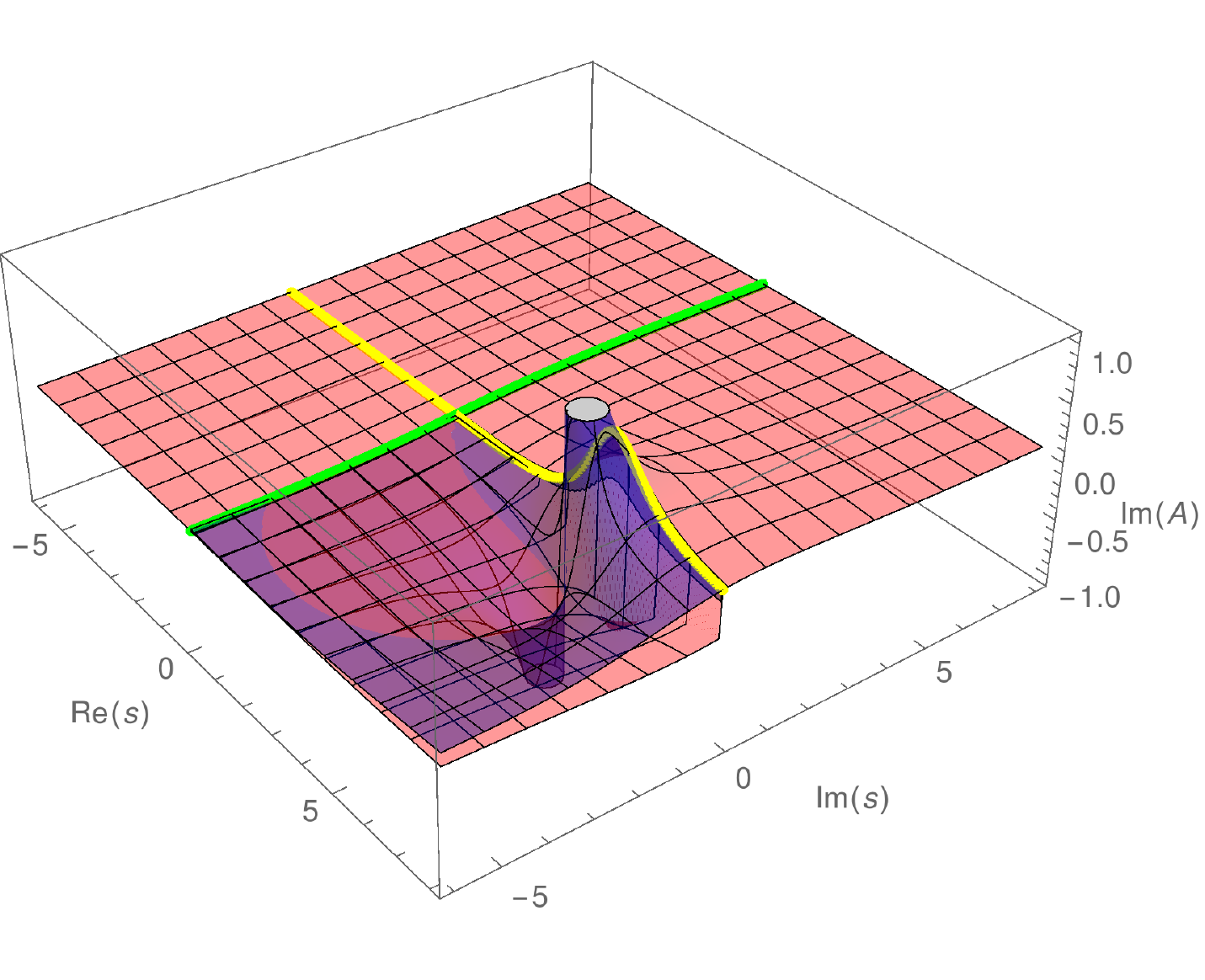}
\caption{\label{fig:poleSecond}%
Example pole of the isoscalar elastic amplitude with $b=a^2$ (only the $\omega\omega\to \omega\omega$ channel is active), $a=0.95$, $a_4=10^{-4}$, and all other NLO parameters set to 0. Pole in the second Riemann sheet (below the physical, real-$s$ axis highlighted in bright yellow online). The lower (salmon online) and upper (blue online) surfaces are, respectively, the first and second Riemann sheets.}
\end{figure}

For broad swipes of $a_4$--$a_5$ parameter space the IAM predicts either isoscalar or isovector resonances or both. In figure~\ref{fig:poleSecond} we show an example of a pole in the second Riemann sheet of elastic $\omega\omega$ scattering in $l=0$, the $A_{00}$ partial wave for one channel only ($b=a^2$). 

Therein the continuation to the second Riemann sheet has been
obtained with Eq.~(\ref{anapro}) and the resonance appears as appropriate below the real, physical s-axis (bright yellow line).
This pole corresponds to the scalar IAM resonance shown for physical $s$ in figure~\ref{fig:dep_a4_a5} (blue solid line there) though $a_4$ is somewhat smaller here.
This serves as illustration of the pole structures in the complex plane (unstable particles or resonances) that accompany our resonant shapes for physical $s$.

A lot of the $a_4$--$a_5$ parameter space represented on~\cite{Espriu:2013fia} is experimentally disfavored because the mass-range where the resonances appear is being covered by LHC data~\cite{searches}, with none found yet, though such experimental bounds are  not very strong because the couplings between the new resonances and the detected SM leptons are quite arbitrary (from the effective theory point of view), so it is difficult to interpret the bounds beyond particular models.

%%%%%%%%%%%%%%
\begin{figure}
\includegraphics[width=0.32\textwidth]{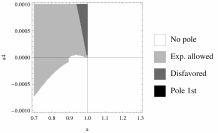}
\includegraphics[width=0.32\textwidth]{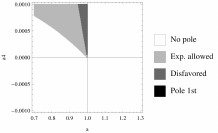}
\includegraphics[width=0.32\textwidth]{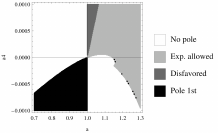}
\caption{\label{fig:poles_a2b_a4}%
From left to right, isoscalar ($IJ=00$), isovector ($IJ=11$) and isotensor ($IJ=20$) channels. $a^2=b$ vs. $a_4$. Note the presence of poles on the first Riemann sheet for certain region of the $a>1,\,a_4<0$ parameter space.}
\end{figure}
On fig.~\ref{fig:poles_a2b_a4} the simultaneous effect of $a$ (with $a^2=b$) and $a_4$ is shown, again swiping the parameter space looking for resonances. Note the presence of a resonance on the first Riemann sheet in the isotensor channel even for $a<1$ and sufficiently negative values of $a_4$. For $a>1$ (and $b=a^2$), there is no resonance on the first Riemann sheet. For $a<1$, we can find a pole in both the isoscalar and isovector channels. For $a>1$, only an isotensor resonance is to be found.

%%%%%%%%%%%%%%%%%%%%%%%%%%%%%%%%%%%%%%%%%%%%%%%%%%%%%
\subsection{Scattering \boldmath $\omega\omega$ in the presence of $b\ne a^2$}\label{subsec:numIAMinelas}
%%%%%%%%%%%%%%%%%%%%%%%%%%%%%%%%%%%%%%%%%%%%%%%%%%%%%

\begin{figure}
\includegraphics[width=0.32\textwidth]{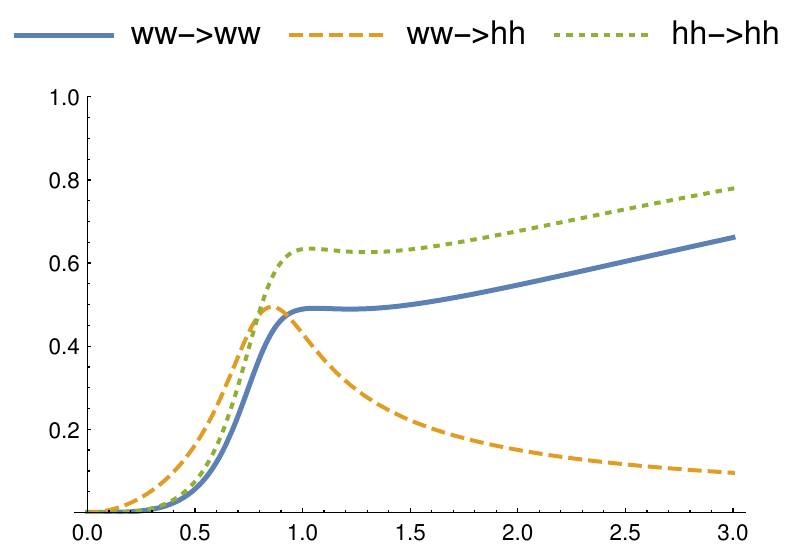}
\includegraphics[width=0.32\textwidth]{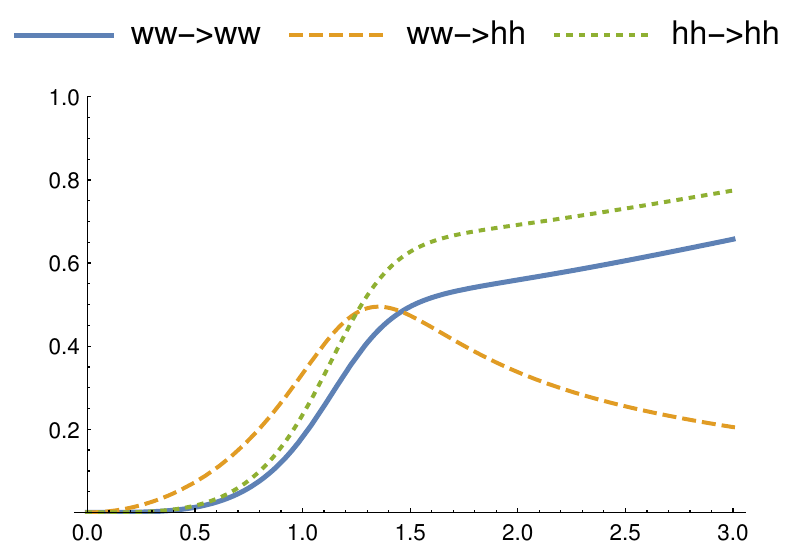}
\includegraphics[width=0.32\textwidth]{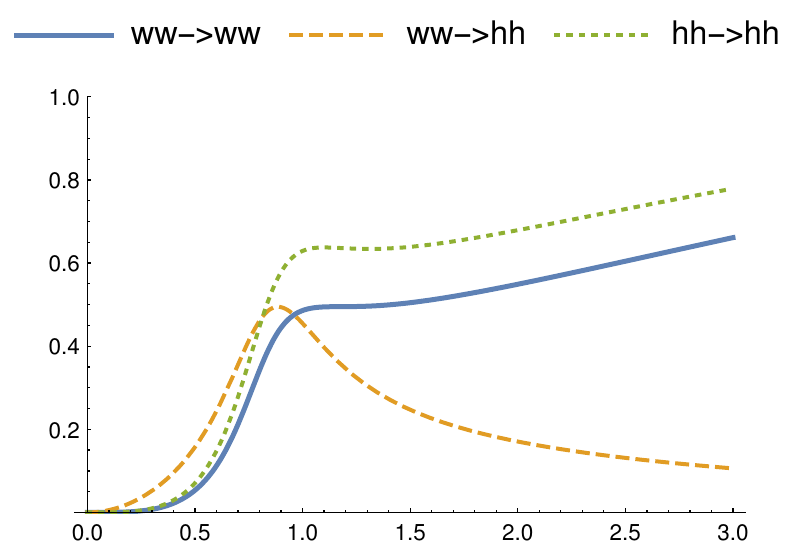}
\caption{\label{fig:inelas_b}%
Moduli of the lowest ($I=J=0$) partial waves in the IAM for $b\ne a^2=1$ (all the strong dynamics comes from the coupled channels). From left to right, $b=-1$, $b=2$, $b=3$ (the first and third are almost equal since they are symmetric respect to $b=1$).  A scalar resonant structure is apparent for $E=1\,{\rm TeV}$; because more extreme values of $b$ lower its mass, we are able to give a bound on the value of $b$, that must be roughly contained in $(-1,3)$, as explained in the companion letter~\cite{Delgado:2014dxa}. We will take the middle plot as reference for the parameter exploration in several of the following graphs.}
\end{figure}

Setting $b\neq a^2=1$ opens the inelastic scattering $\omega\omega\to hh$ channel in the absence of elastic strength. Figure~\ref{fig:inelas_b} shows the dependence on $b$.
Almost all our computed perturbative amplitudes are symmetric around $b=a^2=1$ (see sec.~\ref{sec:partwaves}), with the exception of the scalar-isoscalar $\omega\omega\to hh$ channel-mixing $M_0$ partial wave in Eq.~\ref{Mscalar}; this asymmetry then appears in other channels due to the unitarization (a way of thinking of it is with the image of resumming perturbation theory), but the effect is small, so that the left and right plots are quite similar. 
The scalar-isoscalar resonance shown is very interesting and the object of focuse of the accompanying letter~\cite{Delgado:2014dxa}.

\begin{figure}
\includegraphics[width=0.32\textwidth]{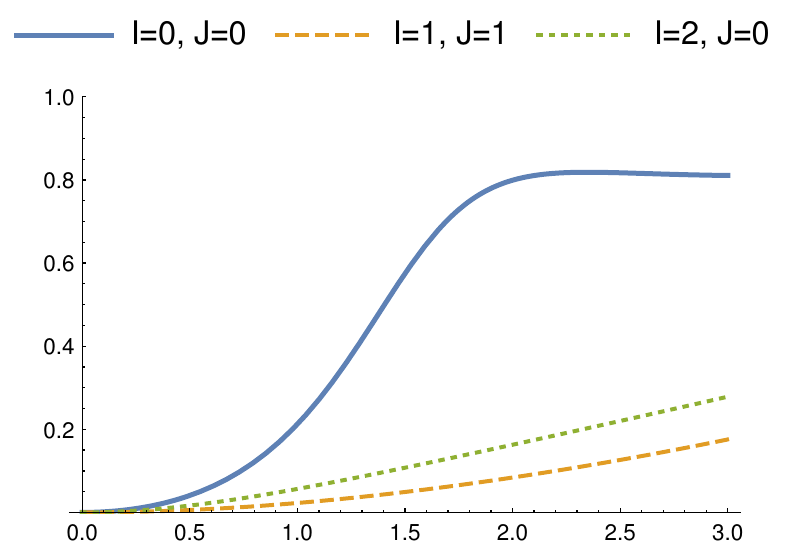}
\includegraphics[width=0.32\textwidth]{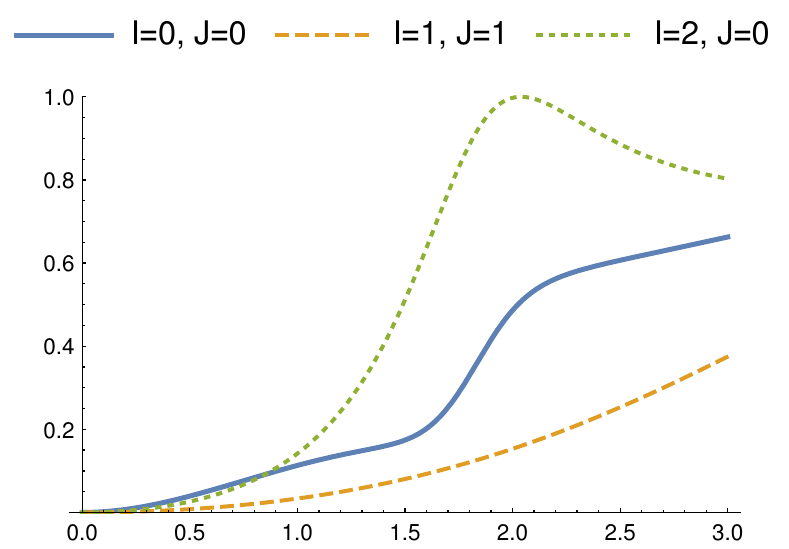}
\includegraphics[width=0.32\textwidth]{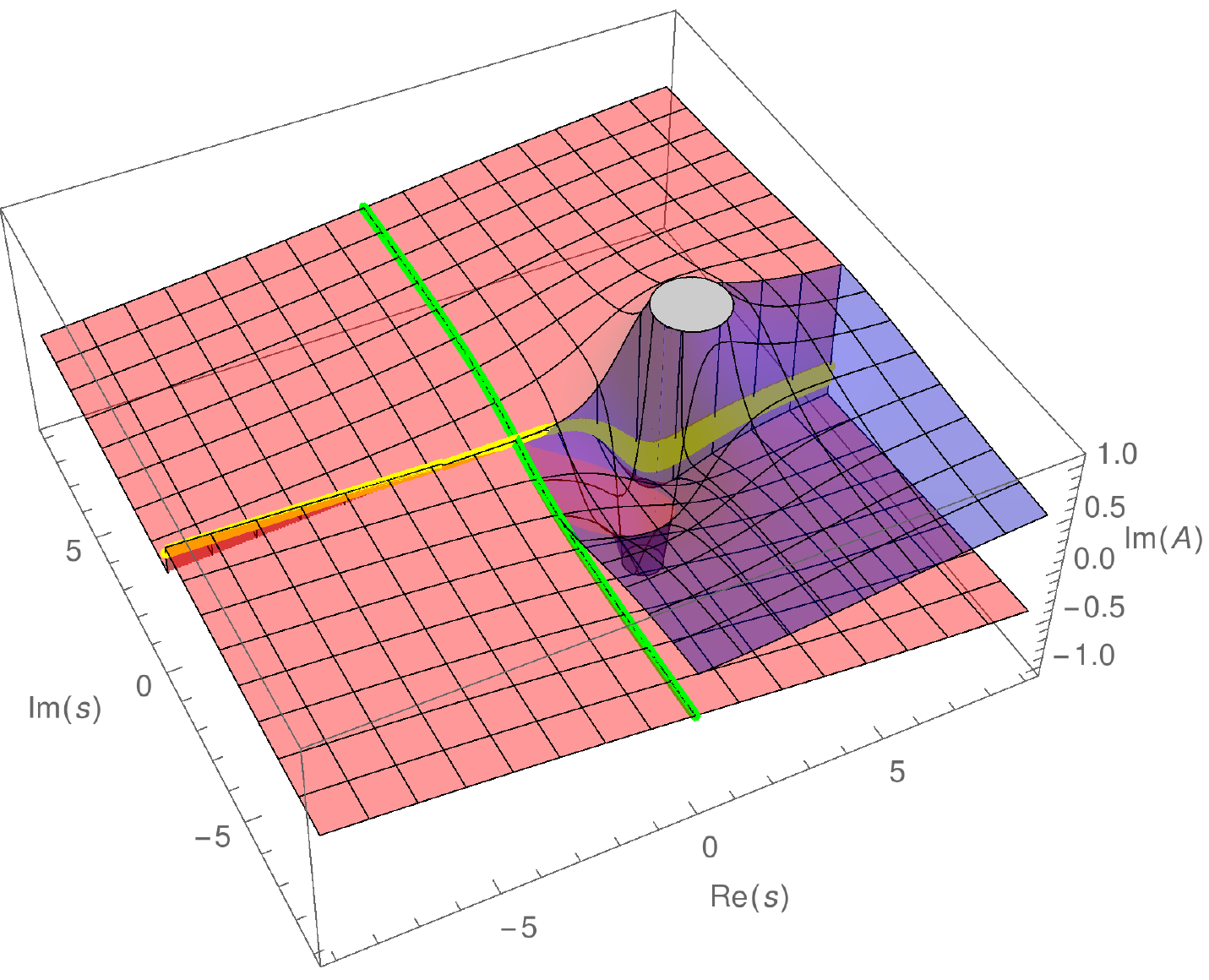}
\caption{\label{fig:a_b}%
Moduli of the lowest elastic $\omega\omega\to \omega\omega$ partial waves 
in the IAM for $b\ne a^2\ne 1$ (strength from both elastic and coupled-channel dynamics).
Left plot: $a=0.75$, $b=0.9$, showing much strength in the scalar channel, presumably due to a $\sigma$ resonance. Center: $a=1.25$, $b=1.1$, showing a pole on the second Riemann sheet in the isotensor channel, clearly seen also on the right plot in the complex plane.}
\end{figure}

\begin{figure}
\includegraphics[width=0.32\textwidth]{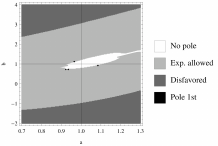}
\includegraphics[width=0.32\textwidth]{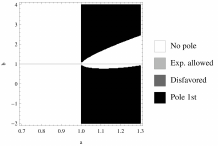}
\includegraphics[width=0.32\textwidth]{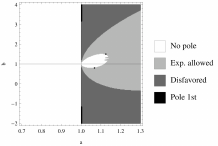}
\caption{\label{fig:poles_ab}%
From left to right, isoscalar ($IJ=00$, isovector ($IJ=11$) and isotensor ($IJ=20$) channels. Note the presence of a pole in the first Riemann sheet of the isovector channel in quite some of the parameter space with $a>1$. All the NLO parameters are set to zero.}
\end{figure}

Figs.~\ref{fig:a_b} shows the lowest elastic $\omega\omega\to\omega\omega$ partial waves in the presence of $a\neq 1$ (as well as $b\neq a^2$), so there is both elastic and inelastic potential strength. The scalar resonance is then more similar to the standard QCD $\sigma$ resonance.

A novelty is the appearance of a pole on the second Riemann sheet of the isotensor channel for $a=1.25$, $b=1.1$. This is very much unlike QCD, where the isotensor channel is weak and repulsive; while there is no $\pi^+\pi^+$ resonance in the hadron spectrum, this is still allowed by current constraints on the $W^+W^+$ one. 

However, as we show on figure~\ref{fig:poles_ab}, this case with $a>1$ is quite critical, because most of the parameter space features an isovector pole on the first Riemann sheet, so that much of this parameter region must be ruled out or declared beyond our validity range. 
Only a small part of the $a>1$ parameter space shows an isotensor pole on the second Riemann sheet while excluding an isovector pole on the first one, and simultaneously remains out of experimentally disfavored values of $a$. 

On the other hand, the behaviour for $a<1$ is more standard, showing a resonance on the second Riemann sheet only in the isoscalar channel. This resonance is quite broad, and only becomes experimentally disfavored for relatively large values of $a^2-b$.

%%%%%%%%%%%%%%

\begin{figure}
\null\hfill%
\includegraphics[width=0.4\textwidth]{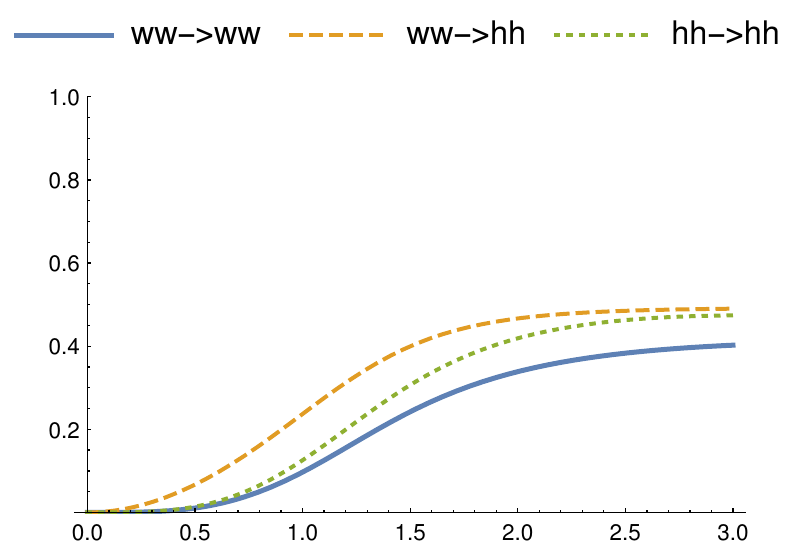}\hfill
\includegraphics[width=0.4\textwidth]{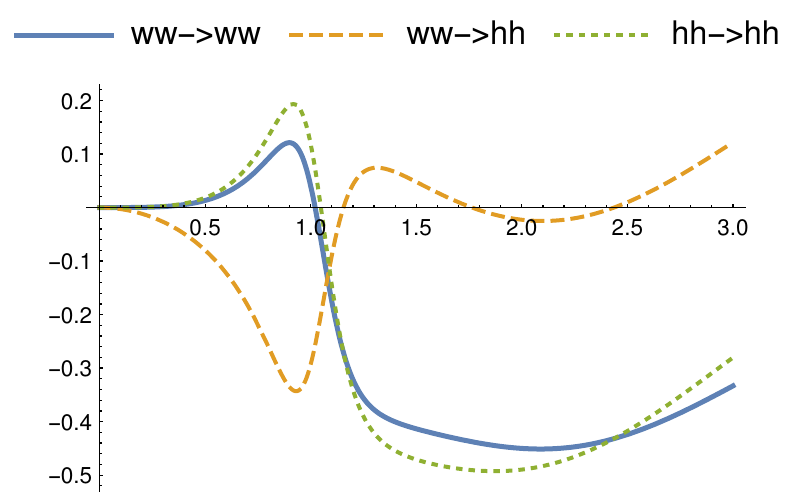}%
\hfill\null\\\null\hfill%
\includegraphics[width=0.4\textwidth]{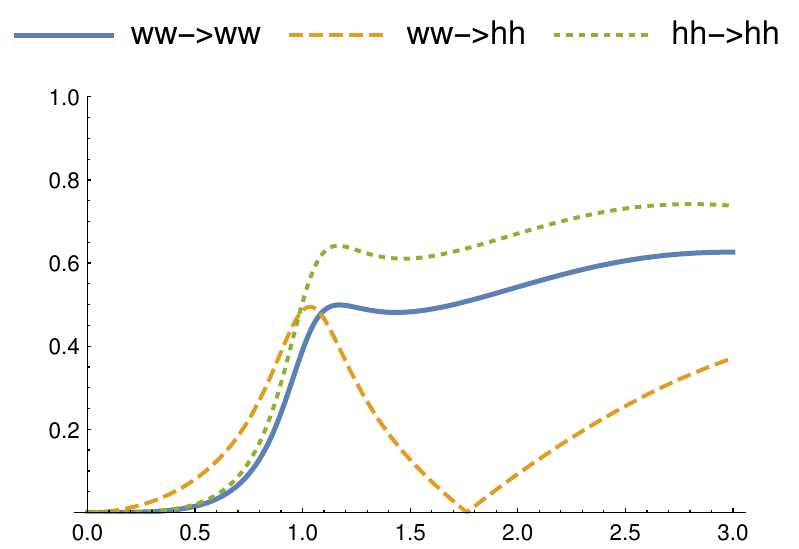}\hfill
\includegraphics[width=0.4\textwidth]{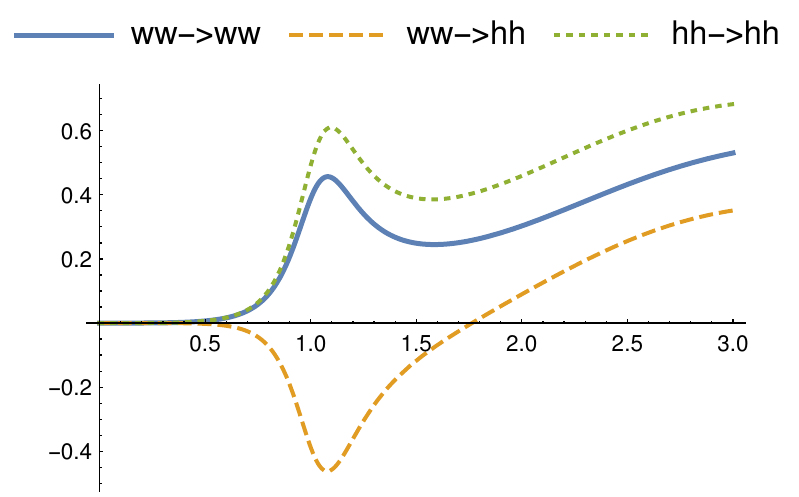}%
\hfill\null
\caption{\label{fig:de}%
Sensitivity to $d$. We depict the lowest ($I=J=0$) partial wave in the IAM for $b=2\ne a^2=1$. Left: moduli of the amplitudes with $d=0.01$ (top) and $d=-0.01$ (bottom). Right: real (top) and imaginary (bottom) value of that partial wave for $d=-0.01$, where we see that the channel-coupling partial wave is analytic but has a zero.}
\end{figure}

\begin{figure}
\null\hfill%
\includegraphics[width=0.4\textwidth]{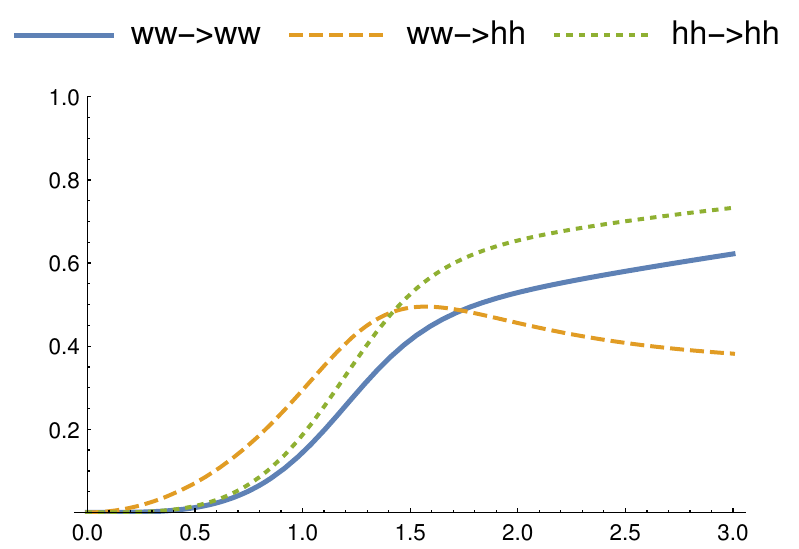}\hfill
\includegraphics[width=0.4\textwidth]{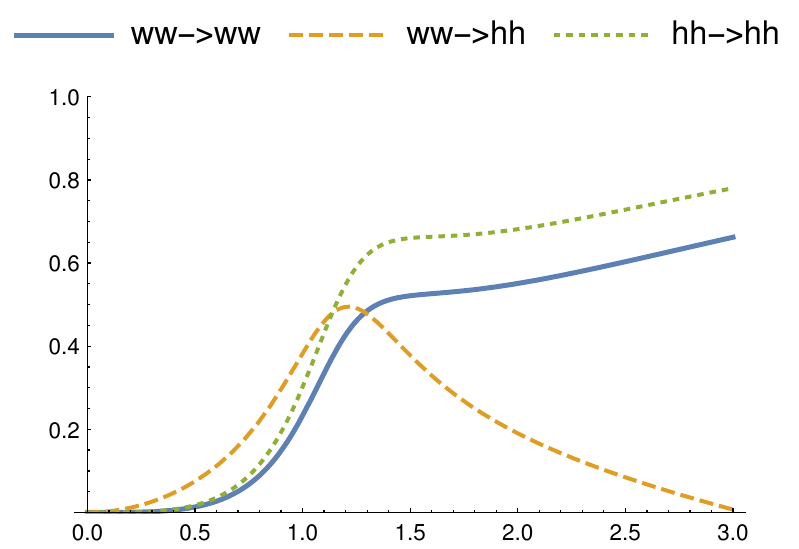}%
\hfill\null
\caption{\label{fig:et}%
Moduli of the lowest ($I=J=0$) partial waves in the IAM for $b=2\ne a^2=1$. Left plot: $e=0.01$. Right plot: $e=-0.01$. The result is similar to fig.~\ref{fig:de}. because, of course, this channel depends only on the parameter combination $d+(e/3)$, which serves as a check.}
\end{figure}

\begin{figure}
\null\hfill\includegraphics[width=0.42\textwidth]{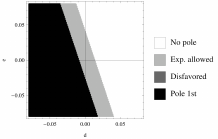}\hfill
\includegraphics[width=0.42\textwidth]{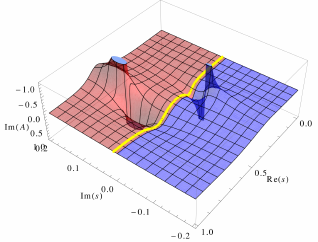}\hfill
\hfill\null
\caption{\label{fig:poles_de_et}%
Scalar-isoscalar channel ($IJ=00$), with $a=1$, $b=2$.
Left: $d$-$e$ parameter map looking for poles.
Right: imaginary part of the elastic $\omega\omega$ scattering ($d=e=-0.005$). The isovector and isotensor channels, not shown, have no poles in the region of interest ($\lvert s \rvert < (3\,{\rm TeV})^2$). 
As discussed above in subsec.~\ref{subsec:spurious}, the black region contains 
a pole on the first Riemann sheet (and a conjugate pole that is outside our circuit).}
\end{figure}

The $d$ and $e$ parameters are studied on figures~\ref{fig:de} and~\ref{fig:et}, respectively. However, note that they appear in the combination $d+(e/3)$ on the lowest partial wave ($IJ=00$), so the IAM applied to any future strongly coupled resonance would be insufficient to separate them and one would need to resort to the $J=2$, $I=0$ resonance in figure~\ref{fig:e_dep} above to obtain $e$ independently of $d$.

We concentrate now on the $I=J=0$, $a=1$, $b=2$ case, which has an isoscalar pole on the second Riemann sheet. A peak on $\omega\omega\to hh$ is shown on figures~\ref{fig:de} (right) and~\ref{fig:et}. This is expected, since $d$ and $e$ accompany four-particle operators  $\omega\omega hh$. In fig.~\ref{fig:poles_de_et} we see that for positive values of $d$ or $e$, the isoscalar pole weakens and then disappears. But for negative values, a pole on the first Riemann sheet emerges. The case of $d=-0.01$ shown in figure~\ref{fig:de} 
is curious because there is no pole on the first Riemann sheet \emph{below} 3TeV so we should not a priori reject all that structure in the corresponding plots of figure~\ref{fig:de}, including a zero of the amplitude at high energies. Of course, we should be cautious: perhaps, for these small negative values of $d$ the pole simply moves to higher energies and 
we should not trust the computation (or discard negative $d$ altogether).

\begin{figure}
\includegraphics[width=0.32\textwidth]{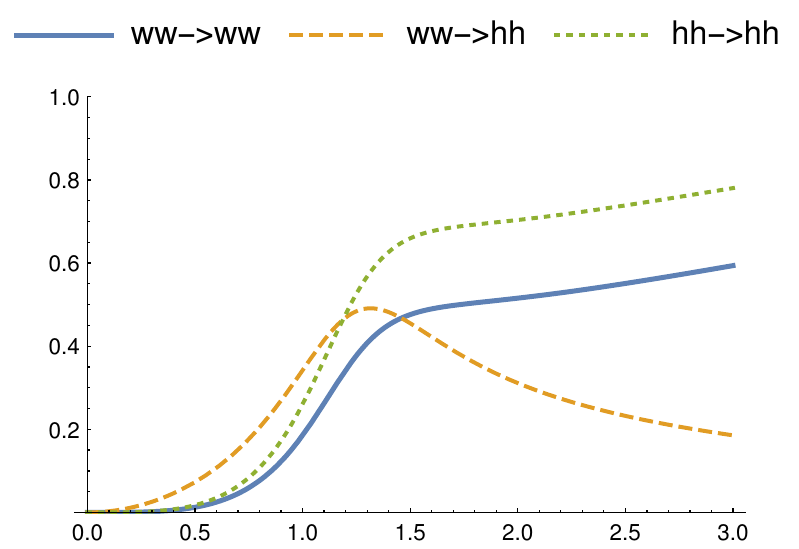}
\includegraphics[width=0.32\textwidth]{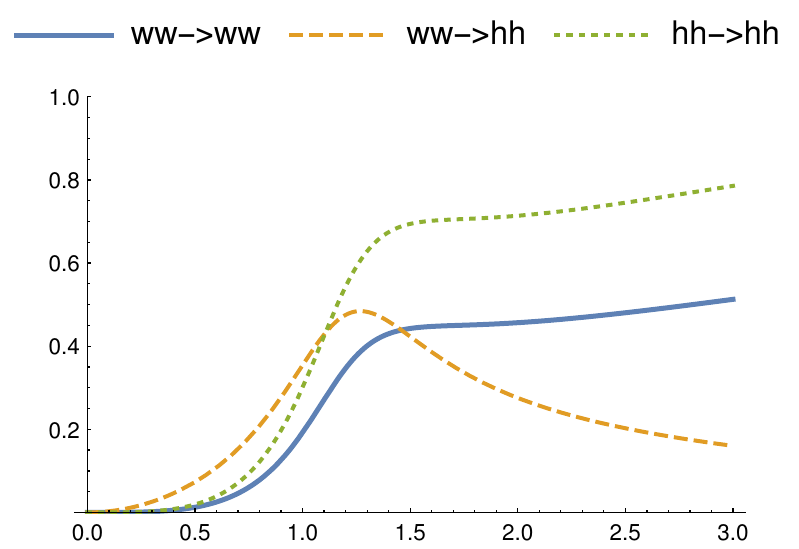}
\includegraphics[width=0.32\textwidth]{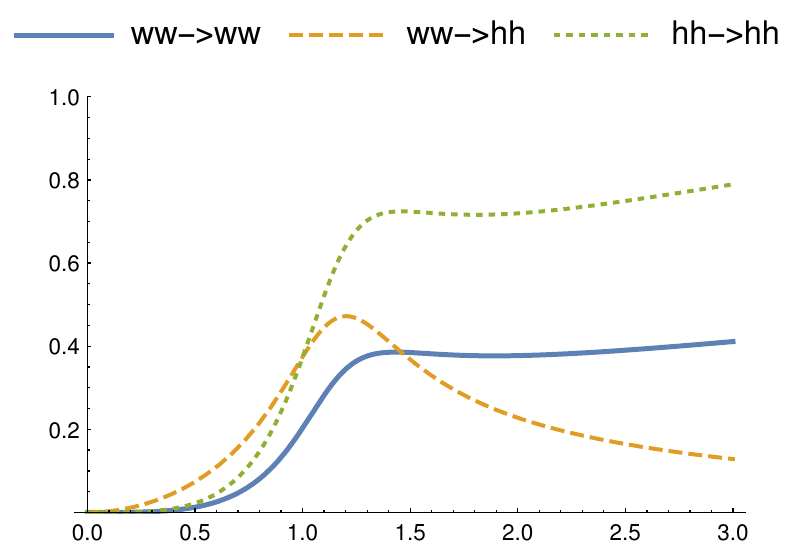}\\
\includegraphics[width=0.32\textwidth]{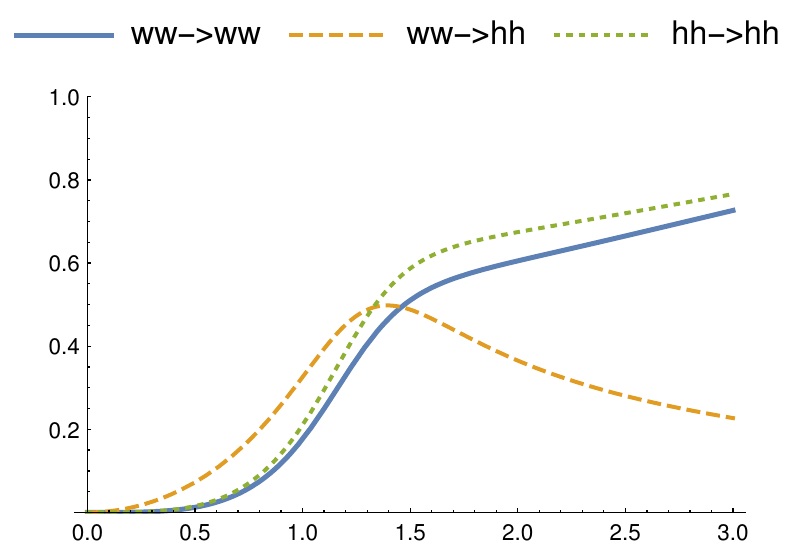}
\includegraphics[width=0.32\textwidth]{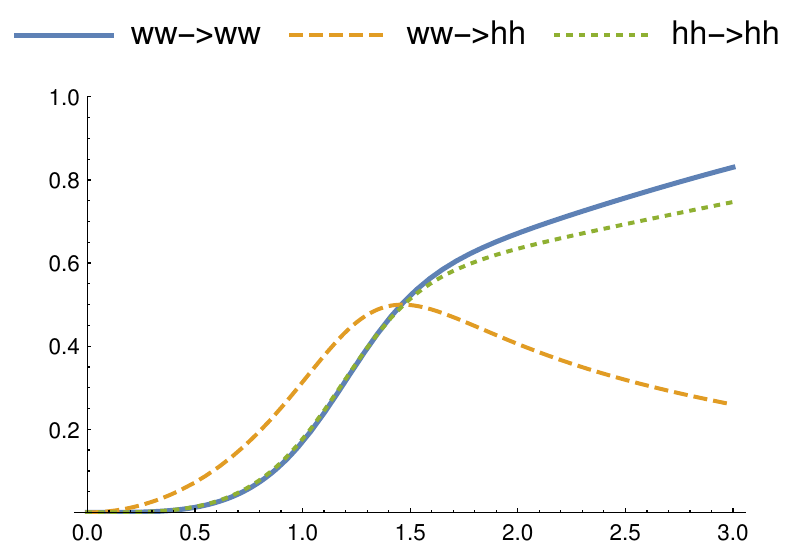}
\includegraphics[width=0.32\textwidth]{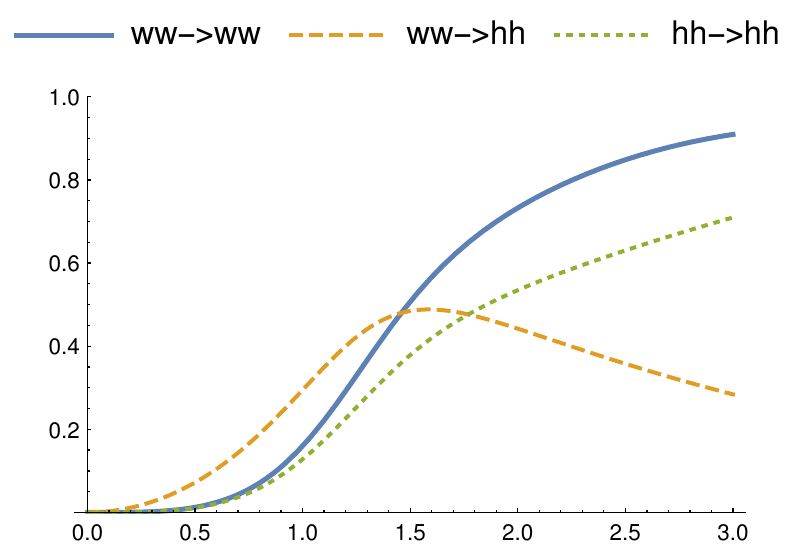}
\caption{\label{fig:ga}%
Dependence on $g$, that we find weak for natural values thereof. Displayed are the moduli of the lowest ($I=J=0$) IAM partial waves for $b=2\ne a^2=1$. Top panel: from left to right, $g=0.002$, $g=0.005$, $g=0.01$. Bottom panel: negative $g$ values of equal magnitude.}
\end{figure}
\begin{figure}
\null\hfill\includegraphics[width=0.42\textwidth]{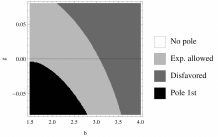}\hfill
\includegraphics[width=0.42\textwidth]{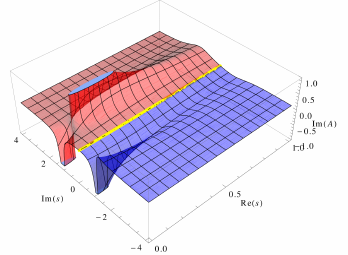}\hfill
\hfill\null
\caption{\label{fig:poles_b_g}%
Left: map of the $b$--$g$ parameter space, seeking poles in the
isocalar channel ($IJ=00$), with $a=1$ fixed and the remaining NLO parameters set to zero. (The isovector and isotensor channels have no poles in the region of interest, $\lvert s \rvert < (3\,{\rm TeV})^2$.) 
In the black regions there are two poles above and below the real axis on the first Riemann sheet, and we capture at least one with Cauchy's theorem, excluding the corresponding parameter swath. Right: explicit plot of these two poles for fixed parameter values $b=2.4$, $g=-0.08$
 (plotting again the imaginary part of the elastic $\omega\omega$ scattering).}
\end{figure}

Finally, we study the dependence of all amplitudes on the $g$ parameter (the only one that we have kept from the pure Higgs scattering sector, as it is needed to renormalize our amplitudes). 

It most directly produces an enhancement of $hh\to hh$ scattering that starts at NLO, as can be seen in figure~\ref{fig:ga}, since it comes from a $(\partial_\mu h\partial^\mu h)^2$ term in the effective Lagrangian. 

In figure~\ref{fig:poles_b_g} we study the parameter combination $a=1$ and $b>1.5$ together with a varying $g$, so we see the interplay of the channel coupling with the Higgs-sector dynamics.

We find a proper isoscalar pole on the second Riemann sheet for positive $g$. 
If either $g$ or $b$ are somewhat large, the isoscalar resonance enters the experimentally disfavored zone where LHC data are having an impact.

On the contrary, negative values of $g$ introduce a pole on the first Riemann sheet, so we must exclude those.

\clearpage

%%%%%%%%%%%%%%%%%%%%%%%%%%%%%%%%%%%%%%%%%%%%%%%%%%%%%
\section{Summary and discussion \label{sec:summary}}
%%%%%%%%%%%%%%%%%%%%%%%%%%%%%%%%%%%%%%%%%%%%%%%%%%%%%

In this article we have presented a thorough study of the unitarization of the 
Effective Lagrangian describing the Electroweak Symmetry Breaking Sector 
in the TeV region. 

The Effective Lagrangian in the massless limit has seven free parameters, namely $a$ and $b$ that respectively provide elastic $\omega\omega\to \omega\omega$ and cross-channel $\omega\omega\to hh$ strength at LO, and five more at NLO: the elastic $a_4$ and $a_5$ (inherited from the old Electroweak Chiral Lagrangian), $d$ and $e$ (that couple the two channels at NLO) and $g$ (in the pure $hh\to hh$ sector). This is the minimum number of parameters necessary to obtain a renormalized theory at NLO for massless $\omega$ and $h$ bosons. The parameter set, the combinations in which they appear, and the experimental reactions useful to extract them are summarized for convenience in table~\ref{tabl:param_extract}.

We have discussed five unitarization methods, aiming at classifying their respective strengths and weaknesses. We argued that three of them satisfy all desirable properties (describe several $IJ$ channels, produce unitary and analytic amplitudes, are independent of the renormalization scale, and agree with perturbation theory at low energy) and provided explicit constructions for them based on exact (elastic) dispersion relations. These are the Inverse Amplitude Method, that we have studied at length, the N/D method and the Improved K-matrix method, that we have also assessed. All three have been compared. 

The three methods are applicable to the $I=J=0$ coupled-channel partial wave, and to the exotic $I=2$, $J=0$ $\omega\omega$ channel. For any given set of parameters in the Lagrangian, the three methods are in qualitative agreement. In particular, they 
all produce a $\sigma$-like resonance when the interactions become strong, and the mass values obtained agree to within a few percent, which is quite remarkable and means that the model dependence is well controlled by imposing all the necessary theory constraints.

We have also unveiled a coupled-channel $f_0$-like scalar-isoscalar resonance that appears even if $a\simeq 1$ as long as $b$ is large enough (to provide coupled-channel strength). We have written a companion letter to this already long article highlighting this resonance. We only remark here that, though the LHC starts imposing relatively significant constraints on the $a$ parameter, it has not made much progress of substance in constraining $b$, so this coupled-channel resonance is one of the most interesting strongly interacting objects that can be sought for at the LHC run-II and beyond, because it may appear at relatively low-energies of $1\,{\rm TeV}$ or less (because of its somewhat large width).

In the $I=1=J$ channel (covering for example the $W'$ and $Z'$ bosons associated to Composite Higgs Models, as long as they be strongly coupled to $\omega\omega$) the IAM is the method of choice because the other two cannot be constructed in a renormalization-scale invariant way.

Finally for the two channels with $J=2$ (where in particular $f_2$-like resonances might appear, as well as exotic ones in $W^+W^+$ same-charge combinations) the IAM cannot be constructed with NLO amplitudes (because the lowest order is $s^2$ for these), but the other two methods do work and are in qualitative agreement.

We have provided extensive numerical analysis of all these amplitudes, for physical, real values of $s$, as well as into the complex $s$ plane. Therein we have searched for poles of the scattering amplitudes in their second Riemann sheet to be interpreted as resonances, as well as poles in the first Riemann sheet that exclude certain regions of parameter space. 
We have then drafted bidimensional maps of the parameter space showing whether the poles are likely to be excluded by LHC searchers, are in violation of causality, or are still viable resonances that can be searched for, and they agree with those in prior literature where available.

To conclude, we believe that we have made a substantive contribution to the discussion of possible strongly-interacting extensions of the Standard Model, which are the currently most natural scenario with the found particles $W$, $Z$ and $h$ in the Electroweak Symmetry Breaking sector. Thus we have extended previous works done long before the discovery of the $125\,{\rm GeV}$ Higgs like boson, that did not include it, as for example those in~\cite{Dobado:1990jy}. Of course, it can still be that the SM exhausts TeV-scale physics, in which case the parameters of the effective Lagrangian become $a=b=1$ (all the NLO ones vanishing). Or it can also be that some of them only slightly deviate from the SM values; this could be suggestive of weakly coupled resonances, as per Eq.~(\ref{couplingsfromresonances}) and the theory would be unitary far from saturation. 
But the strongly interacting regime remains the bulk of the parameter space to be explored by the LHC run-II.

%%%%%%%%%%%%%%%%%%%%%%%%%%%%%%%%%%%%%%%%%%%%%%%%%%%%%
\acknowledgments
%%%%%%%%%%%%%%%%%%%%%%%%%%%%%%%%%%%%%%%%%%%%%%%%%%%%%
The authors thank useful conversations with D. Espriu, M. J. Herrero, J. R. Pel\'aez and J. J. Sanz-Cillero. A. D. thanks the CERN TH-Unit for its hospitality during the time some important parts of this work were done. RLD thanks the hospitality of the NEXT institute and the high energy group at the University of Southampton. The work has been supported by the Spanish grants No. UCM:910309, MINECO:FPA2011-27853-C02-01, MINECO:FPA2014-53375-C2-1-P, and by the grant MINECO:BES-2012-056054 (RLD).

%%%%%%%%%%%%%%%%55

\begin{table}
{\footnotesize\renewcommand{\arraystretch}{1.4}%
\begin{tabular}{|c|c|rcrcr|c|c|} \hline
Parameter    & Combination & \multicolumn{5}{c|}{Simplest reactions} &
Expt. extraction & Resonance type \\\hline
$a$          & $1-a^2$      & $A_{00},A_{11},A_{20}$    &$\propto$& $(1-a^2)s$    & & & Low-E $W_L W_L$  s-wave & $\sigma$-like  \\\hline
$b$          & $a^2-b$      & $M_{0}$                   &$\propto$& $(a^2-b)s$    & &  & Low-E $hh$ s-wave & Coupled-channel $f_0$  \\\hline
$a_4$, $a_5$ & $2a_4+a_5$   & $A_{02}$                  &$\propto$&               &  & $[(\dots) + (\dots) (2a_4+  a_5)]s^2$  &  Low-E $W_L W_L$ d-wave& $f_2$-like \\
             &              & $A_{20}$                  &$\propto$& (\dots) $s$ &+& $[ (\dots)+ (\dots)(2a_4+  a_5)]s^2$ &  &Exotic $W^+ W^+$ \\\cline{2-7}
             & $a_4+2a_5$   & $A_{22}$                  &$\propto$&               & & $[ (\dots) +  (\dots) ( a_4+ 2a_5)]s^2$  &  & \\\cline{2-7}
             & $7a_4+11a_5$ & $A_{00}$                  &$\propto$& (\dots) $s$ &+& $[(\dots)+  (\dots)(7a_4+11a_5)]s^2$  &  & \\\cline{2-7}
             & $a_4-2a_5$   & $A_{11}$                  &$\propto$& (\dots) $s$ &+& $[ (\dots)+ (\dots)( a_4- 2a_5)]s^2$  & Low-E $W_L W_L$ d-wave & $\rho$-like \\\hline
$d$, $e$     & $d+\frac{e}{3}$       & $M_{0}$                   &$\propto$& (\dots) $s$ &+& $[(\dots)+ (\dots)(d+\frac{e}{3})]s^2$  & Low-E $W_L W_L$ s-wave & \\\cline{2-7}
             & $e$          & $M_{2}$                   &$\propto$&               & & $[(\dots)+(\dots)e ]s^2$  &  Low-E $W_L W_L$ d-wave& Coupled-channel $f_2$  \\\hline
$g$          & $g$          & $T_{0}$, $T_{2}$                   &$\propto$&               & & $[(\dots)+(\dots)g]s^2$ 
 & &  Elastic $hh$-$f_0$ \\ \hline
\end{tabular}}
\caption{\label{tabl:param_extract}%
Relevant combination of the free parameters $a$, $b$, $a_4$, $a_5$, $d$, $e$ and $g$, some useful reactions to extract them from the lowest order terms (i.e., $s$ and $s^2$) in a derivative expansion (as well as a few selected resonances with the appropriate quantum numbers for each channel). The numeric coefficients can be found in appendix~\ref{sec:partwaves}, so we gloss them over with ellipsis.}
\end{table}

%%%%%%%%%%%%%%%%%%%%%%%%%%%%%%%%%%%%%%%%%%%%%%%%%%%%%%%
\appendix

%%%%%%%%%%%%%%%%%%%%%%%%%%%%%%%%%%%%%%%%%%%%%%%%%%%%%%%
\section{Further details on the effective Lagrangian and 2-body scattering amplitudes}\label{app:L}
%%%%%%%%%%%%%%%%%%%%%%%%%%%%%%%%%%%%%%%%%%%%%%%%%%%%%%%

%%%%%%%%%%%%%%%%%%%%%%%%%%%%%%%%%%%%%%%%%%%%%%%%%%%%%%%
\subsection{Computation of the amplitudes\label{app:amplitude}}
%%%%%%%%%%%%%%%%%%%%%%%%%%%%%%%%%%%%%%%%%%%%%%%%%%%%%%%
From the Lagrangian in Eq.~(\ref{bosonLagrangian}) the following tree-level elastic $\omega \omega\to \omega\omega$ amplitude results,
\begin{equation} \label{Atree}
A^{(0)}(s,t,u) + A^{(1)}_{\rm tree}(s,t,u) = (1-a^2)\frac{s}{v^2} + \frac{4}{v^4}\left[2a_5 s^2 + a_4(t^2 + u^2)\right].
\end{equation}
The one-loop part computation, rather lengthy because of the number of Feynman diagrams, was automated (refs.~\cite{Alloul:2013bka,Hahn:1998yk,Kuipers:2012rf}), carried out, and reported in~\cite{Delgado:2013hxa}. We obtained
\begin{equation} \label{Aloop}
A^{(1)}_{\rm loop}(s,t,u) = \frac{1}{36 (4\pi)^2 v^4}[f(s,t,u)s^2 +(a^2-1)^2( g(s,t,u) t^2 + g(s,u,t) u^2)]
\end{equation}
with auxiliary functions
\begin{eqnarray}
f(s,t,u) &:=& 
[20 - 40 a^2 + 56 a^4 - 72 a^2 b + 36 b^2] \nonumber\\
& + &% 
[12 - 24 a^2 + 30 a^4- 36 a^2 b + 18 b^2] N_\varepsilon\nonumber\\  \label{ffunction}
& + & %
[-18 + 36 a^2 - 36 a^4 + 36 a^2 b - 18 b^2] \log\left(\frac{-s}{\mu^2}\right) \nonumber\\
& + & %
3 (a^2-1)^2 \left[\log\left(\frac{-t}{\mu^2}\right) +
\log\left(\frac{-u}{\mu^2}\right)\right] \\ \label{gfunction}
g(s,t,u) &:=& 
26 + 12 N_\varepsilon 
-9 \log\left[-\frac{t}{\mu^2}\right]
-3 \log\left[-\frac{u}{\mu^2}\right]
\end{eqnarray}
where in dimensional regularization $D=4-\epsilon$ the pole is contained in
\be
N_\epsilon =\frac{2}{\epsilon} + \log 4\pi -\gamma\ .
\ee
These results coincide with earlier published ones~\cite{Espriu:2013fia} taking the limit of vanishing light scalar mass there.

For the $\omega\omega\to hh$ amplitude we find, in analogy with Eq.~(\ref{loopexpansion}) and at tree level, 
\begin{equation} \label{Mtree}
M^{(0)}_{\rm tree}(s,t,u) + M^{(1)}_{\rm tree}(s,t,u)
= (a^2-b)\frac{s}{v^2}+ 
\frac{2 d}{v^4} s^2+ \frac{e}{v^4}(t^2+u^2)
\end{equation}
that takes a one-loop correction:
\begin{equation} \label{Mloop}
M^{(1)}_{\rm loop}(s,t,u) = \frac{a^2-b}{576\pi^2 v^2}\left[f'(s,t,u)\frac{s^2}{v^2}  +\frac{a^2 - b}{v^2}[ g(s,t,u)t^2 + g(s,u,t)u^2]\right]
\end{equation}
where
\begin{eqnarray}
f'(s,t,u) &=& 
-8 [-9 + 11 a^2 - 2 b]
-6  N_\varepsilon [-6 + 7 a^2 - b]  \\ \nonumber
&  + &
36 (a^2 - 1)\log\left[-\frac{s}{\mu^2}\right] +
3 (a^2 - b) \left(\log\left[-\frac{t}{\mu^2}\right] + 
\log\left[-\frac{u}{\mu^2}\right]\right) 
\end{eqnarray}
and the function $g$ is as defined in Eq.~(\ref{gfunction}).

Finally, the $hh\to hh$ elastic amplitude is, at tree-level and keeping only the operator necessary to renormalize the one-loop part,
\be \label{Ttree}
 T^{(0)}(s,t,u) + T^{(1)}_{\rm tree}(s,t,u)=
\frac{2g}{v^4}(s^2+t^2+u^2)\ ,
\ee
while the one-loop piece may be written in terms of only one function
\begin{equation}
T(s) = 2 + N_\varepsilon - \log\left(-\frac{s}{\mu^2}\right)
\end{equation}
as
\be \label{Tloop}
T^{(1)}_{\rm loop}(s,t,u) = \frac{3(a^2-b)^2}{2(4\pi)^2 v^4}\left[T(s)s^2 + T(t)t^2 + T(u)u^2\right] \ .
\ee

%%%%%%%%%%%%%%%%%%%%%%%%%%%%%%%%%%%%%%%%%%%%%%%%%%%%%%%%%%%%%%%%%%%%%%%%%%%%%%
\subsection{Renormalization of the amplitudes}\label{sec:renorm} 
%%%%%%%%%%%%%%%%%%%%%%%%%%%%%%%%%%%%%%%%%%%%%%%%%%%%%%%%%%%%%%%%%%%%%%%%%%%%%%

Comparing the tree-level amplitudes in Eqs.~(\ref{Atree}), (\ref{Mtree}), (\ref{Ttree}) with the loop ones in Eqs.~(\ref{Aloop}), (\ref{Mloop}), (\ref{Tloop}), we see that the divergences in the one-loop pieces can be absorbed just by redefining the couplings $a_4$, $a_5$, $g$, $d$ and $e$ from the NLO tree-level Lagrangian. Therefore no renormalizations of $a$, $b$, $v$, wave-functions nor (vanishing) masses are needed to obtain finite amplitudes (an advantage of dimensional regularization). 
Our amplitudes are quoted in the $\overline{MS}$ scheme, and the renormalized couplings are
\begin{eqnarray}
a_4^r & = & a_4+ \frac{N_\epsilon}{192\pi^2}(1-a^2)^2   \nonumber \\
a_5^r & = & a_5+\frac{N_\epsilon}{768 \pi^2} (2+5  a^4-4 a^2-6 a^2 b+3 b^2)\nonumber \\
g^r & = & g+\frac{3N_\epsilon}{64 \pi^2 }(a^2-b)^2   \nonumber  \\
d^r & = & d -\frac{N_\epsilon}{192 \pi^2  }(a^2-b)(7 a^2- b-6)  \nonumber  \\
e^r  & = & e+ \frac{N_\epsilon}{48 \pi^2 }(a^2-b)^2.
\end{eqnarray}
As a simple limit, the MSM ($a=b=1$) is renormalizable without any of these additional five couplings (we see that they are unnecessary in this case). The case of the Higgsless EWChL corresponds to  $a=b=0$ and then $g$, $d$ and $e$ do not need any renormalization. We also reproduce the well known results for the constants $a_4$ and $a_5$ \cite{Appelquist}. In more generality, the renormalization of $a_4$ and $a_5$ agrees with \cite{Espriu:2013fia}.
 
The elastic WBGB amplitude reads, in terms of these renormalized couplings 
\begin{eqnarray} \label{Arenorm}
A(s, t, u) & = & \frac{s}{v^2} (1 - a^2)+\frac{ 4}{v^4} [2 a^r_5(\mu) s^2 + a^r_4(\mu) (t^2 + u^2)] \\  \nonumber    
& + &\frac{1}{16 \pi^2 v^4}\left(\frac{1}{9} (14 a^4 - 10 a^2 - 18 a^2 b  + 9 b^2 + 
        5 ) s^2
     + \frac{13}{18} (a^2 - 1)^2 (t^2 + u^2) \right.  \\ \nonumber  
       &  - & \frac{1}{2}  (2 a^4 - 2 a^2 - 
        2 a^2 b  + b^2 + 
        1)   s^2 \log\frac{-s}{\mu^2} \\  \nonumber   
       &  +& \frac{1}{12} (1-a^2 )^2 (s^2 - 3 t^2 - 
        u^2) \log\frac{-t}{\mu^2}        \\ \nonumber  
       &  + & \left.
  \frac{1}{12}   (1-a^2 )^2 (s^2 - t^2 - 3 u^2) \log\frac{-u}{\mu^2}
    \right)
\end{eqnarray}
While the inelastic $\omega\omega  \rightarrow h h$ amplitude is
\begin{eqnarray} \label{Mrenorm}
M(s,t,u)  & = & \frac{a^2-b}{v^2}s  + \frac{2 d^r(\mu)}{v^4}s^2+ \frac{e^r(\mu)}{v^4}(t^2+u^2) \nonumber  \\
& + & 
\frac{(a^2-b)}{576\pi^2v^4}
\left\{\left[72 -  88 a^2+ 16 b  + 36 (a^2-1)\log\frac{-s}{\mu^2}\right.\right. \nonumber \\
& + &
\left.\left. 3  (a^2-b)\left(\log\frac{-t}{\mu^2}+\log\frac{-u}{\mu^2}\right)\right]s^2 \right. \nonumber  \\
& + &  (a^2-b)\left(26-9\log\frac{-t}{\mu^2}-3\log\frac{-u}{\mu^2}\right)t^2  \nonumber  \\
& + & \left. (a^2-b)\left(26-9\log\frac{-u}{\mu^2}-3\log\frac{-t}{\mu^2}\right)u^2 \right\}
\end{eqnarray}
and finally the $h h \rightarrow  h h$ amplitude may be written as
\begin{eqnarray} \label{Trenorm}
T(s,t,u) & = & \frac{2g^r(\mu)}{v^4}(s^2+t^2+u^2) \\ \nonumber
 &+& 
\frac{3(a^2-b)^2 }{32\pi^2v^4}
\left[ 2(s^2+t^2+u^2)-s^2\log\frac{-s}{\mu^2}-t^2\log\frac{-t}{\mu^2}
-u^2\log\frac{-u}{\mu^2}\right]\ .
\end{eqnarray}
Apparently, Eqs.~(\ref{Arenorm}), (\ref{Mrenorm}) and (\ref{Trenorm}) depend on the renormalization scale $\mu$ through the logarithmic terms. But they also depend on this arbitrary $\mu$ through the renormalized couplings $a_4\dots e$. 

However, in the absence of wave or mass renormalization, the amplitudes must be observable, and hence $\mu$-independent; then we may require that their total derivatives with respect to $\log\mu^2$ vanish. Integrating the resulting (very simple) differential equations, we find the renormalization-group evolution equations for the different couplings that allow to change the scale
\begin{eqnarray} \label{RGE}
a_4^r (\mu)& = & a_4^r(\mu_0)- \frac{1}{192 \pi^2}(1-a^2)^2    \log\frac{\mu^2}{\mu_0^2}   \nonumber \\
a_5^r(\mu) & = & a_5^r(\mu_0)- \frac{1}{768 \pi^2}\left[3(a^2-b)^2+2(1-a^2)^2\right] \log\frac{\mu^2}{\mu_0^2}\nonumber \\
g^r(\mu) & = & g^r(\mu_0)-\frac{3}{64\pi^2}(a^2-b)^2  \log\frac{\mu^2}{\mu_0^2}  \nonumber  \\
d^r(\mu) & = & d^r(\mu_0) +\frac{1}{192 \pi^2}(a^2-b)\left[(a^2-b)-6(1-a^2)\right]  \log\frac{\mu^2}{\mu_0^2} \nonumber  \\
e^r(\mu)  & = & e(\mu_0)- \frac{1}{48 \pi^2}(a^2-b)^2 \log\frac{\mu^2}{\mu_0^2}\ .
\end{eqnarray}
These equations are diagonal, so the scale evolution does not mix the couplings at NLO in perturbation theory.
The $\mu$-invariance of all the amplitudes has been checked by
substituting the $\mu$-evolution of the renormalized couplings in Eq.~(\ref{RGE}) into their explicit expressions. 

\begin{figure}
\null\hfill
\includegraphics[width=0.42\textwidth]{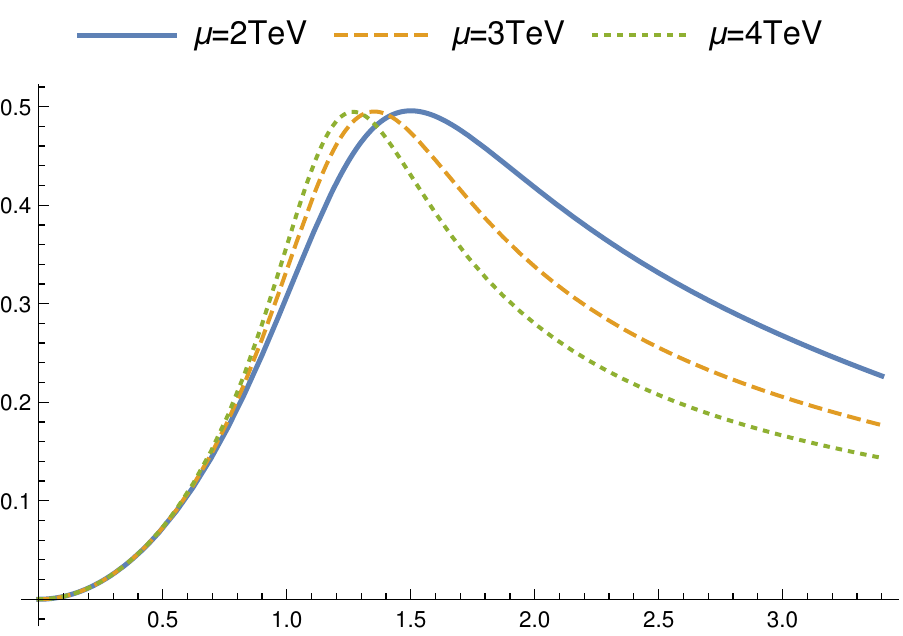}
\hfill\null\\
\caption{\label{fig:depMu}%
The dependence on the renormalization scale is absorbed throughout in the NLO coefficients. But instead of varying them at fixed scale, we can also take the coefficients as fixed (here $a^2=1$, $b=2$ and all the other NLO parameters set to zero) and show the dependence on the election of $\mu$. We take for this example the absolute value of the isoscalar amplitude ($I=J=0$). There is no qualitative difference in adopting one or another scale. So we have used $\mu=3$ TeV throughout the paper.}
\end{figure}

From a practical point of view, we have adopted the values of $a_4^r,\dots,e^r$ to be as quoted for each example in the manuscript at a scale of $\mu=3\,{\rm TeV}$. The dependence on $\mu$ is shown in fig.~(\ref{fig:depMu}) for the $I=J=0$ case, and seen to be rather moderate. Indeed for $a=0.95$, the prefactor of the first Eq.~(\ref{RGE}), say, is $\simeq 5\times 10^{-6}$, so that the scale dependence is small.

%%%%%%%%%%%%%%%%%%%%%%%%%%%%%%%%%%%%%%%%%%%%%%%%%%%%%%%
\subsection{Detailed partial waves}\label{sec:partwaves}
%%%%%%%%%%%%%%%%%%%%%%%%%%%%%%%%%%%%%%%%%%%%%%%%%%%%%%%
Evaluating the partial-wave projection integral in Eq.~(\ref{Jprojection}) by substituting the renormalized amplitude obtained in Eq.~(\ref{Arenorm}) for  $\omega\omega\to\omega\omega$  
provides us with $K$, $D$, $E$ constants and $B(\mu)$ functions.  
 
For the scalar-isoscalar channel with $IJ=00$, the results of~\cite{Delgado:2013hxa}, in terms of the coefficients $a$, $b$, $v$ instead of $\alpha$, $\beta$, $f$, read 
\begin{eqnarray}{}\label{partial00}
\nonumber   K_{00} & = & \frac{1}{16 \pi v^2} (1-a^2) \\
 \nonumber   B_{00}(\mu) & = &\frac{ 1}{9216 \pi^3 v^4}
  \left[101(1-a^2)^2 + 68(a^2-b)^2 + 
   768 \{7 a_4(\mu) + 11 a_5(\mu)\}\pi^2\right]\\
 \nonumber   D_{00} & = & -\frac{1}{4608\pi^3v^4}
  \left[7(1-a^2)^2 + 3(a^2-b)^2\right]\\
 E_{00} & = & -\frac{1}{1024\pi^3v^4} \left[4(1-a^2)^2  + 3(a^2-b)^2\right]\ .
\end{eqnarray}
For the vector isovector $IJ=11$ amplitude,
\begin{eqnarray}{}\label{partial11}
\nonumber   K_{11} & = & \frac{1}{96 \pi v^2} (1-a^2) \\
 \nonumber   B_{11}(\mu) & = & \frac{1}{110592\pi^3 v^4}
  \left[8(1-a^2)^2 - 75(a^2-b)^2 + 
   4608 \{a_4(\mu) - 2 a_5(\mu)\} \pi^2 \right] \\
 \nonumber   D_{11} & = & \frac{1}{9216\pi^3v^4}
  \left[(1-a^2)^2 + 3(a^2-b)^2\right] \\
  E_{11} & = & -\frac{1}{9216\pi^3v^4}  (1-a^2)^2\ .
\end{eqnarray}
For the scalar isotensor $IJ=20$:
\begin{eqnarray}{}\label{partial20}
\nonumber   K_{20} & = & -\frac{1}{32 \pi v^2} (1-a^2) \\
 \nonumber   B_{20}(\mu) & = & \frac{1}{18432 \pi^3 v^4}
 \left[91(1-a^2)^2 + 28(a^2-b)^2 +
  3072 \{2 a_4(\mu) + a_5(\mu)\} \pi^2 \right] \\
 \nonumber   D_{20} & = & -\frac{1}{9216\pi^3v^4}
  \left[11(1-a^2)^2 + 6(a^2-b)^2\right]\\
  E_{20} & = & -\frac{1}{1024\pi^3v^4}(1-a^2)^2  \ 
\end{eqnarray}
and for the tensor isoscalar $IJ=02$,
\begin{eqnarray}{}\label{partial02}
\nonumber   K_{02} & = & 0 \\
 \nonumber   B_{02}(\mu) & = & \frac{1}{921600 \pi ^3 v^4}
 \left[320(1-a^2)^2 + 77(a^2-b)^2 +
  15360\{2 a_4(\mu) + a_5(\mu)\} \pi^2 \right] \\
 \nonumber   D_{02} & = & -\frac{1}{46080\pi^3v^4}
  \left[10(1-a^2)^2 + 3(a^2-b)^2\right] \\
  E_{02} & = & 0\ .
\end{eqnarray}

Next we quote a new calculation of the tensor-isotensor $I=J=2$ partial wave, that to our knowledge has not been reported in the literature. 
\begin{eqnarray}{}\label{partial22}
\nonumber   K_{22} & = & 0 \\
 \nonumber   B_{22}(\mu) & = & \frac{1}{921600 \pi ^3 v^4}
 \left[71(1-a^2)^2 + 77(a^2-b)^2 +
  7680\{a_4(\mu) + 2 a_5(\mu)\} \pi^2 \right] \\
 \nonumber   D_{22} & = & -\frac{1}{46080\pi^3v^4}
  \left[4(1-a^2)^2 + 3(a^2-b)^2\right] \\
  E_{22} & = & 0\ .
\end{eqnarray}
This exhausts the list of elastic partial waves that are non-vanishing at NLO in perturbation theory, since those with angular momentum $J=3$ and higher start at $O(s^3)$ and are NNLO in the derivative counting. Needless to say, they would be tiny at LHC energies.

We now give the equivalent results for the inelastic channel-coupling:
 $\omega\omega\to hh$, with partial waves $M_{J}$, starting by the scalar one,
\begin{eqnarray}{} \label{Mscalar}
\nonumber   K'_{0} & = & \frac{\sqrt{3}}{32 \pi v^2} (a^2-b) \\
 \nonumber   B'_{0}(\mu) & = & \frac{\sqrt{3}}{16\pi v^4} \left[d(\mu)+\frac{e(\mu)}{3}\right]
    +\frac{\sqrt{3}}{18432\pi^3 v^4}(a^2-b)\left[72(1-a^2) + (a^2-b)\right]    \\
 \nonumber   D'_{0} & = & - \frac{\sqrt{3}(a^2-b)^2}{9216\pi^3 v^4}   \\  
  E'_{0} & = &  -\frac{\sqrt{3}(a^2-b)(1-a^2)}{512\pi^3 v^4}                                   
\end{eqnarray}
while for the tensor $M_{2}$ channel
\begin{eqnarray}{} \label{Mtensor}
\nonumber   K'_{2} & = & 0 \\
 \nonumber   B'_{2}(\mu)& = & \frac{e(\mu)}{160\sqrt{3}\pi v^4}  +\frac{83(a^2-b)^2}{307200\sqrt{3}\pi^3 v^4}    \\
 \nonumber   D'_{2} & = & - \frac{(a^2-b)^2}{7680\sqrt{3}\pi^3 v^4} \\ 
  E'_{2} & = &  0\  .                                
\end{eqnarray}

At last we quote the elastic  $hh\to hh$ channel amplitude. The $T_{0}(s)$ scalar partial-wave is given by the set of constants
\begin{eqnarray}{}\label{Tscalar}
\nonumber   K''_{0} & = & 0 \\
 \nonumber   B''_{0}(\mu) & = & \frac{10g(\mu) }{96\pi v^4}   +  \frac{(a^2-b)^2}{96\pi^3 v^4}   \\
 \nonumber   D''_{0} & = & - \frac{(a^2-b)^2}{512\pi^3 v^4}  \\ 
  E''_{0} & = &  -   \frac{3(a^2-b)^2}{1024\pi^3 v^4}   
\end{eqnarray}{}  
while the tensor $T_{2}$ requires
\begin{eqnarray} \label{Ttensor}
\nonumber   K''_{2} & = & 0 \\
 \nonumber   B''_{2}(\mu) & = & \frac{g(\mu) }{240\pi v^4}   +  \frac{77(a^2-b)^2}{307200\pi^3 v^4}   \\
 \nonumber   D''_{2} & = &  -\frac{(a^2-b)^2}{5120\pi^3 v^4}  \\ 
  E''_{2} & = &    0 \ .
\end{eqnarray}         
By using the evolution equations it is possible to check that all the obtained partial waves
are $\mu$ independent.
% \newpage

%%%%%%%%%%%%%%%%%%%%%%%%%%%%%%%%%%%%%%%%%%%%%%%%%%%%%%%
\section{Coupled-channel Inverse Amplitude Method}\label{app:IAM}
%%%%%%%%%%%%%%%%%%%%%%%%%%%%%%%%%%%%%%%%%%%%%%%%%%%%%%%

In this appendix we show how to extend the IAM method to the two-body coupled-channel problem \emph{when all the particle species in the various channels are massless}. Otherwise this cannot be done because of the presence of overlapping cuts. This is well known to happen in the $\pi\pi$ and $K \bar K$ system (see for example~\cite{pipiKK}) where the $K \bar K \rightarrow K \bar K$ left cut terminates at $s=4 M_K^2 -4 M_\pi^2$, which is beyond the $\pi\pi \rightarrow \pi\pi$ threshold branching point located at $s=4 M_\pi^2$ where the $\pi\pi$ RC starts. Thus, the two cuts overlap.

In effective theories we can develop the coupled reaction matrix $F(s)=\{F_{ij}(s)\}$ according to the chiral/derivative expansion
\be
F_{ij}(s)=F^{(0)}_{ij}(s)+F^{(1)}_{ij}(s)+\dots
\ee
where $i$ and $j$ are channel subindices (e.g. $i,j=\omega\omega, hh,..$) but we have omitted the isospin and momentum indices $(I, J)$.
As the interactions are assumed to be time-reversal invariant $F_{ij}=F_{ji}$. Also in the physical region, i.e. on the RC we have:
\be \label{coupledUnitarity}
\Imag F_{ij}= \sum_l F_{il}(F_{lj})^*  
\ee
since coupled-channel unitarity requires the imaginary part of a generic partial wave to receive contributions from all allowed intermediate channels. This equation can be written in slightly 
more compact form as: $\Imag F = F F^\dagger = F^\dagger F$. Now, by using $\Imag F^{-1}= -(F^\dagger)^{-1}\Imag  F F^{-1}$, the unitarity condition on the RC can be written as:
\be
\Imag F^{-1}= -1.
\ee
However in the effective theory this condition is only satisfied perturbatively, so that at one-loop precision
\be \label{pertunit}    
\Imag F_{ij}^{(1)}= \sum_l F_{il}^{(0)}\left(F_{lj}^{(0)}\right)^*\ .
\ee
As we saw in Eq.~(\ref{expandpartialwave}),
 the lowest-$IJ$  NLO partial-waves take the general form
\be
 F^{\rm NLO}_{ij}(s)=F^{(0)}_{ij}(s)+F^{(1)}_{ij}(s) =K_{ij}s + s^2\left( B_{ij}+D_{ij}\log\frac{s}{\mu^2}+E_{ij}\log\frac{-s}{\mu^2}\right)
\ee       
so that the perturbative unitarity of Eq.~(\ref{pertunit}) on the physical RC requires
\be
E_{ij}= -\frac{1}{\pi}\sum_l K_{il}K_{lj}\ .
\ee
Now, by following the same steps as in the single channel case in section~\ref{sec:IAM} we can obtain a twice-subtracted DR for the $F_{ij}^{\rm NLO}(s)$. Next we introduce the {\it inverse amplitude} matrix function as:
\be \label{imG}
W = F^{(0)}F^{-1}F^{(0)}.
\ee
The essential point here is that, as all the particles are massless, the analytical structure of all the matrix elements $F_{ij}(s)$ is the same, namely a LC and a RC starting at the origin. This structure is also shared by each of the $F^{-1}_{ij}(s)$ and $W_{ij}(s)$ matrix elements. Had the masses of the particles appearing in the various channels, and consequently the cut structure, been different, the $W_{ij}(s)$ matrix elements would mix and possibly overlap the different left and right cuts. This would produce spurious contributions to the imaginary part of the partial-waves in the physical region. Considering only massless particles ensures that we will not have this kind of spurious contributions. Extracting the imaginary part from Eq.~(\ref{imG}) on the RC we obtain:
\be 
\Imag W  = F^{(0)} \Imag F^{-1}F^{(0)}=-  F^{(0)} F^{(0)}= - \Imag F^{(1)} 
\ee
where we have used Eq.~(\ref{pertunit}). Then we have:
\be 
\Imag W_{ij}(s)= - \Imag F^{(1)}_{ij}(s)=-E_{ij}\pi s^2\ .
\ee
By using it in a twice-subtracted DR for each $W_{ij}(s)$ matrix element, and assuming that no poles appear when inverting the $F(s)$ matrix, we obtain on the LC at NLO:
\be
\Imag W \simeq - \Imag F^{(1)},
\ee 
as we did in the single channel case. Thus  we finally find:
\be
W_{ij } (s) \simeq K_{ij}s - s^2\left( B_{ij}(\mu)+D_{ij}\log\frac{s}{\mu^2}+E_{ij}\log\frac{-s}{\mu^2}\right)=
F^{(0)}_{ij}(s)-F^{(1)}_{ij}(s)
\ee
and then we arrive to the IAM formula for massless coupled channels in Eq.~(\ref{IAMforF}):
\be
 F^{\rm IAM}= F^{(0)}\left(F^{(0)}-F^{(1)}\right)^{-1}F^{(0)}.
\ee
As discussed already in subsection~\ref{subsec:coupledIAM} this matrix is exactly unitary, 
i.e. $\Imag  F^{\rm IAM} = F^{\rm IAM}\left(F^{\rm IAM}\right)^\dagger$ on the RC and it is also compatible with the NLO approximation as
\be
F^{\rm IAM}_{ij}(s)=   F^{(0)}_{ij}(s)+F^{(1)}_{ij}(s)+ O(s^3).
\ee
In addition all the elements have the same proper analytical structure (left and right cuts). This makes the analytical continuation to the second Riemann sheet of the different amplitudes possible, and eventually the presence of poles there, that could be understood as dynamical resonances for some regions of the parameter space.

Once again, this construction is possible only because all the particles are assumed to be massless. This is a good approximation because we are using the ET and the Landau gauge in our computations and also because the physical $h$ mass $M_h \simeq 125\,{\rm GeV}$ is close to $M_W$ and $M_Z$. However, if we had taken $M_h$ different from zero the $hh\to hh$ channel would have had a LC ending at the positive value $4 M_h^2$. In this situation the IAM multichannel method produces a spurious imaginary part in the physical region of the channel $\omega\omega\to\omega\omega$ ranging from $s=0$ to $s=4M^2_h$ thus spoiling unitarity in that  region.

%%%%%%%%%%%%%%%%%%%%%%%%%%%%%%%%%%%%%%%%%%%%%%%%%%%%%%%%%%%%%%%%%%%%%%%%%%%%%%
\section{The N/D solution}\label{app:NoverD}
%%%%%%%%%%%%%%%%%%%%%%%%%%%%%%%%%%%%%%%%%%%%%%%%%%%%%%%%%%%%%%%%%%%%%%%%%%%%%%

In this appendix we detail some computations needed to construct an approximation to the N/D method in section~\ref{sec:NoverD}. The dispersion relation for the denominator $D_0$ in Eq.~(\ref{NsobreD}), under the approximation $N_0(s)\simeq A^{(0)}(s)+A_L(s)$, results in
\be
D_0(s)  =  1+ h_1 s + h_2 s^2- \frac{s^3}{\pi}\left(KI_{2}+\frac{D}{D+E}B(\mu)I_{1}+DI'_{1}\right)
\ee
where the IR and UV regularized $I_n$ and  $I'_n$ integrals are respectively defined as
\be \label{integralcitas1}
 I_n(s; m,\Lambda)=\int_{m^2}^{\Lambda^2}  \frac{ds'}{s'^n(s'-s-i\epsilon)}
\ee
and
\be \label{integralcitas2}
 I'_n(s; m,\Lambda,\mu)=\int_{m^2}^{\Lambda^2}  \frac{ds' \log\frac{s'}{\mu^2}}{s'^n(s'-s-i\epsilon)} \ .
\ee
These integrals can be computed and are UV convergent thanks to $n$ being positive because of the subtractions in the dispersion relation. Thus, taking $\Lambda\to \infty$ directly, one finds 
\begin{eqnarray}{}
\nonumber
I_{2}(s;m,\infty)  & = & -\frac{1}{s^2}\left[\frac{s}{m^2}+\log\left(1-\frac{s}{m^2}\right)\right]
\\
\nonumber
I_{1}(s;m,\infty)  & = & -\frac{1}{s}\log\left(1-\frac{s}{m^2}\right)
\\
I'_{1}(s;m,\infty,\mu)  & = & \frac{1}{s}\left[
-\frac{1}{2}\log ^2 \left(\frac{-s}{\mu^2}\right)  -\frac{\pi^2}{3}
-\log \frac{m^2}{\mu^2}\log\left(1-\frac{m^2}{s}\right)
-Li_2\left(\frac{m^2}{s}\right)+\frac{1}{2}\log ^2 \frac{m^2}{\mu^2}\right]
\end{eqnarray}
where $Li_2(\eta)$ is the dilogarithm  function. Therefore we can write:
 \be
A(s)\simeq A_{\rm N/D}(s)\simeq \frac{A^{(0)}(s)+A_L(s)}{1+h_1s+h_2s^2-\frac{s^3}{\pi}T(s)}
\ee 
where:
\be 
T(s)=KI_{2}+\frac{B(\mu)}{D+E}DI_{1}+DI'_{1}. 
\ee 
Now it is not difficult to check that for small $m$
\be
D_0(s)=1+h_1s+\frac{A^{(0)}(s)}{\pi}\log\left(\frac{-s}{m^2}\right)+O(s^2)
\ee
so that matching the dispersion relation to perturbation theory sets the $h_1$ subtraction constant, the correct choice being
\be
h_1=h_1(m)=\frac{K}{\pi(D+E)} B(m)
\ee
so we have
\be
D_0(s)=1-\frac{A_R(s)}{A^{(0)}(s)}+O(s^2)
\ee
and then
\be
 A_{\rm N/D}(s)=  \frac{N_0(s)}{D_0(s)}=A^{(0)}(s)+A_L(s)+A_R(s)+O(s^3)
\ee
which reproduces the NLO computation. From the integrals above in Eq.~(\ref{integralcitas1}) and~(\ref{integralcitas2}) it is also  not difficult to show that for small enough IR cutoff  $m$:
\be 
\Imag T(s) = \frac{\pi}{s^3}\left[A^{(0)}(s)+A_L(s)\right]
\ee
on the RC so that the $A_{\rm N/D}(s)$ partial waves fullfil exact elastic unitarity.

In order to have a clearer mathematical description of the amplitude obtained it is useful to introduce an additional subtraction constant. Thus we define 
\be
H=H(m)\equiv h_2(m)\pi+\frac{K}{m^2}+D\frac{\pi ^2}{3}\ .
\ee
Then it is very easy to show that 
\be
D_0(s)=1-\frac{A_R(s)}{A^{(0)}(s)}+\frac{s^2}{\pi}\left[H(m)+\frac{D}{D+E}B(\mu) \log \frac{-s}{m^2}+\frac{D}{2}\left(\log^2\frac{-s}{\mu^2}-\log^2\frac
{m^2}{\mu^2}\right)\right] .
\ee
As usual $ H=H(m)$ must be considered a renormalized parameter at the scale $m$.
By demanding $D_0(s)$ to be independent of this scale we find the renormalization equation:
\be
m^2\frac{dH(m)}{dm^2}=\frac{D}{D+E}B(m) 
\ee
which upon integration leads to an evolution equation characteristic of an NNLO parameter in perturbation theory,
\be \label{evolutionh}
H(\mu)=H(\mu_0)+\frac{D}{D+E}B(\mu_0)\log\frac{\mu^2}{\mu^2_0}+\frac{D}{2}\log^2\frac{\mu^2}{\mu_0^2}\ .
\ee
Finally we can remove the IR cutoff $m$ from $D_0(s)$ to find the $\mu$ independent equation:
\be \label{approxD0}
D_0(s)=1-\frac{A_R(s)}{A^{(0)}(s)}+\frac{s^2}{\pi}\left[H(\mu)+\frac{D}{D+E}B(\mu)\log
\frac{-s}{\mu^2}+\frac{D}{2}\log^2\frac {-s}{\mu^2}\right]. 
\ee 
The first term, 1 here, corresponds to LO in perturbation theory (the $s$ power being contained in the numerator $N$). The $A_R$ term contains the NLO physics, and finally the method has generated an NNLO piece that is necessary to have the correct analytic properties. Thus the renormalized constant $H(\mu)$ can contain contributions from the NNLO chiral couplings. However it is possible to neglect these contributions in a consistent way by choosing:
\be
H(\mu)=\left(\frac{B(\mu)}{D+E}\right)^2\frac{D}{2}
\ee 
which, at it is easy to check, satisfies the above evolution equation. With this choice the partial wave denominator takes the simpler form:
\be
D_0(s)=1-\frac{A_R(s)}{A^{(0)}(s)}+\frac{1}{2}\pi [g(s)]^2 D s^2 \ ,
\ee
which we have used in the main text [Eq.~(\ref{N/D__D0_g})]. By using the the $A_L(s)$ and $A_R(s)$ definitions this denominator can also be written as:
\be
D_0(s)=1-\frac{A_R(s)}{A^{(0)}(s)}-\frac{A_L(-s)A_R(s)}{2(A^{(0)})^2}.
\ee
Notice that this denominator has not any LC as it must be the case.

%%%%%%%%%%%%%%%%%%%%%%%%%%%%%%%%%%%%%%%%%%%%%%%%%%%%%%%%%%%%%%%%%%%%%%%%%%%%%%
\section{Numerical extraction of coupled-channel poles in complex \boldmath $s$}
\label{app:numeric}
%%%%%%%%%%%%%%%%%%%%%%%%%%%%%%%%%%%%%%%%%%%%%%%%%%%%%%%%%%%%%%%%%%%%%%%%%%%%%%

Here we describe very briefly the numeric finding of resonances and their parameter extraction as poles of the amplitude in the complex-$s$ plane, and also to assess when violations of causality occur (finding poles in the first Riemann sheet instead of the second).

An accurate method is the use of a Cauchy line integral around a closed path; a finite value indicates that some pole has been enclosed. The difficulty comes from having two coupled-channels, though it is not severe since we have taken all particles as massless so that the cuts of the two channels start at the same point (the origin in the complex $s$ plane).

We find convenient to use an integration contour shaped as a half-circle out at a radius 
$R=\left(3\,{\rm TeV}\right)^2$ (roughly the range of validity of the unitarization methods considered in this work), %Integral de contorno en s, con s en TeV^2
closed by a segment of the imaginary axis, and parametrized in terms of a dummy integration variable $t$,
\be\label{ec_integr_reg}
\gamma(t) = %
\begin{cases}
R\exp\left[i\left(\frac{\pi (2t-1)}{2}\right)\right] & \mbox{ for } t\in [0,1] \\
iR(3-2t) & \mbox{ for } t\in (1,2]
\end{cases},
\ee
Cauchy's theorem states that  if a function $A(s)$ has $N$ poles on points $s_i$ ($i=1,\dots,N$) within the region enclosed by $\gamma$, the value of the line integrals 
\be\label{ec_integr}
\int_{\gamma(t)} dt\, A(t) t^k = 2\pi i\sum_{i=1}^N s_i^k A_0(s_i),
\ee
is given by the respective pole residues $A_0(s_i)$ (there is no reason to expect double poles in our NLO-based computation). The $A(s)$ function generically stands for any of the considered scattering amplitudes. Since the low-energy perturbative interactions are weak, we do not look for bound states and thus restrict $\Real s\geq 0$. For $\Imag s>0$, the logarithms in $A(s)$ are evaluated on the first Riemann sheet; for $\Imag s<0$, on the second.

Now, the circuit in Eq.~\ref{ec_integr} is taken on the second Riemann sheet. Thus, it immediately captures all poles on its lower half-plane and also on the upper half-plane (common to both first and second sheets). Poles in the lower half plane of the \emph{first} Riemann sheet are outside the contour. Still, we can detect them because they occur simultaneously with a pole on the upper half plane, as we now argue.

Because of the analytical properties of scattering amplitudes $A_I(s)$ on the first Riemann sheet (analyticity on the upper half plane plus cut along the positive real-$s$ axis), Schwarz reflection applies and $\left[A_I(s^*)\right]^*=A_I(s)$. So, every pole on the first Riemann sheet below the real axis ($\Imag s<0$) implies the presence of a pole at $s'=s^*$ over the real axis ($\Imag s'>0$).

In contrast, by definition, in the second Riemann sheet $A(s)$ will be analytic on $\Imag s=0,\,\Real s > 0$, so the pole on the lower-half plane of the second sheet does not reflect on the upper half-plane.

Thus, the path given by Eq.~(\ref{ec_integr_reg}) is sufficient to detect all poles generated with the IAM method within its range of validity, \emph{on both the first and second Riemann sheets} (those with $\Imag s<0$ are on the second, those with $\Imag s>0$ tag a pair on the first sheet, respectively).

For each studied parameter set, three integrals ($I_k$) of the family in Eq.~\ref{ec_integr} have been computed, with $k=0,\,1,\,2$. 
If no pole lies inside the contour, the value of all these integrals is zero.
Next, if we have only one pole at position $\tilde{s}$, we can equate two ratios of these $I_k$ integrals,
\be
 \tilde{s} = \frac{I_1}{I_0},\,\tilde{s}^2=\frac{I_2}{I_0}\Longrightarrow I_1^2 = I_0 I_2\ .
\ee
For a larger contained-pole count, $N>1$ (\emph{e.g.} one on the first and one on the second Riemann sheets), it wouldn't be generally true that $I_1^2=I_0 I_2$, so we can use this relation as a check of whether there is exactly one pole there.
Should it fail, a more detailed study would be necessary. However, it would still be possible to compute the position of an arbitrary (but finite) number of poles by computing integrals of increasing order and solving the non--linear equation system $I_k = 2\pi i\sum_{i=1}^N s_i^k A_0(s_i)$. For the particular case of $N=2$, the expressions would still be analytic. And, in particular,
$$
s_{1,2} = \frac{(I_1 I_2 - I_0 I_3) \pm\sqrt{(I_1 I_2 - I_0 I_3)^2-4(I_1^2-I_0 I_2)(I_2^2-I_1 I_3)}}{2(I_1^2-I_0 I_2)}\ .
$$

We also record here some analytical expressions to find the location of 3 poles from the $I_k$, $k=0,\dots,5$ integrals. It is best quoted in terms of several auxiliary quantities, namely 
\ba
  \Delta &=& -I_3^3 + 2 I_2 I_3 I_4 - I_1 I_4^2 - I_2^2 I_5 + I_1 I_3 I_5 \\
  \Delta\cdot\hat{A} &=& -I_3^2 I_4 + I_2 I_4^2 + I_2 I_3 I_5 - I_1 I_4 I_5 - I_2^2 I_6 + I_1 I_3 I_6 \\
  \Delta\cdot\hat{B} &=& -I_3 I_4^2 + I_3^2 I_5 + I_2 I_4 I_5 - I_1 I_5^2 - I_2 I_3 I_6 + I_1 I_4 I_6 \\
  \Delta\cdot\hat{C} &=& -I_4^3 + 2 I_3 I_4 I_5 - I_2 I_5^2 - I_3^2 I_6 + I_2 I_4 I_6 \\
  \Lambda &=& -\hat{A}^2 \hat{B}^2 + 4 \hat{B}^3 + 4 \hat{A}^3\hat{C}-18\hat{A}\hat{B}\hat{C} + 27\hat{C}^2 \\
  \Gamma &=& \left( -2 \hat{A}^3 + 9\hat{A}\hat{B} - 27\hat{C} + 3\sqrt{3\Lambda} \right)^{1/3}\ .
\ea
Then the pole locations become
\ba
	s_1 &=& \frac{\hat{A}}{3}+\frac{2^{1/3}(3\hat{B}-\hat{A}^2)}{3\Gamma}-\frac{\Gamma}{3\cdot 2^{1/3}} \\
    s_2 &=& \frac{\hat{A}}{3}-\frac{3(1+i\sqrt{3})(\hat{B}+\hat{A}^2)}{3\cdot 2^{2/3}\Gamma}-\frac{(1-i\sqrt{3})\Gamma}{6\cdot 2^{1/3}} \\
    s_3 &=& \frac{\hat{A}}{3}-\frac{3(1-i\sqrt{3})(\hat{B}+\hat{A}^2)}{3\cdot 2^{2/3}\Gamma}-\frac{(1+i\sqrt{3})\Gamma}{6\cdot 2^{1/3}}\ .
\ea

%%%%%%%%%%%%%%%%%%%%%%%%%%%%%%%%%%%%%%%%%%%%%%%%%%%%%%

%%%%%%%%%%%%%%%%%%%%%%%%%%%%%%%%%%%%%%%%%%%%%%%%%%%%%%%%%%%%%%%%%%%%%%%%

\end{document}